\documentclass[journal,comsoc]{IEEEtran}
\usepackage{gensymb}
\usepackage{cite}
\usepackage{graphicx}
\usepackage{url}
\usepackage{amsmath}
\usepackage{latexsym,amssymb,epsfig,color,verbatim,bm}
\usepackage{algorithm}
\usepackage{algpseudocode}
\usepackage{pifont}
\usepackage{slashbox}
\usepackage[tight,footnotesize,center]{subfigure}
\usepackage{color}
\usepackage{cancel}
\usepackage{multirow}

\newcommand{\BEQA}{\begin{eqnarray}}
\newcommand{\EEQA}{\end{eqnarray}}

\newtheorem{rem}{{\bf Remark}}
\newtheorem{lemma}{\bf Lemma}
\newtheorem{theorem}{\bf Theorem}
\newtheorem{corollary}{\bf Corollary}
\newtheorem{definition}{\bf Definition}

\newtheorem{eg}{\bf Example}

\newenvironment{remark}[1][Remark]{\begin{trivlist}
\item[\hskip \labelsep {\bfseries #1}]}{\end{trivlist}}

\begin{document}

\title{Accurate Learning or Fast Mixing? \\Dynamic Adaptability of Caching Algorithms}

\author{ Jian Li$^*$, Srinivas Shakkottai$^\dagger$, John C.S. Lui$^\S$ and Vijay Subramanian$^\ddagger$\\ \small{${}^*$University of Massachusetts Amherst,  ${}^{\dagger}$Texas A\&M University, \\${}^{\S}$The Chinese University of Hong Kong, ${}^{\ddagger}$University of Michigan\\ Email: ${}^*$jianli@cs.umass.edu, ${}^{\dagger}$sshakkot@tamu.edu, ${}^{\S}$cslui@cse.cuhk.edu.hk, ${}^{\ddagger}$vgsubram@umich.edu}}

\maketitle

\begin{abstract}
Typical analysis of content caching algorithms using the metric of {steady state hit probability under a stationary request process} does not account for  performance loss under a variable request arrival process. In this work, we consider adaptability of caching algorithms from two perspectives: (a) the accuracy of learning a fixed popularity distribution; and (b) the speed of learning items' popularity.   In order to attain this goal, we compute the distance between the stationary distributions of several popular algorithms with that of a genie-aided algorithm that has knowledge of the true popularity ranking, which we use as a measure of learning accuracy. We then characterize the mixing time of each algorithm, i.e., the time needed to attain the stationary distribution, which we use as a measure of learning efficiency.  We merge both measures above to obtain the ``learning error'' representing both how quickly and how accurately an algorithm learns the optimal caching distribution, and use this to determine the trade-off between these two objectives of many popular caching algorithms.  Informed by the results of our analysis, we propose a novel hybrid algorithm, Adaptive-LRU (A-LRU), that learns both faster and better the changes in the popularity.   We show numerically that it also outperforms all other candidate algorithms when confronted with either a dynamically changing synthetic request process or using real world traces. 
\end{abstract}


%
\IEEEpeerreviewmaketitle

\section{Introduction}\label{sec:intro}

The dominant application in today's Internet is streaming of content such as video and music.  This content is typically streamed by utilizing the services of a content distribution network (CDN) provider such as Akamai or Amazon.  Streaming applications often have stringent conditions on the acceptable latency between the content source and the end-user, and CDNs use caching as a mean to reduce access latency and bandwidth requirements at a central content repository.  The fundamental idea behind caching is to improve performance by making information available at a location close to the end-user.   Managing a CDN requires policies to route requests from end-users to near-by distributed caches, as well as algorithms to ensure the availability of the requested content in the cache that is polled. 

While the request routing policies are optimized over several economic and technical considerations, they end up creating a request arrival process at each cache.  Caching algorithms attempt to ensure content availability by trying to learn the distribution of content requests in some manner.  Typically, the requested content is searched for in the cache, and if not available, a miss is declared, the content is then retrieved from the central repository (potentially at a high cost in terms of latency and transit requirements), stored in the cache, and served to the requester.  Since the cache is of finite size, typically much smaller than the total count of content, some content may need to be evicted in order to cache the new content, and caching algorithms are typically described by the eviction method employed.

Some well known content eviction policies are Least Recently Used (LRU) \cite{coffman73operating}, First In First Out (FIFO), RANDOM \cite{coffman73operating}, CLIMB \cite{coffman73operating, starobinski01}, LRU($\boldsymbol m$) \cite{gast16asym}, k-LRU \cite{martina2014},  and Adaptive Replacement Cache (ARC) \cite{megiddo03arc}; these will be described in detail later on.  {Performance analysis typically consists of determining the hit probability at the cache either at steady-state under a synthetic arrival process (usually with independent draws of content requests following a fixed Zipf popularity distribution, referred to as the Independent Reference Model (IRM)), or using a data trace of requests observed in a real system. }It has been noted that performance of an eviction algorithm under synthetic versus real data traces can vary quite widely \cite{martina2014}.  For instance, 2-LRU usually does better than LRU when faced with synthetic traffic, but LRU often outperforms it with a real data trace. The reason for this discrepancy is usually attributed to the fact that while the popularity distribution in a synthetic trace is fixed, real content popularity changes with time \cite{cha09analyzing,zink08}. Thus, it is not sufficient for a caching algorithm to learn a fixed popularity distribution accurately, it must also learn it \emph{quickly} in order to track the changes on popularity that might happen frequently.

Underlying restrictions on memory usage and computation forces caching algorithms to adopt a low complexity finite-state automata scheme. A key contribution of this paper is to argue that caching algorithm design should be viewed as the design of online distribution learning algorithms but using low complexity finite-state automata schemes. Note that the low complexity finite-state automata restriction precludes the use of complex dictionary constructions from universal source coding, the use of (dynamic) index policies from bandit problems or even the use of accurate empirical distribution estimation procedures.  {Additionally, as the complexity is held fixed (determined by cache size), no causal algorithm can learn perfectly, and hence, it places a distribution on all possible cache configurations. Therefore, the correct measure of absolute performance of an algorithm is the closeness of the distribution it assigns to the distribution of an ideal algorithm.}

\vspace{0.05in}
\noindent\textbf{\textit{Caching Algorithms as Markov Chains:} } Given the finite-state automata structure, each caching algorithm generates a Markov process over the occupancy states of the cache.  Suppose there are a total of $n$ content items in a library, and the cache size is $m<n$.  Then each state $\boldsymbol x$ is a vector of size $m$ indicating the content in each cache spot; we call the state space of all such vectors $\mathcal{S}.$  Each cache request generates a state transition based on the caching algorithm used via item entrances and evictions.  Hence, for a given request arrival process, a caching algorithm is equivalent to a state transition matrix over the cache states.  Since there is one state transition per request, time is discrete and measured in terms of the number of requests seen.

The typical performance analysis approach is then to determine the stationary distribution of the Markov process of occupancy states, and from it derive the hit probability.  However, this approach loses all notion of time, and also does not allow us to compare the performance of each algorithm with the best possible (assuming causal knowledge\footnote{ The provably optimal B\'{e}l\'{a}dy's algorithm \cite{belady96} uses the entire sequence of future requests, and is neither causal nor Markovian.} of the request sequence).  {A major goal of this paper is to define an \emph{error function} that captures the online distributional learning perspective, and hence accounts for both the error due to time lag of learning, as well as the error due to the inaccuracy of learning the popularity distribution}.    Such an error function would allow us {to understand better} the performance of existing algorithms, as well as {aid in developing} new ones.

\vspace{0.05in}
\noindent\textbf{\textit{{Main Contributions:}}}
{Our goal is to design a metric that accounts for both the accuracy of and the lag in learning.  To develop a good accuracy metric, we need to characterize the nearness of the stationary distribution of an algorithm to the best-possible cache occupancy distribution.  If the statistics of the cache request process are known, the obvious approach to maximizing the hit probability (without knowing the realization of requests) is to simply cache the most popular items as constrained by the cache size, creating a fixed vector of cached content (a point mass).  How do we compare the stationary distribution generated by a caching algorithm with this vector?  A well known approach to comparing distributions is to determine the Wasserstein distance between them \cite{villani08optimal}.  However, since we are dealing with distributions of  permutations of vectors, we need to utilize a notion of a cost that depends on the ordering of elements.  Such a notion is provided by a metric called the generalized Kendall's tau \cite{kumar10}.  Coupling these two notions together, we define a new metric that we call the ``$\tau$-distance", which correctly represents the accuracy of learning the request distribution.  Being a distance between cache occupancy distributions, the $\tau$-distance can also be mapped back to hit probability or any other performance measure that depends on learning accuracy.   The closest existing work to our approach is \cite{bitner79}, in which distributions over permutations of $n$ items over $n$ spots are studied with a cost function resembling Kendall's tau. We also emphasize that the $\tau$-distance formalizes the conceptual remark made earlier on assessing the performance of a caching algorithm by comparing it to an ideal algorithm in terms of the distance between the distributions they engender. Additionally, this is not meant to be calculated for any realistic cache parameters; in fact, in such settings even the stationary hit probability of an algorithm cannot be calculated.}

{To characterize lag, we need to study the evolution of the Markov chain associated with caching algorithm to understand its rate of convergence to stationarity.  The relevant concept here is that of \emph{mixing time}, which is the time (number of requests in our case) required for a Markov process to reach within $\epsilon$ distance (in Total Variation (TV) norm) of its stationary distribution.  To the best of our knowledge, except for \cite{fill96} that studies LRU, there is comparatively little work on analyzing the mixing time of caching algorithms, although there has been some brief commentary on the topic \cite{ bitner79, starobinski01}.  However, this metric is crucial to understanding algorithm performance, as it characterizes the speed of learning.  }

{Once we have both the $\tau$-distance and the mixing time for a caching algorithm, we can determine algorithm performance as a function of the number of requests. Using the triangle inequality and combining the $\tau$-distance and mixing time (with appropriate normalization), we define a metric that provides an upper bound on the performance at any given time. We call this metric the \emph{learning error,}  which effectively combines accuracy and learning lag. Whereas comparisons of learning error may not reflect the true performance differences between two algorithms, nevertheless it correctly determines the trade-off achieved by separately calling out the speed and accuracy of learning achieved.}

{If we know the time constant of the changes in the requests process by studying the arrival process over time, we can use this knowledge to pick a caching algorithm that has the least learning error, and hence the highest hit probability over a class of caching algorithms.   Could we also design an optimal caching algorithm for a dynamic arrival process?  While this is a difficult problem to solve optimally, in this paper,  we first characterize the performance of an isolated cache through $\tau$-distance and mixing time to study the adaptability of the candidate caching algorithms with simple and meta caches.  We use the insights gained in this process to develop an algorithm that operates over the hybrid paradigm. 
  We call the resulting algorithm as Adaptive-LRU (A-LRU).  In particular, we focus on a two-level version of A-LRU, and are able to ensure that its learning error at a given time can be made less than either LRU or $2$-LRU.  We also show that it has the highest hit probability over a class of algorithms that we compare it with using both synthetic requests generated using a Markov-modulated process, as well as trace-based simulations using traces from YouTube and the IRCache project.}

\vspace{0.05in}
\noindent\textbf{\textit{Related Work:}} Caching algorithms have mostly been analytically studied under the IRM Model.  Explicit results for stationary distribution and hit probability for LRU, FIFO, RANDOM, CLIMB \cite{king71, coffman73operating, gelenbe73unified,starobinski01} have been derived under IRM, however, these results are only useful for small caches due to the computational complexity of solving for the stationary distribution.  Several approximations have been proposed to analyze caches of reasonably large sizes \cite{rosensweig2010}, and a notable one is the Time-To-Live (TTL) approximation, which was first introduced for LRU under IRM {\cite{fagin77,che02}.}  It has been further generalized to other cache settings \cite{berger14exact, martina2014, rosensweig2010, gast16asym}. Theoretical support for the accuracy of TTL approximation was presented in \cite{berger14exact}.  Closest in {spirit} to our work is \cite{basu2017adaptive} that studies TTL caching under non-stationary arrivals, but does not consider mixing times.  A rich literature also studies the performance of caching algorithms in terms of hit probability based on real trace simulations, e.g., \cite{martina2014,zink08,megiddo03arc}, and we do not attempt to provide an overview here.

\vspace{0.05in}
\noindent\textbf{\textit{Paper Organization:}} The next section contains some technical preliminaries and caching algorithms.   We consider our new notions of learn error, $\tau$-distance and mixing time in Section~\ref{sec:error}.   We derive steady state distributions of caching algorithms in Section~\ref{sec:steady-state-analy} and analyze the mixing time in Section~\ref{sec:mixing-analysis}.  We characterize the performance of different algorithms in terms of permutation distance and learning error in Section~\ref{sec:evaluation}.  Finally, we provide trace-based numerical results in Section~\ref{sec:trace-sim}. We conclude in Section~\ref{sec:conclusion}.  Some additional discussions and proofs are provided in Appendices.

\section{Preliminaries}\label{sec:prelim}

\noindent\textbf{\textit{Traffic Model:}} We assume that there is a library of $n$ items.  The request processes for distinct content are described by independent Poisson processes with arrival rate $\lambda_i$ for content $i=1,\cdots, n.$   Without loss of generality (w.l.o.g.), we assume that the aggregate arrival rate is $1$, then the popularity of content $i$ satisfies $p_i=\lambda_i$.    W.l.o.g., we assume that the reference items are numbered so that the probabilities are in a non-increasing order, i.e., $p_1\geq p_2\geq\cdots\geq p_n$.

\noindent\textbf{\textit{Popularity Law:}} Whereas our analytical results are not for any specific popularity law, for our numerical investigations we will use a Zipf distribution as this family has been frequently observed in real traffic measurements, and is widely used in performance evaluation studies in the literature \cite{cha09analyzing}.  For a Zipf distribution, the probability to request the $i$-th most popular item is $p_i=A/i^{\alpha}$, where $\alpha$ is the Zipf parameter depending on the application considered,  and $A$ is the normalization constant so that $\sum_{i=1}^n p_i=1$ if there are $n$ unique items in total.

\begin{figure}
\centering
\includegraphics[width=0.95\linewidth]{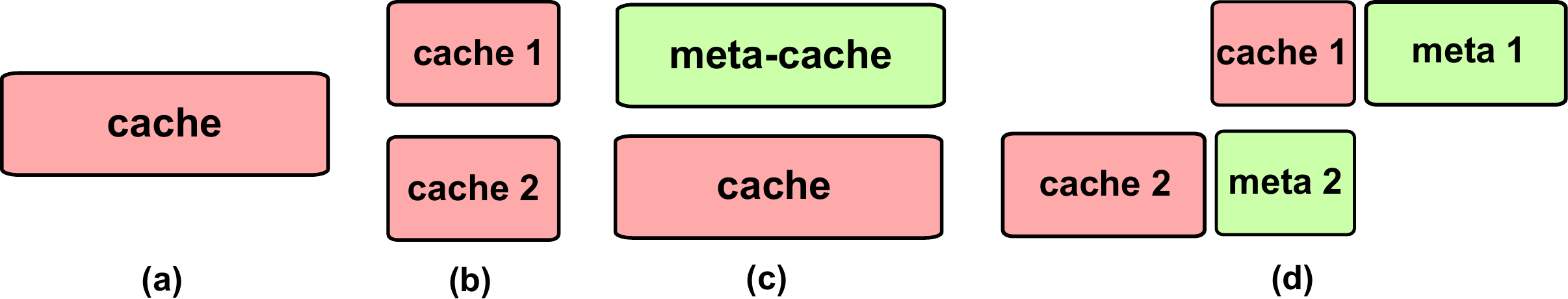}
\caption{{Dimensions of caching.}}
\label{fig:example}
\vspace{-0.2in}
\end{figure}

\noindent\textbf{\textit{{Dimensions of Caching:}}} {A cache is fundamentally a block of memory that can be used to store data items that are frequently requested.  Over the years, different paradigms have evolved on how best to utilize the available memory.  
Most conventional caching algorithms, such as LRU, RANDOM and FIFO, have been designed and analyzed on a simple (isolated) cache, as shown in Figure~\ref{fig:example} (a).   New caching algorithms have been proposed that have been shown to have better performance than the classical paradigm, often through numerical studies.  The different dimensions that have been explored are two fold.  On the one hand, the memory block can be divided into two or more levels, with a hierarchical algorithm attempting to ensure that more popular content items get cached in the higher levels.  For example, a simple $2$-level cache (also called a linear cache network) is shown in Figure~\ref{fig:example} (b), and it has been empirically observed that under an appropriate caching algorithm, it could display a higher hit probability than that of a simple cache of the same size.  On the other hand, a meta-cache that simply stores content identities can be used to better learn popularity without wasting memory to cache the actual data item.  The idea is illustrated in Figure~\ref{fig:example} (c) with one level of meta caching.   A concept that we will explore further in this paper is to mix both ideas, shown in Figure~\ref{fig:example} (d).  However, in all cases, it is not clear how the different dimensions enhance the hit probability, and how they impact convergence to stationarity.  }

\noindent{\textit{\textbf{{Caching Algorithms:}}}}\label{sec:alg} {There exist a large number of caching algorithms, with the difference being in their choice of insertion or eviction rules.  
 In this paper, we consider the following three dimensions of caching algorithms as illustrated in Figure~\ref{fig:example}.  First, we consider the conventional \emph{signle-level caching algorithms} to manage a single cache shown in Figure~\ref{fig:example} (a),  including LRU \cite{coffman73operating,fill96,king71}, FIFO \cite{coffman73operating,king71}, RANDOM \cite{coffman73operating} and CLIMB \cite{coffman73operating, starobinski01}.   Second, we consider \emph{meta-cache caching algorithms} to manage caches as shown in Figure~\ref{fig:example} (c), including $k$-LRU \cite{martina2014,gast16asym}.  Third, we consider \emph{multi-level caching algorithms} to manage caches as shown in Figure~\ref{fig:example} (b), including LRU($\boldsymbol m$) \cite{gast16asym} and ARC.  Due to space constraints, a detailed explanation on the operations of these algorithms are available in Appendix~\ref{app:algorithms}.}

{Based on the learning error analysis of these algorithms in the following sections, we also propose a novel hybrid algorithm, Adaptive-LRU (A-LRU).  Detailed operation of A-LRU is presented in Section~\ref{sec:alru}.}

\section{Learning Error}\label{sec:error}

We desire a notion of error that accounts for the tradeoff between accuracy and speed of learning.   The accuracy can be characterized by the nearness of the stationary distribution of an algorithm to the best-possible cache occupancy distribution.  The speed of learning can be characterized by the mixing time of the caching algorithm.  Clearly, a figure of merit of this kind is the distance 
{\begin{align}\label{eq:error1}
\delta_A(t)& = \sup_{{\boldsymbol x} \in \mathcal{S}} |\pi_A({\boldsymbol x},t) - \mathbf{c}^*|_\tau
\leq |\pi_A^* - {\mathbf{c}^*}|_\tau+ \sup_{{\boldsymbol x} \in \mathcal{S}} |\pi_A({\boldsymbol x},t) - \pi^*_A|_\tau,
\end{align}}
measured in some metic $\tau$ after $t$ requests, where $\pi_A^*$ is the stationary distribution of algorithm $A,$ ${\mathbf{c}^*}$ is the best-possible occupancy vector and  $\pi_A(\boldsymbol x,t)$ is  the row corresponding to state $\boldsymbol x\in \mathcal{S}$ of the $t$-step transition matrix of algorithm $A$; these will be described in detail later.  The first term above is the error in the (eventual) learning of the algorithm (accuracy), and the second term is the error due to time lag of learning (efficiency).  We could then argue that if the time-constant of the change in the request distribution is $t,$ then the caching algorithm $A$ that has a low value of the RHS would have attained some fraction of optimality by that time.  {We pause to reemphasize that the utility of the learning error is in revealing the trade-off achieved by an algorithm by disentangling the speed and accuracy of learning.} In the rest of this section, we will posit an appropriate metric $\tau$ and characterize both the error terms.

\subsection{Permutation Distance}\label{sec:distance}

We seek a refinement that would allow us to determine ``how close'' the stationary performance of an algorithm is to the best-possible. If we have full knowledge of the popularity distribution at any time, we could simply cache the most popular items in the available cache spots, placing the most popular element in first cache spot, and then proceeding onwards until the $m$-th spot.  This approach would maximize the hit probability, as well as any other metric that yields better performance when more popular items are cached.  We denote this ideal occupancy vector as $\mathbf{c}^*$; this ideal distribution is then a point-mass at $\mathbf{c}^*$.  {As a means to determine  closeness of the stationary performance of an algorithm to the ideal scheme, we start by discussing an appropriate distance between the ideal occupancy vector and any possible cache occupancy state.}

\subsubsection{\bf Generalized Kendall's Tau Distance}
Let $[n]=\{1, \cdots, n\}$ be a library of items and $[n]_m$ be the set of $m$ items chosen from $[n]$. 
Let $S_n^m$ be the set of permutations on $[n]_m.$ Consider a permutation $\sigma\in S_n^m,$ we interpret $\sigma(i)$ as the position of item $i$ in $\sigma$, and we say that $i$ is ahead of $j$ in $\sigma$ if $\sigma(i)<\sigma(j).$ W.l.o.g, we take $\sigma(i)=0$ for $i\in [n]/[n]_m,$ i.e., all items absent in the cache have position $0.$

The classical Kendall's tau distance \cite{fagin03comparing} \footnote{We consider $p=0$ for the definition given in \cite{fagin03comparing}, which is an ``optimistic approach" that corresponds to the intuition that we assign a nonzero penalty to the pair $\{i, j\}$ only if we have enough information to know that $i$ and $j$ are in the opposite order in the two permutations $\sigma_1$ and $\sigma_2$.} is given by
\begin{equation}\label{eq:Kendalltau}
K(\sigma_1, \sigma_2)=\sum_{(i, j): \sigma_1(i)>\sigma_1(j)} 1_{\{\sigma_2(i)<\sigma_2(j)\}},
\end{equation}
where $1_{\mathcal{A}}$ is the indicator function and $1_{\mathcal{A}}=1$ if the condition $\mathcal{A}$ holds true, otherwise $1_{\mathcal{A}}=0$.

However, this conventional definition does not take into account the item relevance and positional information, which are crucial to evaluating the distance metric in a permutation. 
Since we wish to compare with $\mathbf{c}^*,$ in which the most popular items are placed in lower positions, the errors in lower positions in the permutation need to be penalized more heavily than those in higher positions.  Many alternative distance measures have been proposed to address these shortcomings.  In the following, we consider the generalized Kendall's tau distance proposed in \cite{kumar10} that captures the importance of each item as well as the positions of the errors.

{Let $w_i>0$ be the \emph{element weight} for $i\in[n].$ For simplicity, we assume that $w_i\in\mathbb{Z}^+;$ all the subsequent results hold for non-integral weights as well. In addition to the element weight, as discussed earlier, we wish to penalize inversions early in the permutation more than inversions later in the permutations. In order to achieve this, we define \emph{position weights} to differentiate the importances of positions in the permutation. We first consider the cost of swapping between two adjacent positions. Let $\zeta_j\geq 0$ be the cost of swapping an item at position $j-1$ with an item at position $j$,  and let $q_0=0,$ $q_1=1$ and $q_j=q_{j-1}+\zeta_j$ for $1<j\leq m$. Define $\bar{q}^i_{\sigma_1, \sigma_2}=\frac{q_{\sigma_1(i)}-q_{\sigma_2(i)}}{\sigma_1(i)-\sigma_2(i)}$ to be the average cost that item $i$ encountered in moving from position $\sigma_1(i)$ to position $\sigma_2(i),$ with the understanding that $\bar{q}^i_{\sigma_1, \sigma_2}=1$ if $\sigma_1(i)=\sigma_2(i)$. We set the value of $ \bar{q}^j_{\sigma_1, \sigma_2}$ similarly. We then define the generalized Kendall's tau distance as follows: 
\begin{equation}\label{eq:general-kendall}\small
K_{w, \zeta}(\sigma_1, \sigma_2)=\sum_{\sigma_1(i)<\sigma_1(j)} w_i w_j\bar{q}^i_{\sigma_1, \sigma_2} \bar{q}^j_{\sigma_1, \sigma_2}1_{\{\sigma_2(i)>\sigma_2(j)\}}.
\end{equation}}

\begin{rem}
{Note that if we are interested only in cache hits and misses (eg., if there is no search cost within the cache), the ordering is irrelevant and only content presence or absence matters.  The Kendall's tau still applies with weights being just $0$ and $1,$ to indicate presence and absence, respectively.   Further, we now need only consider distances between {equivalent classes of cache states,} where two cache states with identical content are equivalent. }
\end{rem}

\subsubsection{\bf Wasserstein Distance}
While the generalized Kendall's tau distance is a way of comparing two permutations, the algorithms that we are interested in do not converge to a single permutation, but yield distributions over permutations with more elements in their support.  Hence, we should compare the stationary distribution $\pi^*_A$ of an algorithm $A,$ with $\mathbf{c}^*$ using a distance function that accounts for the ordering of content in each state vector.  Given a metric on permutations, the Wasserstein distance \cite{villani08optimal} is a general way of comparing distributions on permutations.

Let $(\mathcal{S}, d)$ be a Polish space, and consider any two probability measures $\mu$ and $\nu$ on $\mathcal{S}$, then the Wasserstein distance\footnote{W.l.o.g., we are interested in the $L^1$-Wasserstein distance, which is also commonly called the Kantorovich-Rubinstein distance \cite{villani08optimal}. For convenience, we express Wasserstein distance by means of couplings (joint distributions) between random variables with given marginals.} between $\mu$ and $\nu$ is defined as
\begin{equation}\label{eq:wasserstein-distance}\small
W(\mu, \nu)=\inf_{P_{X,Y}(\cdot, \cdot)}\Big\{\mathbb{E} [d(X, Y)], \quad P_X(\cdot)=\mu, P_Y(\cdot)=\nu\Big\},
\end{equation}
which is the minimal cost between $\mu$ and $\nu$ induced by the cost function $d.$

\subsubsection{\bf $\tau$-distance}
We are now ready to define the specific form of Wasserstein distance between distributions on permutations that is appropriate to our problem.  We define the $\tau$-distance as the Wasserstein distance taking the generalized Kendall's distance in~(\ref{eq:general-kendall}) as the cost function in~(\ref{eq:wasserstein-distance}). 

Since the ideal occupancy vector $\mathbf{c}^*$ is unique and fixed, the infimum in~(\ref{eq:wasserstein-distance}) over all the couplings is trivially given by the average distance, i.e., 
\begin{equation}\label{eq:tau-distance}
|\pi^*_A -\mathbf{c}^*|_\tau:=W(\pi^*_A,\delta_{\mathbf{c}^*}) =\sum_{\boldsymbol x}K_{w, \zeta}(\boldsymbol x_{k(A)}, \mathbf{c}^*)\pi^*_A(\boldsymbol x),
\end{equation}
where $K_{w, \zeta}(\cdot, \cdot)$ is the generalized Kendall's tau distance defined in~(\ref{eq:general-kendall}), and $\boldsymbol x_{k(A)}$ is the component of state under algorithm $A$ corresponding to real items. Thus, $\boldsymbol x_{k(A)}=\boldsymbol x$ for all the conventional algorithms considered in Section~\ref{sec:sta-distri}, and the multi-level caching algorithms in Section~\ref{sec:multi-level-cache}, but $\boldsymbol x_{k(A)}=\boldsymbol x_k$ for $k$-LRU as discussed in Section~\ref{sec:meta-cache}.

\subsection{Computation of Mixing Time}\label{sec:mixing}

While comparing the $\tau$-distance of the stationary distance of an algorithm to the ideal occupancy vector does provide insight into the algorithm's accuracy of learning, it says nothing about how quickly the algorithm can respond to changes in the request distribution---a critical shortcoming in developing and characterizing the ideal algorithm for a given setting. How does one come up with a metric that accounts for both accuracy and speed of learning?  It seems clear that one ought to study the evolution of the caching process over time to understand how quickly the distribution evolves. Within the Markovian setting of our algorithms, the metric of relevance in this context is called \emph{mixing time}, which is the time required for a Markov process to reach within $\epsilon$ distance (in Total Variation (TV) norm) of the eventual stationary distribution.   If we denote the row corresponding to state $\boldsymbol x\in \mathcal{S}$ of the $t$-step transition matrix of algorithm $A$ by $\pi_A(\boldsymbol x,t),$ then the mixing time is the smallest value of $t$ such that
\begin{align}
\sup_{\boldsymbol x\in \mathcal{S}} |\pi_A(\boldsymbol x,t) - \pi_A^*|_{TV} \leq \epsilon,
\end{align} 
for a given $\epsilon >0$ \cite{levin09markov}; denote it as $t_{\text{mix}}(\epsilon).$ As mentioned earlier, we will also think of $\pi_A(\boldsymbol x,t)$ and $\pi^*_A$ as distributions on permutations of the $n$ objects.

Mixing times can be estimated using many different procedures.  Here, we use \emph{conductance}  to build bounds on the mixing time through \emph{Cheeger's inequality}. We first introduce these techniques and then characterize the mixing time of various caching algorithms in Section~\ref{sec:mixing-analysis}. These bounds will allow us to compare all the algorithms on an equal footing, and also to determine the first-order dependence on algorithm parameters.

\subsubsection{\bf Spectral Gap and Mixing Time}
Let $\gamma^*$ be the \emph{spectral gap} of any Markov chain with transition matrix $P$, and denote $\pi^*_P$ as its corresponding stationary distribution.  {Then defining $\pi_{\text{min}}:=\min_{\boldsymbol x\in\mathcal{S}}\pi^*_P(\boldsymbol x)$, an upper bound on the mixing time in terms of spectral gap and the stationary distribution of the chain is given as follows \cite{levin09markov, montenegro06}: 
\begin{equation}\label{eq:eig-mixing}
 t_{\text{mix}}(\epsilon)<1+\frac{1}{\gamma^*}\ln\left(\frac{1}{\epsilon\pi_{\text{min}}}\right).
\end{equation}}

\subsubsection{\bf Conductance and Mixing Time}
For a pair of states $\boldsymbol x, \boldsymbol y\in\mathcal{S},$ we define the transition rate $Q(\boldsymbol x, \boldsymbol y)=\pi(\boldsymbol x)P(\boldsymbol x, \boldsymbol y).$ Let $Q(S_1, S_2)=\sum_{\boldsymbol x\in S_1}\sum_{\boldsymbol y\in S_2}Q(\boldsymbol x, \boldsymbol y),$ for two sets $S_1, S_2\in\mathcal{S}.$
Now, for a given subset $S\in\mathcal{S},$ we define its conductance as $\Phi(S)=\frac{Q(S, \bar{S})}{\min(\pi(S),\pi(\bar{S}))},$ where $\pi(S)=\sum_{i\in S}\pi_i.$ and $(S,\bar{S})$ is a cut of the graph. Note that $Q(S, \bar{S})$ represents the ``ergodic flow" from $S$ to $\bar{S}$.  Finally, the \emph{conductance of the chain $P$} is the conductance of the ``worst" set, i.e., 
\begin{equation}\label{eq:conductance}
\Phi=\min_{S\subset\mathcal{S}, \pi(S)\leq\frac{1}{2}} \Phi(S).
\end{equation}

The relationship between the conductance and the mixing time of a Markov chain (the spectral gap) is given by the \emph{Cheeger inequality} \cite{chung05,montenegro06}:
\begin{equation}\label{eq:cheeger}
\frac{\Phi^2}{2}\leq \gamma^*\leq 2\Phi.
\end{equation}

Combining~(\ref{eq:cheeger}) with the previous result in~(\ref{eq:eig-mixing}), we can relate the conductance directly to the mixing time as follows:
\begin{equation}
 t_{\text{mix}}(\epsilon)\leq \frac{2}{\Phi^2}\left(\ln\frac{1}{\pi_{\text{min}}}+\ln\frac{1}{\epsilon}\right).
\end{equation}

While the spectral gap and conductance of a Markov chain can provide tight bounds on the mixing time of the chain, these values are often difficult to calculate accurately. If we are interested in proving rapid mixing\footnote{A family of ergodic, reversible Markov chain with state space of size $|\mathcal{S}|$ and conductance $\Phi_{|\mathcal{S}|}$ is \emph{rapidly mixing} if and only if $\Phi_{|\mathcal{S}|}\geq\frac{1}{\mathcal{P}(|\mathcal{S}|)}$for some polynomial $\mathcal{P}$ \cite{mihail89conductance}. This result is commonly used to show rapid mixing of Markov chains.}, we can provide a lower bound on the conductance.  {Canonical paths and congestion are used in this regard as they are easier to compute, and bound the conductance from below.}  For any pair $\boldsymbol x, \boldsymbol y\in\mathcal{S}$, we can define a canonical path $\psi_{\boldsymbol x\boldsymbol y}=(\boldsymbol x=\boldsymbol x_0, \cdots, \boldsymbol x_l=\boldsymbol y)$ running from $\boldsymbol x$ to $\boldsymbol y$ through adjacent states in the state space $\mathcal{S}$ of the Markov chain. Let $\Psi=\{\psi_{\boldsymbol x\boldsymbol y}\}$ be the family of canonical paths between all pairs of states.  The \emph{congestion} of the Markov chain is then defined as
\begin{equation}\label{eq:congestion}\small
\rho=\rho(\Psi)=\max_{(\boldsymbol u, \boldsymbol v)}\left\{\frac{1}{\pi(\boldsymbol u)P_{\boldsymbol u\boldsymbol v}}\sum_{\substack{\boldsymbol x, \boldsymbol y\in\mathcal{S} \\ \exists\; \psi_{\boldsymbol x\boldsymbol y} \text{ using } (\boldsymbol u,\boldsymbol v)}}\pi(\boldsymbol x)\pi(\boldsymbol y)\right\},
\end{equation}
where the maximum runs over all pairs of states in the state space, and the number of canonical path is of the order of $\Gamma^2,$ {with $\Gamma$ being the number of possible states.} Therefore, high congestion corresponds to a lower conductance, and as demonstrated in \cite{sinclair92improved}
\begin{equation}\label{eq:congestion-conductance}
\Phi\geq\frac{1}{2\rho}.
\end{equation}
Note that the above result applies to all possible choices of canonical paths, for example, no requirement is made that the shortest path between two states has be chosen.

\subsubsection{\bf Mixing Time Bound}
Based on the above analysis, we are ready to present an explicit general bound on mixing time as follows. 
\begin{lemma}\label{lem:mixing-methodology}
Suppose the Markov chain associated with caching algorithm $A$ has a reversible transition matrix $P,$ and denote the corresponding stationary distribution as $\pi_P^*.$ Based on the analysis of the relations between conductance and mixing time in (\ref{eq:conductance}) -~(\ref{eq:congestion-conductance}), we can directly characterize the mixing time of a caching algorithm\footnote{We omit the superscript $A$ for brevity.}:
\begin{small}
\begin{align}\label{eq:mixing-reversible}
t_{\text{mix}}(\epsilon)\leq \frac{8(\pi^*_{P, \text{max}})^4\cdot\Gamma^4}{(\pi^*_{P, \text{min}}P_{\text{min}})^2}\left(\ln\frac{1}{\pi^*_{P, \text{min}}}+\ln\frac{1}{\epsilon}\right),
\end{align}
\end{small}
where $\pi^*_{P, \text{max}}=\max_{\boldsymbol x\in\mathcal{S}}\pi_P^*(\boldsymbol x),$ $\pi^*_{P, \text{min}}=\min_{\boldsymbol x\in\mathcal{S}}\pi_P^*(\boldsymbol x),$ 
 and $P_{\text{min}}$ is the minimal transition probability from one state to another state.
\end{lemma}
The proof of this lemma is straightforward given (\ref{eq:conductance}) -~(\ref{eq:congestion-conductance}), since we only need to find an upper bound on the congestion $\rho$ through characterizing $\pi^*_{P, \text{max}}$ and $\pi^*_{P, \text{min}}$ given the steady state distribution.  As the analysis for different algorithms varies considerably in terms of the characterization of the spectral gap, we will use the above result for an equal footing comparison.

\subsubsection{\bf Non-Reversibility and Mixing Time}\label{sec:reversibility}
Many results on mixing times have been developed in the context of a reversible Markov chain.  However, many of the popular caching algorithms such as the LRU family generate non-reversible Markov chains.  From \cite{montenegro06} it follows that several of results that apply to reversible Markov chains hold even without reversibility but with the modifications described below.

For any non-reversible Markov chain with transition matrix $P,$ first determine $P^*,$ which is the time-reversed transition matrix:
\begin{equation}\label{eq:reversal1}
\pi^*_P(\boldsymbol x)P^*(\boldsymbol x, \boldsymbol y)=\pi^*_P(\boldsymbol y)P(\boldsymbol y, \boldsymbol x),
\end{equation}
where $\boldsymbol x, \boldsymbol y\in\mathcal{S}.$ Then we can construct a reversible Markov chain with transition matrix $\frac{P+P^*}{2},$ for which the following result holds.
\begin{lemma}\label{lem:nonreversible}
Let $P$ be the transition matrix of a non-reversible Markov chain, and $P^*$ be the time-reversal. Denote $\pi^*_P$ and $\gamma^*_P$ be the corresponding stationary distribution and spectral gap, then we have 
\begin{align}
&\pi^*_P(\boldsymbol x)=\pi^*_{P^*}(\boldsymbol x)=\pi^*_{\frac{P+P^*}{2}}(\boldsymbol x), \forall \boldsymbol x\in\mathcal{S},\text{and}
\gamma^*_P=\gamma^*_{P^*}=\gamma^*_{\frac{P+P^*}{2}}.
\end{align}
\end{lemma}
{This result was originally presented in \cite{Fill91}, and the proof in our context is presented in Appendix~\ref{app:revers-mixing} for completeness.}  
Then from~(\ref{eq:eig-mixing}), we can equivalently use the reversible Markov chain $\frac{P+P^*}{2}$ to bound the mixing time of the non-reversible Markov chain $P$ through applying existing results on reversible Markov chains.  This procedure will be utilized in the following subsections.

Given the results in Lemma~\ref{lem:mixing-methodology} and Lemma~\ref{lem:nonreversible}, and the fact that $\frac{(P + P^*)}{2} (\boldsymbol x, \boldsymbol y) \geq P (\boldsymbol x, \boldsymbol y)/2$ we immediately have the following result:
\begin{corollary}\label{cor:mixing-methodology}
Suppose the Markov chain associated with caching algorithm $A$ has a non-reversible transition matrix $P.$   Then we have 
\begin{small}
\begin{align}\label{eq:mixing-non-reversible}
t_{\text{mix}}(\epsilon)\leq\frac{8(\pi^*_{P, \text{max}})^4\cdot\Gamma^4}{(\pi^*_{P, \text{min}}\frac{P_{\text{min}}}{2})^2}\left(\ln\frac{1}{\pi^*_{P, \text{min}}}+\ln\frac{1}{\epsilon}\right),
\end{align}
\end{small}
where $\pi^*_{P, \text{max}}=\max_{\boldsymbol x\in\mathcal{S}}\pi_P^*(\boldsymbol x),$ $\pi^*_{P, \text{min}}=\min_{\boldsymbol x\in\mathcal{S}}\pi_P^*(\boldsymbol x),$ $\Gamma$ is the number of possible states and $P_{\text{min}}$ is the minimal transition probability from one state to another state.
\end{corollary}

\begin{rem}
From~(\ref{eq:mixing-reversible}) and~(\ref{eq:mixing-non-reversible}), it is clear that for both reversible and irreversible Markov chains, we need to characterize $\pi^*_{P, \text{max}},$ $\pi^*_{P, \text{min}}$ and $P_{\text{min}}$ in order to obtain mixing time bounds.  Thus, identification of the steady state distribution $\pi^*_P$ (discussed in Section~\ref{sec:steady-state-analy}) in a {way} 
that allows us to determine these bounds is crucial in obtaining mixing time bounds.
\end{rem}

\subsection{Learning Error}
Since the space of all permutations on $n$ objects is finite, it has a finite diameter in terms of the generalized Kendall's tau distance. Let this diameter be denoted as $\kappa_{\tau}.$  
Then ~(\ref{eq:error1}) can be bounded using the triangle inequality as 
{\begin{align}\label{eq:error2}
\delta_A(t) \leq|\pi_A^* -{\boldsymbol c^*}|_\tau +  \kappa_{\tau} \sup_{{\boldsymbol x} \in \mathcal{S}} |\pi_A({\boldsymbol x},t) - \pi^*_A|_{TV} \triangleq e_A(t).
\end{align}}
We refer to $e_A(t)$ as the \emph{learning error} of algorithm $A$ at time $t,$ which is now only a function of the accuracy and mixing time of the algorithm.  In order to compute the learning error $e_A(t)$ in~(\ref{eq:error2}), we need the stationary distribution of algorithm $A$.  In the following sections, we first characterize the stationary distributions of different caching algorithms in Section~\ref{sec:steady-state-analy},  then analyze their mixing time  in Section~\ref{sec:mixing-analysis}, and finally evaluate their performance in Section~\ref{sec:evaluation}.

\section{Steady State Distribution}\label{sec:steady-state-analy}

We consider the question of determining the stationary distribution of the contents of a cache based on the caching algorithm used.  Each (known) caching algorithm $A$ under any Markov modulated request arrival process {(IRM too)} 
results in a Markov process over the occupancy states of the cache.

\subsection{Classical Results {on Single-Level Caching Algorithms}}\label{sec:sta-distri}
Suppose there are a total of $n$ content items in a library $\mathcal{L}$, and the cache size is $m<n$.  Then each cache state $\boldsymbol x$ is a vector of length $m$ that indicates the content in each cache spot.   We denote $x_j\in\mathcal{L}$ as the identity of the item at position $j$ in the cache, i.e., $\boldsymbol x=(x_1, \cdots, x_m).$  As mentioned earlier, we call the state space of all such vectors $\mathcal{S}.$   Our notation for state is consistent with respect to the algorithm descriptions in Section~\ref{sec:alg} and represents motion from ``left-to-right'' under our candidate algorithms.  For example, under LRU $x_j$ has been requested more recently than $x_k$ if $j<k.$ 

One can potentially determine the stationary distribution of the Markov process generated by a particular algorithm $A$, denoted $\pi^*_A.$   This procedure has been carried out for several classical caching algorithms in the literature \cite{coffman73operating}, but the results are not available in the desired form (viewed in terms of permutations) {so we present them for the Markov chains generated by FIFO, RANDOM, CLIMB and LRU.}

\begin{theorem}\label{thm:stationary}
Under the IRM, the steady state probabilities $\pi_{\text{FIFO}}^*(\boldsymbol x)$, $\pi_{\text{RANDOM}}^*(\boldsymbol x)$, $\pi_{\text{CLIMB}}^*(\boldsymbol x)$ and $\pi_{\text{LRU}}^*(\boldsymbol x)$, with $\boldsymbol x\in \mathcal{S}$ are as follows:
\begin{small}
\begin{align}\label{eq:steadystate}
&\pi_{\text{FIFO}}^*(\boldsymbol x)=\frac{\Pi_{i=1}^m p_{x_i}}{\sum_{\boldsymbol y\in\mathcal{S}} \Pi_{i=1}^m p_{y_i}},\nonumber\displaybreak[0]\\
&\pi_{\text{RANDOM}}^*(\boldsymbol x)=\frac{\Pi_{i=1}^m p_{x_i}}{\sum_{\boldsymbol y\in\mathcal{S}^\prime} \Pi_{i=1}^m p_{y_i}},\nonumber\displaybreak[1]\\
&\pi_{\text{CLIMB}}^*(\boldsymbol x)=\frac{\Pi_{i=1}^m p^{m-i+1}_{x_i}}{\sum_{\boldsymbol y\in\mathcal{S}} \Pi_{i=1}^m p^{m-i+1}_{y_i}},\nonumber\displaybreak[2]\\
&\pi_{\text{LRU}}^*({\boldsymbol x})=\frac{\Pi_{i=1}^m p_{x_i}}{(1-p_{x_1})(1-p_{x_1}-p_{x_2})\cdots(1-p_{x_1}-\cdots-p_{x_{m-1}})},
\end{align}
\end{small}
where $\mathcal{S}^\prime$ denotes the set of all combinations of elements of $\{1,\cdots, n\}$ taken $m$ at a time. Note that elements of $\mathcal{S}^\prime$ are subsets of $\{1,\cdots, n\}$, while elements of $\mathcal{S}$ are ordered subset of $\{1,\cdots, n\}$, satisfying
$\sum_{\boldsymbol y\in\mathcal{S}}\Pi_{j=1}^{m} p_{y_j}=m! \sum_{\boldsymbol y\in\mathcal{S}^\prime}\Pi_{j=1}^{m} p_{y_j}$.
\end{theorem}
These are the well-known steady state probabilities for FIFO, RANDOM, CLIMB and LRU \cite{king71,coffman73operating,aven87,gelenbe73unified,starobinski01}. The result for $\text{LRU}$ is obtained by a probabilistic argument~\cite{hendricks76}. We detail this in Appendix~\ref{app:station-lru} for completeness, and since we will use the method for other related algorithms.


\subsection{{Meta-cache Caching Algorithms}}\label{sec:meta-cache}

A closed form result of the stationary distribution of the Markov chain generated by $k$-LRU (see Figure ~\ref{fig:example} (c) for typical configuration) is not available currently.   Motivated by the approach to determining the stationary distribution of LRU \cite{hendricks76}, we use a probabilistic argument to obtain the general form of the stationary distribution of $k$-LRU.  While the expression that we obtain is complex from the perspective of numerical computation, it will provide us with the necessary structure to obtain mixing time bounds in Section~\ref{sec:mixing}.

Consider a cache system with $k-1$ levels of meta-cache, followed by a level of real cache.  We denote $x_{(i, j)} \in \mathcal{L}$ as the identity of the item at position $j$ in cache $i.$  Here, $i\in\{1,\cdots, k-1\}$ refer to meta caches, while $i=k$ refers to the real cache.   We denote the state of level $j$ as $\boldsymbol x_j=(x_{(j,1)}, \cdots, x_{(j, m)}),$ and the state of the whole cache as $\boldsymbol x=(\boldsymbol x_1, \cdots, \boldsymbol x_k).$  Note that only $\boldsymbol x_k$ caches the real items, while all other levels cache only meta-data.  Finally,  let $\mathcal{X}\subset \mathcal{L}$ be the set of items present in $\boldsymbol x.$

\begin{definition}
{\bf Sample Path:} {A \emph{sample path} $\gamma({\boldsymbol x})$ for state $\boldsymbol x$ is a sequence of requests that leads to the state $\boldsymbol x$ under the $k$-LRU algorithm starting from any fixed initial state (such as the empty cache). }
\end{definition}

{For any item $y\in \mathcal{X},$ let $h(y)=\max\{i: x_{(i,j)}=y\},$ i.e., $h(y)$ is highest cache level at which item $y$ is present.   Each sample path $\gamma(\bm{x})$ must contain a set of the final $h(y)$ requests for each $y \in \mathcal{X}.$  Call the union of all these requests as $\hat{\mathcal{X}}.$  Hence, $\hat{\mathcal{X}}$ will contain exactly $h(y)$ copies of each item in $\mathcal{X}.$  Note that $|\hat{\mathcal{X}}|$ is at most $\sum_{j=1}^kjm,$ which occurs when all the items in $\boldsymbol x$ are distinct from each other.  }

{Let $\bm{\hat{x}}$ represent an arrangement of the items in $\hat{\mathcal{X}}.$   Again, note that this arrangement has exactly $h(y)$ copies of each item in $\mathcal{X}.$   Then an arbitrary request sequence $\hat{\gamma}(\bm{\hat{x}})$ following $\bm{\hat{x}}$ interspersed with any other items drawn from $\mathcal{L},$ which does not violate the condition that $\bm{\hat{x}}$ is the ordering of the final $h(y)$ requests for each $y \in \mathcal{X}$, is a candidate sample path leading to $\boldsymbol x.$  However, $\hat{\gamma}(\bm{\hat{x}})$  might not be consistent with $k$-LRU, i.e., not all arrangements of $\hat{\mathcal{X}}$ can be used to generate sample paths using  $k$-LRU.  Hence, we denote a sample path consistent with $k-$LRU by $\gamma(\bm{\hat{x}}),$ which gives rise to the following definition.}

\begin{definition}
{\bf Class of Sample Paths:} 
 We define a {valid class of sample paths} $\Lambda(\bm{\hat{x}})$ as a set of sample paths $\gamma(\bm{\hat{x}})$ each of which follows arrangement $\bm{\hat{x}},$ and is consistent with the operation of $k$-LRU.  Let $\Upsilon(\boldsymbol x)$ be the set of the classes of sample paths associated with state $\boldsymbol x,$ i.e., $\Upsilon(\boldsymbol x)=\{\Lambda(\bm{\hat{x}})\}.$
\end{definition}

\begin{definition}
{\bf Subclass of Sample Paths:}
We define a valid subclass of sample paths $\tilde{\Lambda}(\bm{\hat{x}})$ as a set of sample paths $\gamma(\bm{\hat{x}})\in\Lambda(\bm{\hat{x}})$ and the set of items that can be requested between any two items on the arrangement $\bm{\hat{x}}$ are identical. Let $\tilde{\Upsilon}_{\Lambda(\bm{\hat{x}})}(\boldsymbol x)$ be the set of the subclasses of sample paths associated with class $\Lambda(\bm{\hat{x}})$.
\end{definition}

Thus, each valid sample path $\gamma(\boldsymbol x)$ leading to state $\boldsymbol x$ under $k$-LRU belongs to a valid subclass $\tilde{\Lambda}(\bm{\hat{x}})\in\Lambda(\bm{\hat{x}}).$

{We present one illustrative example to explain the above definitions with a more simple one ($k=1$) detailed in Appendix~\ref{app:example}.  We consider $n=5, m=2$, and denote the items as $1, 2, 3, 4, 5,$ with popularities $p_1>p_2>p_3>p_4>p_5$ and $\sum_{i=1}^5p_i=1.$  }

\begin{eg}\label{exm2}
{Consider $k=2,$ i.e., the $2$-LRU algorithm.  Consider state $\boldsymbol x=((34), (12)),$ i.e., $(12)$ are in the second level and $(34)$ are in the first level.  So there are totally $2*1+2*2=6$ items that need to be fixed to obtain a sample path, i.e., items $\hat{\mathcal{X}} =\{2, 2, 1, 1, 3, 4\}.$  Based on the $2$-LRU policy, there are totally $9$ valid arrangements (and hence classes) over these $6$ items, i.e., $|\Upsilon(\boldsymbol x)|=9.$  It can be verified that these valid arrangements are:
$2\rightarrow 2\rightarrow 1\rightarrow 1\rightarrow 4\rightarrow 3;$ $2\rightarrow 2\rightarrow 1\rightarrow 4\rightarrow 1\rightarrow 3;$ $2\rightarrow 2\rightarrow 4\rightarrow 1\rightarrow 1\rightarrow 3;$ $2\rightarrow 4\rightarrow 2\rightarrow 1\rightarrow 1\rightarrow 3;$ $4\rightarrow 2\rightarrow 2\rightarrow 1\rightarrow 1\rightarrow 3;$ $2\rightarrow 1\rightarrow 2\rightarrow 1\rightarrow 4\rightarrow 3;$ $2\rightarrow 1\rightarrow 2\rightarrow 4\rightarrow 1\rightarrow 3;$ $1\rightarrow 2\rightarrow 2\rightarrow 1\rightarrow 4\rightarrow 3;$ $1\rightarrow 2\rightarrow 2\rightarrow 4\rightarrow 1\rightarrow 3.$ 
 Each arrangement $\bm{\hat{x}}$ above can be used to generate infinite sample paths that belong to the corresponding class $\Lambda(\bm{\hat{x}}).$  {In each class $\Lambda(\bm{\hat{x}})$, the items can be requested between any two fixed items on arrangement $\bm{\hat{x}}$ define the subclass $\tilde{\Lambda}(\bm{\hat{x}}).$ }}
\end{eg}

Consider a valid sample path {$\gamma(\bm{x})\in \tilde{\Lambda}(\bm{\hat{x}})\in \Lambda(\bm{\hat{x}}).$}  It can be split into the requests following $\bm{\hat{x}}$ and all other requests.  Define $\Xi({\gamma{(\boldsymbol x)}})$ as the probability of requesting all these other items.  Hence, the probability of any sample path is the product of $\Xi({\gamma{(\boldsymbol x)}})$ and the probability of requesting the items in $\bm{\hat{x}}.$  Now, consider an item $x_{(i,j)} \in \bm{x}.$  According to our notation $h(x_{(i,j)})$ is the highest cache level in which the item appears.  Define a function
\begin{align}
    \mathcal{J}_{x_{(i,j)}}= 
\begin{cases}
    p_{x_{(i,j)}},& \text{if } i= h(x_{(i,j)})\\
    1,              & \text{otherwise.}
\end{cases}
\end{align}

We then have a  characterization of the steady state distribution of $k$-LRU. 
\begin{theorem}\label{eq:thm-klru}
Under the IRM, the steady state probabilities of $k$-LRU satisfies the following form 
\begin{small}
\begin{align}\label{eq:steady-state-klru}
\pi_{\text{$k$-LRU}}^*(\boldsymbol x)=&\prod_{i=1}^k\left(\prod_{j=1}^{m}\mathcal{J}_{x_{(i,j)}}\right)^i \sum_{\Lambda(\bm{\hat{x}})\in\Upsilon(\boldsymbol x)}\sum_{\tilde{\Lambda}(\bm{\hat{x}})\in\Lambda(\boldsymbol x)}\sum_{\gamma{(\boldsymbol x)}\in\tilde{\Lambda}(\bm{\hat{x}})} \Xi({\gamma{(\boldsymbol x)}}),
\end{align}
\end{small}
\end{theorem}
The proof is presented in Appendix~\ref{app:station-k-lru}.  {Here, we work through Example~\ref{exm2} to illustrate the above approach for calculating the stationary probabilities; again a more illustrative and simple example ($k=1$) is in Appendix~\ref{app:example-k-lru}.  }

\begin{eg}\label{exm4}
{Consider $k=2,$ i.e., $2$-LRU algorithm.  Consider state $\boldsymbol x=((34), (12)),$ i.e., $(12)$ are in the second level and $(34)$ are in the first level.  From Example~\ref{exm2},  $|\Upsilon(\boldsymbol x)|=9.$  Consider a the set of sample paths $\Lambda(2\rightarrow 2\rightarrow 1\rightarrow 1\rightarrow 4\rightarrow 3)$.  Each sample path $\gamma(\boldsymbol x)$ in this set must satisfy one of the following conditions:
(a) There is no request between any fixed items in $\gamma(\bm{x}).$ In this case, $\Xi(\gamma(\boldsymbol x))=1;$ 
(b) In general, between the two fixed instances of $2\rightarrow 2,$ we can request one of the items in  $\{1, 3, 4, 5\}$ any number of times.  Give these requests, between the two fixed items $2\rightarrow 1$, we can request item $1$ any number of times if it has been requested between $2\rightarrow 2$, or request one item in $\{1, 3, 4, 5\}$ once that has not been requested between $2\rightarrow 2.$ Given these requests, between the fixed items $1\rightarrow 1$, we can request one item in $\{3, 4, 5\}$ once that has not been requested before.  Given these requests, between the two fixed items $1\rightarrow 4$, we can request one item from $\{3, 4, 5\}$ once that has not been requested before.  Given these requests, between the two fixed items $4\rightarrow 3,$ we cannot request any other item.  {Note that the way of choosing different items between any fixed two items on the arrangement $2\rightarrow 2\rightarrow 1\rightarrow 1\rightarrow 4\rightarrow 3$ define different subclass of sample paths $\tilde{\Lambda}(2\rightarrow 2\rightarrow 1\rightarrow 1\rightarrow 4\rightarrow 3)$.  In this way, we consider all the possible $\gamma(\boldsymbol x)$ for the subclass of sample path $\tilde{\Lambda}(2\rightarrow 2\rightarrow 1\rightarrow 1\rightarrow 4\rightarrow 3)\in\Lambda(2\rightarrow 2\rightarrow 1\rightarrow 1\rightarrow 4\rightarrow 3).$}  Computing the corresponding probabilities, we can obtain $\Xi(\gamma(\boldsymbol x))$ for each $\gamma(\boldsymbol x)$.  Then summing them all up, we get $\sum_{\gamma(\boldsymbol x)\in\Lambda(\bm{\hat{x}})}\Xi(\gamma(\boldsymbol x))$ for this particular class $\Lambda(2\rightarrow 2\rightarrow 1\rightarrow 1\rightarrow 4\rightarrow 3).$}

{Similarly, we can consider sample paths corresponding to all the classes in $\Upsilon(\boldsymbol x)$ to obtain the steady state probability $\pi_{\text{$2$-LRU}}^*(\boldsymbol x)$ for state $\boldsymbol x.$}
\end{eg}

\subsection{{Multi-level Caching Algorithms}}\label{sec:multi-level-cache}
{Suppose there are $h$ caches in the linear cache network (see Figure ~\ref{fig:example} (b)).  Each item enters the cache network via cache $1$ and moves up one cache upon a cache hit.  For simplicity, {we denote $x_{(i, j)}$ as the identity of the item at position $j$ in cache $i$}, where $i=1,\cdots, h$ and $j=1,\cdots, m_i.$  Denote $\pi^*_A(\boldsymbol x)$ as the stationary probability of state $\boldsymbol x=(\boldsymbol x_1, \cdots, \boldsymbol x_h),$ {with $\boldsymbol x_i=(x_{(i, 1)}, \cdots, x_{(i, m_i))}.$}}  Following a similar sample path argument as discussed for $k$-LRU in Section~\ref{sec:meta-cache}, we have
\begin{theorem}\label{thm:thm-lrum}
Under the IRM, the steady state probabilities of LRU($\boldsymbol m$) satisfies the following form $\pi_{\text{LRU($\boldsymbol m$)}}^*(\boldsymbol x)=$
\begin{small}
\begin{align}\label{eq:steady-state-lrum}
&\prod_{i=1}^h\left(\prod_{j=1}^{m_i}p_{x_{(i,j)}}\right)^i \sum_{\Lambda(\bm{\hat{x}})\in\Upsilon(\boldsymbol x)}\sum_{\tilde{\Lambda}(\bm{\hat{x}})\in\Lambda(\boldsymbol x)}\sum_{\gamma{(\boldsymbol x)}\in\tilde{\Lambda}(\bm{\hat{x}})}\Xi({\gamma{(\boldsymbol x)}}).
\end{align}
\end{small}
\end{theorem}
The proof is presented in Appendix~\ref{app:station-lru-m}.   This form allows us to analyze the mixing time of LRU($\boldsymbol m$)\footnote{{CLIMB is LRU($\boldsymbol{m}$) with $h=m$ and size 1 for all levels.}} through the conductance argument in Section~\ref{sec:mixing}.

\begin{rem}
We can directly compute the hit probability of each algorithm once we have the stationary distribution.  Due to space constraints,  we relegate the details to Appendix~\ref{sec:hit-prob}.
\end{rem}

\section{Analysis of Mixing time}\label{sec:mixing-analysis}

We characterize the mixing times of LRU, FIFO, RANDOM, CLIMB, $k$-LRU, LRU($\boldsymbol m$) using the result in Lemma~\ref{lem:mixing-methodology} and Corollary~\ref{cor:mixing-methodology}.  As it is convention to present mixing time results with  $\epsilon=\frac{1}{2},$ here onwards we will omit $(\epsilon)$ in $t_{\text{mix}}$.  {Due to space constraints, we only present the proof for $k$-LRU and relegate all other proofs to Appendices~\ref{app:mixing-lru}, ~\ref{app:mixing-rand-fifo}, ~\ref{app:mixing-climb}, ~\ref{app:mixing-k-lru} and~\ref{app:mixing-lrum}.}

\subsection{Mixing Time of Single-Level Cache}
\subsubsection{Mixing Time of LRU}
We consider the IRM arrival process and denote the probability of requesting item $i$ by $p_i.$  It is easily shown that the Markov chain associated with the LRU algorithm is non-reversible, e.g., using the Kolmogorov condition \cite{kelly11reversibility}.   Hence, as discussed in Section~\ref{sec:reversibility}, we first need to construct the time reversal $P^{\text{LRU}, *}$, given the transition matrix $P^{\text{LRU}}$ of LRU.  Then the Markov chain with transition matrix $\frac{P^{\text{LRU}}+P^{\text{LRU}, *}}{2}$ is reversible.

\begin{theorem}\label{thm:upper-bound-lru}
The mixing time of LRU satisfies 
\begin{small}
\begin{align}
t_{\text{mix}}^{\text{LRU}}=&O\Bigg(\frac{n^{4m}\left(\prod_{j=1}^m p_j\right)^4\prod_{j=1}^{m-1}\left(1-\sum_{l=1}^{j}p_{n-l+1}\right)^2}{p_n^2\left(\prod_{j=1}^{m-1}\left(1-\sum_{l=1}^{j}p_{l}\right)\right)^4\prod_{j=n-m+1}^n p_j^2}\cdot\ln\left(\frac{\prod_{j=1}^{m-1}\left(1-\sum_{l=1}^{j}p_{n-l+1}\right)}{\prod_{j=n-m+1}^n p_j}\right)\Bigg).
\end{align}
\end{small}
\end{theorem}

\begin{corollary}\label{cor:upper-bound-lru}
If the popularity distribution is Zipf$(\alpha)$,  then the mixing time of LRU satisfies 
\begin{align}
t_{\text{mix}}^{\text{LRU}}=O(n^{(4\alpha+2) m+2}\ln n).
\end{align}
\end{corollary}

\subsubsection{Mixing Time of RANDOM and FIFO}
Since these algorithms have reversible Markov chains, we can use \eqref{eq:mixing-reversible} in Lemma~\ref{lem:mixing-methodology} to bound the mixing times.

\begin{theorem}\label{thm:upper-bound-rand}
The mixing times of RANDOM and FIFO satisfy
\begin{small}
\begin{align}
t_{\text{mix}}^{(\text{RANDOM},\ \text{FIFO})}=O\left(\frac{n^{2m}\left(\prod_{i=1}^m p_i\right)^6}{p_n^2\left(\prod_{i=n-m+1}^n p_i\right)^6}\ln\left(\frac{n^m\prod_{i=1}^m p_i}{\prod_{i=n-m+1}^n p_i}\right)\right).
\end{align}
\end{small}
\end{theorem}

\begin{corollary}\label{cor:upper-bound-rand}
If the popularity follows a Zipf distribution,  then the mixing times of RANDOM and FIFO satisfies 
\begin{align}
t_{\text{mix}}^{(\text{RANDOM},\ \text{FIFO})}=O(n^{(6\alpha+2)m+2}\ln n).
\end{align}
\end{corollary}

\subsubsection{Mixing Time of CLIMB}
We now turn to the CLIMB algorithm. Again we have a reversible Markov chain, and \eqref{eq:mixing-reversible} in Lemma~\ref{lem:mixing-methodology} is {applied}.

\begin{theorem}\label{thm:upper-bound-climb}
The mixing time of CLIMB satisfies 
\begin{small}
\begin{align}
t_{\text{mix}}^{\text{CLIMB}}=O\left(\frac{n^{2m}\left(\prod_{i=1}^m p_i^{m-i+1}\right)^6}{p_n^2\left(\prod_{i=n-m+1}^n p_i^{i-n+m}\right)^6}\ln\left(\frac{n^m\prod_{i=1}^m p_i^{m-i+1}}{\prod_{i=n-m+1}^n p_i^{i-n+m}}\right)\right).
\end{align}
\end{small}
\end{theorem}

\begin{corollary}\label{cor:upper-bound-climb}
If the popularity distribution is Zipf$(\alpha)$, then the mixing time of CLIMB satisfies 
\begin{align}
t_{\text{mix}}^{\text{CLIMB}}=O(n^{3\alpha m(m+1)+2m+2}\ln n).
\end{align}
\end{corollary}
{In the worst case, the most likely item takes $m-1$ steps to attain its final position, the next item takes $m-2$ steps, etc., giving the quadratic exponent term in the  mixing time bound of CLIMB.}

\subsubsection{Comparison of Mixing Times}
We are now in a position to compare the bounds on mixing times of our candidate algorithms for the simple cache system under a Zipf$(\alpha)$ distribution.  While the results are all upper bounds on mixing time, they should allow us to make a judgement on the worst case behaviors of each algorithm, and are a conservative estimate on likely performance in practice.  From Corollaries~\ref{cor:upper-bound-lru}, \ref{cor:upper-bound-rand}, and \ref{cor:upper-bound-climb}, for any $m\geq 2,$ the expected ordering in mixing times from smallest to largest is likely to be 
\begin{align}
t_{\text{mix}}^{\text{LRU}} \quad \leq \quad t_{\text{mix}}^{(\text{RANDOM},\ \text{FIFO})} \quad \leq\quad t_{\text{mix}}^{\text{CLIMB}}.
\end{align}
The phenomenon of LRU mixing faster than CLIMB has been observed in earlier in simulation studies \cite{hester85, bitner79}.  However, analytical characterization and comparison of mixing times of single and multi-level caching algorithms has not been done earlier.

Finally, we note that under the Zipf distribution, the congestion of LRU, FIFO, RANDOM, CLIMB are all polynomial in the size of the state space, i.e., the corresponding associated Markov chains are all rapid mixing.  The same phenomenon holds for a uniform distribution.  However, if we consider {non-heavy-tailed} distributions, such as a geometric distribution, mixing time is likely to be severely degraded.  This observation for LRU algorithm is discussed in \cite{fill96}.

\subsection{Mixing Time of Meta-Cache: $k$-LRU}\label{sec:k-lru-mixing}
We next characterize the mixing time of the meta-cache algorithm, $k$-LRU.  Again the Markov chain associated with the $k$-LRU algorithm is non-reversible (using the Kolmogorov condition \cite{kelly11reversibility}). Hence, using Corollary~\ref{cor:mixing-methodology} and the general form of the stationary distribution of $k$-LRU presented in Theorem~\ref{eq:thm-klru}, we should obtain mixing time bounds for this algorithm.

\begin{theorem}\label{eq:thm-mixing-klru}
The mixing time of $k$-LRU satisfies 
\begin{small}
\begin{align}\label{eq:mixing-time-klru}
&t_{\text{mix}}^{\text{$k$-LRU}}=O\Bigg(\frac{n^{4km+4(m-1)[k(k+1)m/2-1]}}{\left(\prod_{i=1}^k\left(\prod_{l=1+(i-1)m}^{im}p_{n-l+1}\right)^{k-i+1}\right)^2\cdot p_n^2}\left(\left(\prod_{j=1}^mp_j\right)^k\left(\frac{1}{1-\max_{\boldsymbol x\in\mathcal{S}, \delta}\left(\sum_{j\in\chi^{\Lambda^{\text{max}}(\boldsymbol x)}_{(\delta, \delta+1)}} p_j\right)}\right)^{\frac{k(k+1)m}{2}-1}\right)^4\nonumber\displaybreak[0]\\
&\qquad\qquad\qquad\qquad\cdot\ln\frac{1}{\prod_{i=1}^k\left(\prod_{l=1+(i-1)m}^{im}p_{n-l+1}\right)^{k-i+1}}\Bigg).
\end{align}
\end{small}
\end{theorem}

\begin{IEEEproof}
{Consider the stationary probability $\pi_{\text{$k$-LRU}}^*(\boldsymbol x)$ given in Equation~(\ref{eq:steady-state-klru}).  We will first obtain an upper bound $\pi^*_{\text{max}}$ and a lower bound $\pi^*_{\text{min}}$.  We omit the subscript $k$-LRU for brevity. }

{First, we characterize $\pi^*_{\text{max}}.$   Recall that we denote the sequence of items fixed by the arrangement $\bm{\hat{x}}$ by  $\xi_1, \cdots, \xi_\delta, \cdots, \xi_{|\hat{\mathcal{X}}|},$ where $\xi_\delta$ stands for the identity of the $\delta$-th item in this arrangement.  A sample path contains other items from $\mathcal{L}$ between these fixed items in such a way that the end result is $\bm{\hat{x}}$ under $k$-LRU, with the probability of requesting these other items being $\Xi({\gamma{(\boldsymbol x)}}).$ }

{We obtain an upper bound on $\pi^*_{\text{max}}$ by taking the maximum of each part in (\ref{eq:steady-state-klru}) as follows: (i) We take the maximum of the product form $\prod_{i=1}^k\left(\prod_{j=1}^{m}\mathcal{J}_{x_{(i,j)}}\right)^i$ over all states $\boldsymbol x\in\mathcal{S};$ (ii) We consider the maximum number of subclasses of sample paths and classes of sample paths; 
and  (iii) We maximize the sum of  $\Xi({\gamma{(\boldsymbol x)}})$ over all states, sample paths and classes. 
Then we have 
\begin{small}
\begin{align}
\pi^*(\boldsymbol x)&\leq\max_{\boldsymbol x\in\mathcal{S}}\left(\prod_{i=1}^k\left(\prod_{j=1}^{m}\mathcal{J}_{x(i,j)}\right)^i\right)\cdot|\tilde{\Upsilon}_{\Lambda(\bm{\hat{x}})}(\boldsymbol x)| \cdot |\Upsilon(\boldsymbol x)|\cdot \max_{\boldsymbol x\in\mathcal{S}}\left(\max_{\Lambda(\bm{\hat{x}})}\left(\sum_{\gamma{(\boldsymbol x)}\in\Lambda(\bm{\hat{x}})} \Xi({\gamma{(\boldsymbol x)}})\right)\right).
\end{align}
\end{small}
There are three terms in the above expression.  We upper bound the first term by using our assumption that $p_1\geq\cdots\geq p_n.$  }

{It is obvious $\max_{\boldsymbol x}\left(\prod_{i=1}^k\left(\prod_{j=1}^{m}\mathcal{J}_{x(i,j)}\right)^i\right)=\left(\prod_{j=1}^mp_j\right)^k.$  }

{The second term is upper bounded by $(\sum_{j=1}^j jm)!=(k(k+1)m/2)!,$ and the third term is upper bounded by $\left(\binom{n}{m-1}\right)^{k(k+1)m/2-1}$ since at most $m-1$ unique items can be requested between any two fixed items on the arrangement.} 

{We then consider the fourth term in the expression.  Now, for any $\bm{x}$
\begin{small}
\begin{align}\label{eq:third-term}
\sum_{\gamma{(\boldsymbol x)}\in\Lambda(\bm{\hat{x}})} \Xi({\gamma{(\boldsymbol x)}})
= \prod_{\delta=1}^{|\hat{\mathcal{X}}|} \left(\sum_{\zeta=0}^{\infty} \left(\sum_{j\in\chi^{\Lambda(\bm{\hat{x}})}_{(\delta, \delta+1)}} p_j\right) ^{\zeta}\right),
\end{align}
\end{small}
where $\chi^{\Lambda(\bm{\hat{x}})}_{(\delta, \delta+1)}$ is the set of items that can be requested between the $\delta$-th and $(\delta+1)$-th fixed items on the class of sample path $\Lambda(\bm{\hat{x}}).$ }

{It is clear that this term is maximized when the number of product terms is in maximum, since each term in the product is greater than or equal to one.  The maximum number of items that need to be fixed in a sample path leading to a state $\bm{\hat{x}}$ is $\sum_{j=1}^k jm=k(k+1)m/2$ (which happens when all the items in state $\boldsymbol x$ are distinct from each other).  }

{Denote $\Lambda^{\text{max}}(\boldsymbol x)=\arg\max_{\Lambda(\bm{\hat{x}})}\left(\sum_{\gamma{(\boldsymbol x)}\in\Lambda(\bm{\hat{x}})} \Xi({\gamma{(\boldsymbol x)}})\right)$ as the class of sample path that achieves the maximum in the third term~(\ref{eq:third-term}).   Let  $\chi^{\Lambda^{\text{max}}(\boldsymbol x)}_{(\delta, \delta+1)}$ be the set of items that can be requested between the $\delta$-th and $(\delta+1)$-th fixed items on this class of sample path. }

{Then we have 
\begin{small}
\begin{align}\label{eq:upper-stationary-prob}
&\max_{\boldsymbol x\in\mathcal{S}}\left(\max_{\Lambda(\bm{\hat{x}})}\left(\sum_{\gamma{(\boldsymbol x)}\in\Lambda(\bm{\hat{x}})} \Xi({\gamma{(\boldsymbol x)}})\right)\right)
= \max_{\boldsymbol x\in\mathcal{S}}\left(\sum_{\gamma{(\boldsymbol x)}\in\Lambda^{\text{max}}(\boldsymbol x)}\Xi(\gamma{(\boldsymbol x))}\right)\nonumber\displaybreak[0]\\
\stackrel{(a)}{=} &\max_{\boldsymbol x\in\mathcal{S}}\left( \prod_{\delta=1}^{k(k+1)m/2-1}\left(\sum_{\zeta=0}^{\infty} \left(\sum_{j\in\chi^{\Lambda^{\text{max}}(\boldsymbol x)}_{(\delta, \delta+1)}} p_j\right) ^{\zeta}\right)\right)\nonumber\displaybreak[1]\\
\stackrel{(b)}{\leq}&\prod_{\delta=1}^{k(k+1)m/2-1}\left(\sum_{\zeta=0}^{\infty}\left(\max_{\boldsymbol x\in\mathcal{S}, \delta}\left(\sum_{j\in\chi^{\Lambda^{\text{max}}(\boldsymbol x)}_{(\delta, \delta+1)}} p_j\right)\right)^{\zeta}\right)\nonumber\displaybreak[2]\\
=&\left(\frac{1}{1-\max_{\boldsymbol x\in\mathcal{S}, \delta}\left(\sum_{j\in\chi^{\Lambda^{\text{max}}(\boldsymbol x)}_{(\delta, \delta+1)}} p_j\right)}\right)^{k(k+1)m/2-1}.
\end{align}
\end{small}
Note that in~(\ref{eq:upper-stationary-prob}), (a) follows from the discussion above, and (b) is true since we take the maximum of the probabilities over all the items that can be requested between any two fixed items over all possible states, sample paths and classes. }

{Hence, we have 
\begin{small}
\begin{align}
&\pi^*(\boldsymbol x)\leq\left(\prod_{j=1}^mp_j\right)^k \left(\frac{k(k+1)m}{2}\right)!\left(\binom{n}{m-1}\right)^{\frac{k(k+1)m}{2}-1}\cdot\left(\frac{1}{1-\max_{\boldsymbol x\in\mathcal{S}, \delta}\left(\sum_{j\in\chi^{\Lambda^{\text{max}}(\boldsymbol x)}_{(\delta, \delta+1)}} p_j\right)}\right)^{k(k+1)m/2-1}\triangleq\pi^*_{\text{max}}.
\end{align}
\end{small}}

{Next, we characterize $\pi^*_{\text{min}}$. We obtain $\pi^*_{\text{min}}$ by taking the minimum of each part in Equation~(\ref{eq:steady-state-klru}): (i) We take the minimum over the product form $\prod_{i=1}^k\left(\prod_{j=1}^{m}\mathcal{J}_{x(i,j)}\right)^i$ over all states $\boldsymbol x\in\mathcal{S};$  (ii) {We consider the minimum number of subclass of sample path and the minimum number of class of sample path, i.e., $|\tilde{\Upsilon}_{\Lambda(\bm{\hat{x}})}(\boldsymbol x)|=1$ and $|\Upsilon(\boldsymbol x)|=1;$} and (iii) We minimize the sum of  $\Xi({\gamma{(\boldsymbol x)}})$ over all states, sample paths and classes. }

{Then we have
\begin{small}
\begin{align}
\pi^*(\boldsymbol x)&\geq\min_{\boldsymbol x\in\mathcal{S}}\left(\prod_{i=1}^k\left(\prod_{j=1}^{m}\mathcal{J}_{x(i,j)}\right)^i\right)\cdot 1\cdot 1 \cdot \min_{\boldsymbol x\in\mathcal{S}}\left(\min_{\Lambda(\bm{\hat{x}})}\left(\sum_{\gamma{(\boldsymbol x)}\in\Lambda(\bm{\hat{x}})} \Xi({\gamma{(\boldsymbol x)}})\right)\right).
\end{align}
\end{small}}

{Again, there are three terms in the above expression. We may lower bound the first term by using our assumption that $p_1\geq\cdots\geq p_n.$
It is obvious that this term is lower bounded when the least popular $km$ distinct items are stored in the cache, following the sequence of popularity decreasing with increasing levels.  Thus, we have the least popular $m$ items in the $k$-th level, the next least popular items $m+1$ to $2m$ items in the $(k-1)$-level, and so on until the whole cache is filled with the least popular $km$ distinct items, i.e.,
\begin{small}
\begin{align}
\min_{\boldsymbol x\in\mathcal{S}}\left(\prod_{i=1}^k\left(\prod_{j=1}^{m}\mathcal{J}_{x(i,j)}\right)^i\right)
&=\left(\prod_{j=1}^{m}p_{n-j+1}\right)^k\cdots\left(\prod_{j=(k-1)m+1}^{km}p_{n-j+1}\right)^1=\prod_{i=1}^k\left(\prod_{l=1+(i-1)m}^{im}p_{n-l+1}\right)^{k-i+1}. 
\end{align}
\end{small}
The second and third terms are both already lower bounded by $1$.}

{We then consider the fourth term in the expression.  Again, by~(\ref{eq:third-term}), we know that this term is greater than or equal to one (which happens when all terms equal to one). Since we only consider one class of sample path, we fix the items of the current state (which leads to the first term), and then consider all the possible requests between each two fixed items.  Since we want to lower bound the third term, we consider the case where there are no further requests between any two fixed items, i.e., $ \min_{\boldsymbol x\in\mathcal{S}}\left(\min_{\Lambda(\bm{\hat{x}})}\left(\sum_{\gamma{(\boldsymbol x)}\in\Lambda(\bm{\hat{x}})} \Xi({\gamma{(\boldsymbol x)}})\right)\right)=1.$}

{Hence, we have 
\begin{small}
\begin{align}
\pi^*(\boldsymbol x)&\geq\min_{\boldsymbol x\in\mathcal{S}}\left(\prod_{i=1}^k\left(\prod_{j=1}^{m}\mathcal{J}_{x(i,j)}\right)^i\right)\geq\prod_{i=1}^k\left(\prod_{l=1+(i-1)m}^{im}p_{n-l+1}\right)^{k-i+1}\triangleq\pi^*_{\text{min}}.
\end{align}
\end{small}}

{Therefore, given that $\Gamma=O(n^{km}),$ we have 
\begin{small}
\begin{align}
\rho\leq\frac{1}{\pi^*_{\text{min}}P_{\text{min}}}\cdot\Gamma^2\cdot(\pi^*_{\text{max}})^2=&O\Bigg(\frac{n^{2km}\cdot {n^{2(m-1)[k(k+1)m/2-1]}}}{\prod_{i=1}^k\left(\prod_{l=1+(i-1)m}^{im}p_{n-l+1}\right)^{k-i+1}\cdot p_n}\Bigg(\Bigg(\prod_{j=1}^mp_j\Bigg)^k\nonumber\displaybreak[0]\\
&\cdot\left(\frac{1}{1-\max_{\boldsymbol x\in\mathcal{S}, \delta}\Bigg(\sum_{j\in\chi^{\Lambda^{\text{max}}(\boldsymbol x)}_{(\delta, \delta+1)}} p_j\Bigg)}\right)^{\frac{k(k+1)m}{2}-1}\Bigg)^2\Bigg).
\end{align}
\end{small}}

{Then follow Corollary~\ref{cor:mixing-methodology}, we have~(\ref{eq:mixing-time-klru}).}
\end{IEEEproof}

\begin{corollary}\label{cor:upper-bound-klru}
If the popularity distribution is Zipf$(\alpha)$, then the mixing time of $k$-LRU satisfies 
\begin{align}\label{eq:mixing-klru-zip}
t_{\text{mix}}^{\text{$k$-LRU}}=O(n^{(k+1)k(2m-1)m+4(k\alpha-1)m+6}\ln n).
\end{align}
\end{corollary}

\begin{rem}
From~(\ref{eq:mixing-klru-zip}), it is clear that increasing the number of levels $k,$ the upper bound of mixing time increases.  We will see in Figure~\ref{fig:tau-distance-hit-prob-syn-multi} in Section~\ref{sec:distance-evaluation} that the meta-cache enhances the hit probability.   However, as $2$-LRU has a larger mixing time upper bound than LRU, this accuracy is at the expense of an increased mixing time.
\end{rem}

\subsection{Mixing time of Multi-level Cache: LRU($\boldsymbol m$)}\label{sec:lru-m-mixing}
Finally, we characterize the mixing time of the multi-level caching algorithm, LRU($\boldsymbol m$).  The LRU($\boldsymbol m$) algorithm  Markov chain is also non-reversible (by using the Kolmogorov condition \cite{kelly11reversibility}).  Hence, we should obtain mixing time bounds for this algorithm based on Corollary~\ref{cor:mixing-methodology}, by using the general form of the stationary distribution of LRU($\boldsymbol m$) presented in Theorem~\ref{thm:thm-lrum}. 

\begin{theorem}\label{thm:upper-bound-lrum}
The mixing time of LRU($\boldsymbol m$) satisfies 
\begin{small}
\begin{align}\label{eq:upper-bound-lrum}
&t_{\text{mix}}^{\text{LRU($\boldsymbol m$)}}=O\Bigg(\frac{n^{4m+4(m-1)(m_1+\cdots+hm_h-1)}}{\prod_{i=1}^h\left(\prod_{k=1+\sum_{j=1}^{i-1}m_{j}}^{\sum_{j=1}^i m_{j}}p_{n+k-m}\right)^{2i}\cdot p_n^2} \left(\prod_{i=1}^h\left(\prod_{k=1+\sum_{j=1}^{i-1}m_{h-j+1}}^{\sum_{j=1}^i m_{h-j+1}}p_k\right)^{h-i+1} \right)^4\nonumber\displaybreak[0]\\
&\cdot\left( \left(\frac{1}{1-\max_{\boldsymbol x\in\mathcal{S}, \delta}\left(\sum_{j\in\chi^{\Lambda^{\text{max}}(\boldsymbol x)}_{(\delta, \delta+1)}} p_j\right)}\right)^{\sum_{j=1}^h jm_j-1} \right)^4\ln\frac{1}{\prod_{i=1}^h\left(\prod_{k=1+\sum_{j=1}^{i-1}m_{j}}^{\sum_{j=1}^i m_{j}}p_{n+k-m}\right)^i}\Bigg).
\end{align}
\end{small}
where $m=m_1+m_2+\cdots+m_h$ and define $\sum_{j=1}^{i-1}m_{h-j+1}=0$ for $i=1.$
\end{theorem}

\begin{corollary}\label{cor:upper-bound-lrum}
If the popularity distribution is Zipf$(\alpha)$, then  the mixing time of LRU($\boldsymbol m$) satisfies 
\begin{align}\label{eq:upper-bound-lrum}
t_{\text{mix}}^{\text{LRU($\boldsymbol m$)}}= O(n^{(4m+4\alpha-6)(m_1+2m_2+\cdots+hm_h)+6}\ln n),
\end{align}
where $m=m_1+m_2+\cdots+m_h.$
\end{corollary}

\begin{rem}
{The mixing time bounds obtained in the above analysis are not as tight as ones that could be obtained directly by the characterization of the spectral gap of a Markov Chain and using the bound presented in (\ref{eq:eig-mixing}).  However, accurate determination of the spectral gap using the eigenvalues of a Markov chain (if real and positive) is typically difficult.  In the special case of LRU, \cite{phatarfod94,fill96} provide all the eigenvalues, {which can then be used to tighten the mixing time bound to $O(m\ln n).$}   However, such eigenvalue-based results are not generalizable to $2$-LRU and above, or indeed to any of the other algorithms considered in this paper. Hence, we use the simpler and uniform bound that, nevertheless, brings out first-order dependence on algorithm parameters: e.g., the exponent of the mixing time upper bound depends quadratically in $k$ and $h$ for $k$-LRU and for $h$-level LRU($\boldsymbol{m}$) paralleling the dramatic increase in mixing time observed in practice/numerically.}
\end{rem}

\section{A-LRU}\label{sec:alru}

{Our analysis thus far shows that different caching algorithms choose a different trade-off between speed and accuracy of learning. LRU has been widely used due to its speed of learning 
and ease of implementation. FIFO and RANDOM have been used to replace LRU in some scenarios since they are easier to implement with tolerable performance degradation. 
CLIMB has been numerically shown to have a higher hit ratio than LRU, at the expense of {increased time to reach this steady state in comparison to} LRU.  ARC is an online algorithm with a self-tuning parameter, which has good performance in some real systems but with complex implementation. $k$-LRU has relatively low complexity, which requires just one parameter, i.e., the number of meta caches $k-1$. We will see that these meta caches will provide a significant improvement in hit probability over LRU even for small values of $k$, with most of the gain achieved by $k=2$. LRU($\boldsymbol{m}$) too has relatively low complexity, and provides much higher hit probability over LRU. Both schemes pay for the higher hit probability in terms of much slower speed of learning.}

{Based on the previous analysis, we propose a novel hybrid algorithm, Adaptive-LRU (A-LRU), which captures advantages of LRU, 2-LRU and LRU(2), i.e., it learns both faster and better about the changes in the popularity.  A $k$-level version of A-LRU will allow for an interpolation between LRU, $2$-LRU, $\cdots,$ $k$-LRU, while at the same time incorporating multi-levels as in LRU($\boldsymbol{m}$).}

\begin{figure}
\centering
\includegraphics[width=0.9\linewidth]{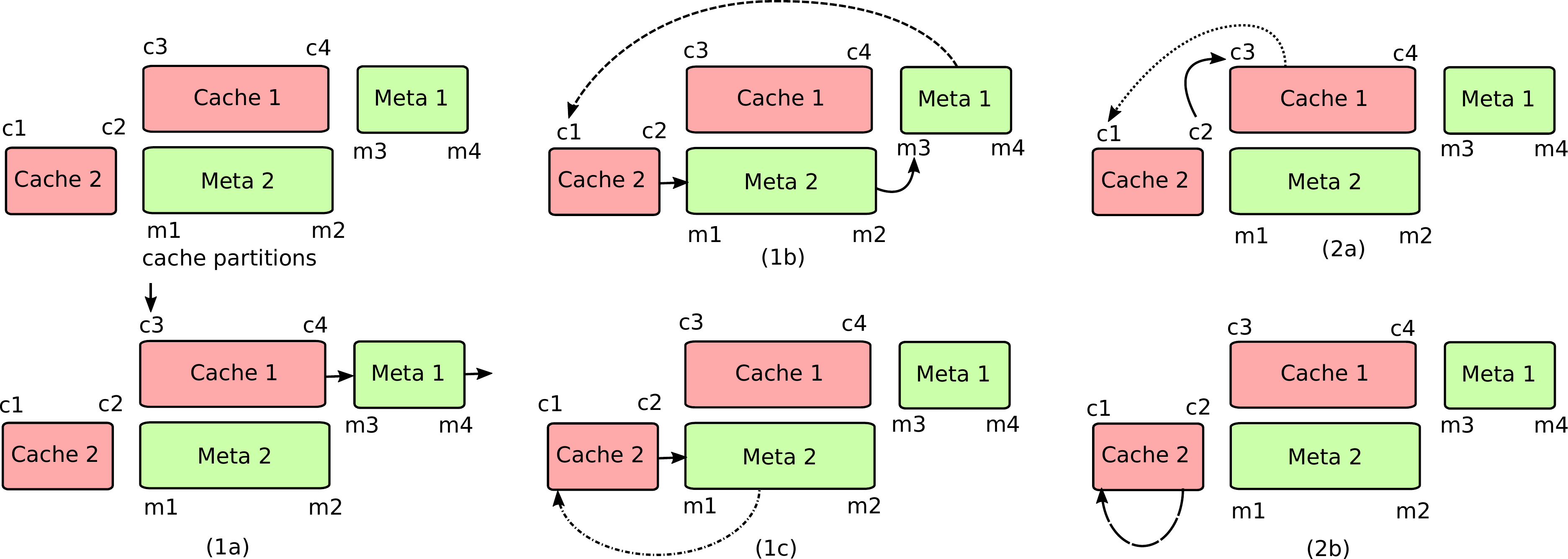}
\caption{Operation of the A-LRU algorithm.}
\label{fig:hybrid}
\vspace{-0.2in}
\end{figure}

\noindent{\textbf{{Adaptive-LRU (A-LRU):}}} {We define the quantities $c_1 = \min(1,\lfloor(1-\beta)m\rfloor),$  $c_2 = \lfloor(1-\beta)m\rfloor,$ $c_3=\lfloor(1-\beta)m\rfloor +1$ and $c_4=\max(m,\lfloor(1-\beta)m\rfloor +1)$, where  $\beta\in[0, 1]$ is a parameter.  We partition the cache into two parts with $C2$ defined as the positions from $c_1\cdots c_2$ and $C1$ as the positions from $c_3\cdots c_4.$  We also define the quantities  $m_1 = \min(1,\lfloor \beta m\rfloor),$  $m_2 = \lfloor\beta m\rfloor,$ $m_3=\lfloor \beta m\rfloor +1$ and $m_4=\max(m,\lfloor\beta m\rfloor +1).$  We associate positions $m_1\cdots m_2$ with meta cache\footnote{{Meta cache is also called virtual cache, which only stores meta-data.}} $M2$ and $m_3\cdots m_4$ with meta cache $M1.$  Note that value $m_1=m_2=0$ is an extreme point that yields behavior similar to 2-LRU, while $m_3=m_4=m+1$ yields LRU.   {See Section~\ref{sec:evaluation}.}  The cache partitions are shown in Figure~\ref{fig:hybrid}.}

{Let us denote the meta-data associated with a generic item $i$ by $M(i).$  If item $i$ is requested, the operation of A-LRU is illustrated in Figure~\ref{fig:hybrid}.
 There are two possibilities:\\
(1) {\bf Cache miss}, then there are three cases to consider:}

{(1a) $M(i) \not\in M1 \cup M2:$ If $c_3 \neq m+1,$ $i$ is inserted into cache position $l=c_3,$ else (extreme case similar to 2-LRU) $M(i)$ is inserted into meta cache position $l=m_3.$  Cache/meta cache items in positions  greater than $l$ move back one position, and the last meta-data is evicted;}

{(1b) $M(i) \in M1:$ Item $i$ is inserted into position $c_1$, all other items in $C2$ move back one position, the meta data of item in cache position $c_2$ is placed in  position $m_1,$ all other meta-data items move back one position, and the meta data in position $m_2$ moves to position $m_3$;}

{(1c) $M(i) \in M2:$ If $c_1 = 1, $ item $i$ is inserted into position $l=c_1.$  All other items in $C2$ move back one position, and the meta data of item in cache position $c_2$ is placed in  position $m_1.$   Note that this situation cannot occur in the extreme case of LRU, since $M2$ is always empty for LRU;}

\noindent {(2) {\bf Cache hit}, then there are two cases to {consider}}

{(2a) $i \in C1$  (suppose in position $j$): If $c_1=1,$ then item $i$ moves to cache position $l=c_1$, else (extreme case of LRU) item $i$ moves to cache position $l=c_3.$  If $l =c_1,$  all other items in $C2$ move back one position, the item in cache position $c_2$ is placed in position $c_3,$ all other items in $C1$ upto position $j$ move back one position.  If $l=c_3$ (extreme case of LRU),  all other items in $C1$ upto position $j$ move back one position.}

{(2b) $i \in C2$ (suppose in position $j$): Item $i$ moves to cache position $c_1,$ and all other items in positions $\min(2,c_2)$ to $j-1$ move back one position.}

{Finally, note that the A-LRU setup can be generalized to as many levels as desired by simply ``stacking up'' sets of real and meta caches, and following the same caching and eviction policy outlined above (where (1a) would apply to the top level, while (1c) and (2b) would apply to the bottom level).  In that case, it would be parameterized by $\beta_1, \cdots, \beta_k$ with $\sum_{i=1}^k\beta_i=1,$ for a $k$-level A-LRU. The cache size at level $i$ is $m-\lfloor(1-\beta_i) m\rfloor$, and meta-cache size at level $i$ is $\lfloor(1-\beta_i)m\rfloor.$  {Note that small corrections have to be made to the above selections for a particular value of $k$  (as we did in the case of $k=2$ described above) to ensure that the total amount of cache space is limited to $m$.} }

\textbf{{Dynamic A-LRU:}}
{Whereas in our description of A-LRU, we use a fixed partitioning parameter $\beta$, the algorithm (and an implementation of it) can easily consider time-varying $\beta$ values.  For concreteness, we consider a $k$ levels A-LRU with a sequence $\{\chi_1(t), \chi_2(t), \cdots, \chi_{k-1}(t)\}_{t=1}^\infty$, with each term going to $0$ as the number of requests go to infinity, satisfying (i) $\sum_t \chi_i(t)\rightarrow\infty$; (ii) $\sum_t\chi_i^2(t)<\infty$; and (iii) $\chi_i(t)/\chi_{i+1}(t)\rightarrow 0.$  Here, $\chi_1$ stands for the proportion of LRU to the rest, $\chi_2$ stands for the proportion between $2$-LRU and $3$-LRU to the rest, etc.  A typical choice of sequences will be $\chi_i(t)=m/(m+\max(0,t-T_i)^{\frac{i+1}{2i}}/c_i),$ where $t$ counts the number of requests, and $T_i, c_i>0$ are parameters to be varied.  Under such setting, the $\beta$\/s in the previous definition of A-LRU satisfy that (i) at level $i\leq k-1$,  it is $(1-\chi_1(t))(1-\chi_2(t))\cdots(1-\chi_{i-1}(t))\chi_i(t),$ and (ii) at level $k$, it is $(1-\chi_1(t))(1-\chi_2(t))\cdots(1-\chi_{k-1}(t)).$}

{In particular,  we consider the $2$-level A-LRU shown in Figure~\ref{fig:hybrid}.  Here, the $\beta$\/s are $\beta_1(t)=m/(m+\max\{0, t-T\}/c)$ and $\beta_2(t)=1-\beta_1(t)$, where $T$ and $c$ are parameters.  With such a sequence of $\beta$\/s, A-LRU will start at $1$ (LRU) and (slowly) decrease to $0$ (2-LRU). }

\begin{remark}
{If the popularity distribution changes with time (in Section~\ref{sec:trace-sim}), we should only consider constant $\beta$ algorithms.  These two distinctions follow from stochastic approximation ideas where while decreasing step-size algorithms can converge to optimal solutions in stationary settings, constant step-size algorithms provide good tracking performance for non-stationary settings.  Therefore, we denote A-LRU(f) and A-LRU(d) to distinguish A-LRU with a fixed or dynamic $\beta$, respectively.}
\end{remark}

\section{Performance Evaluations}\label{sec:evaluation}
\subsection{Permutation Distance}\label{sec:distance-evaluation}

\begin{figure*}[ht]
\centering
\begin{minipage}{.32\textwidth}
\centering
\includegraphics[width=1\columnwidth]{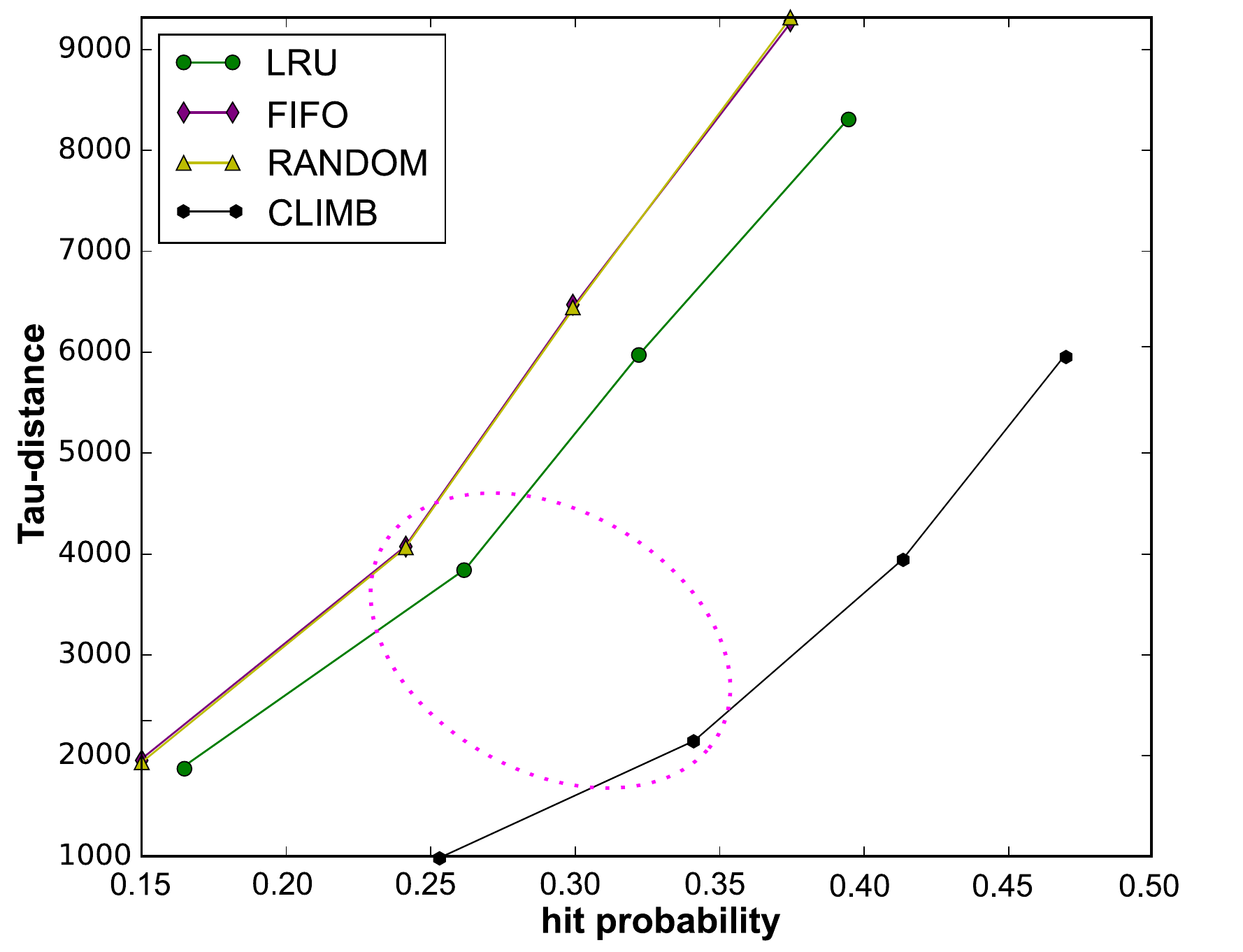}
\caption{{Stationary $\tau$-distance vs. hit probability for {single-level} caching with IRM arrivals. }}
\label{fig:tau-distance-hit-prob-syn-single}
\end{minipage}\hfill
\begin{minipage}{.32\textwidth}
\centering
\includegraphics[width=1\columnwidth]{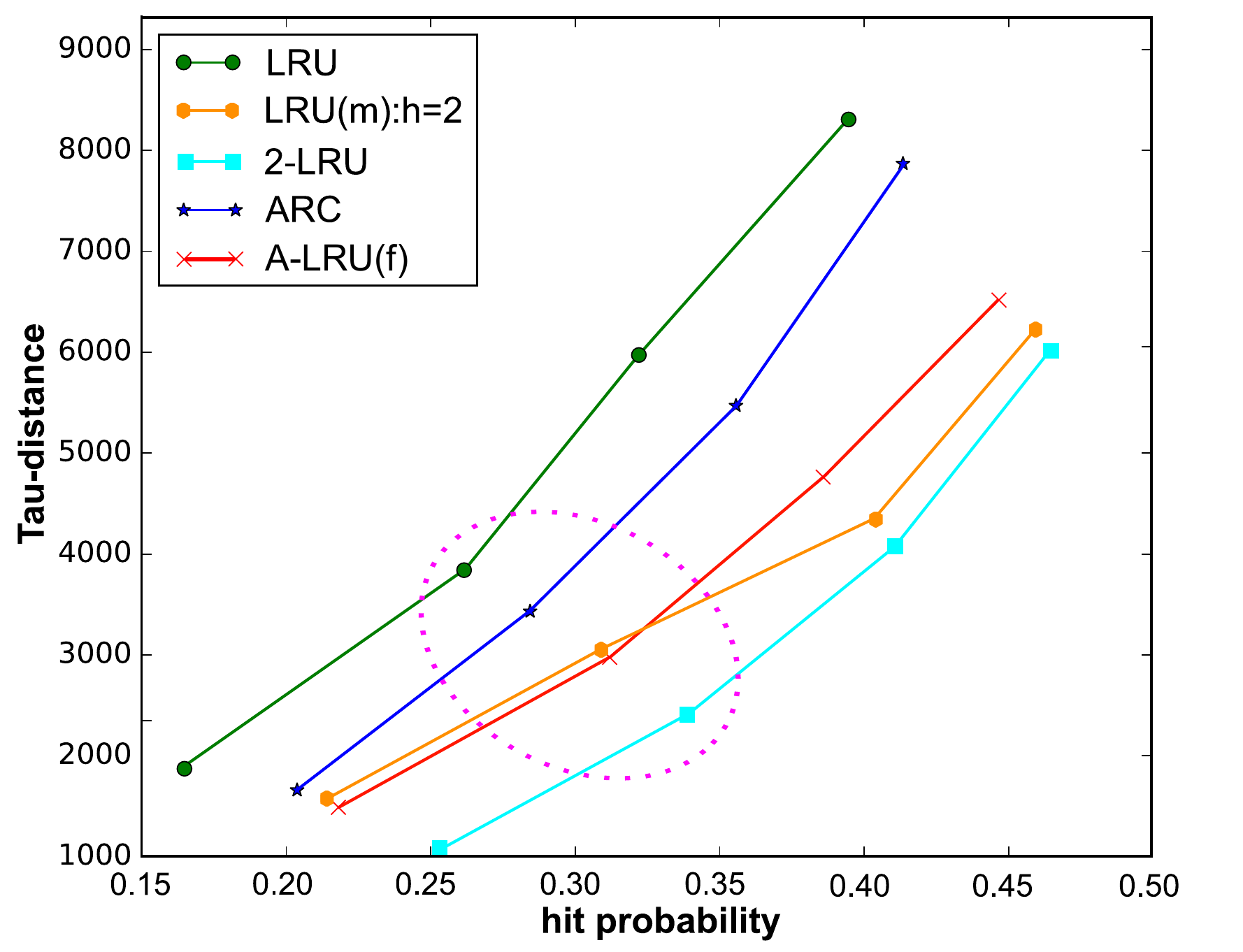}
\caption{{Stationary $\tau$-distance vs. hit probability for {multi-level} caching with IRM arrivals.} }
\label{fig:tau-distance-hit-prob-syn-multi}
\end{minipage}\hfill
\begin{minipage}{.32\textwidth}
\centering
\includegraphics[width=1\columnwidth]{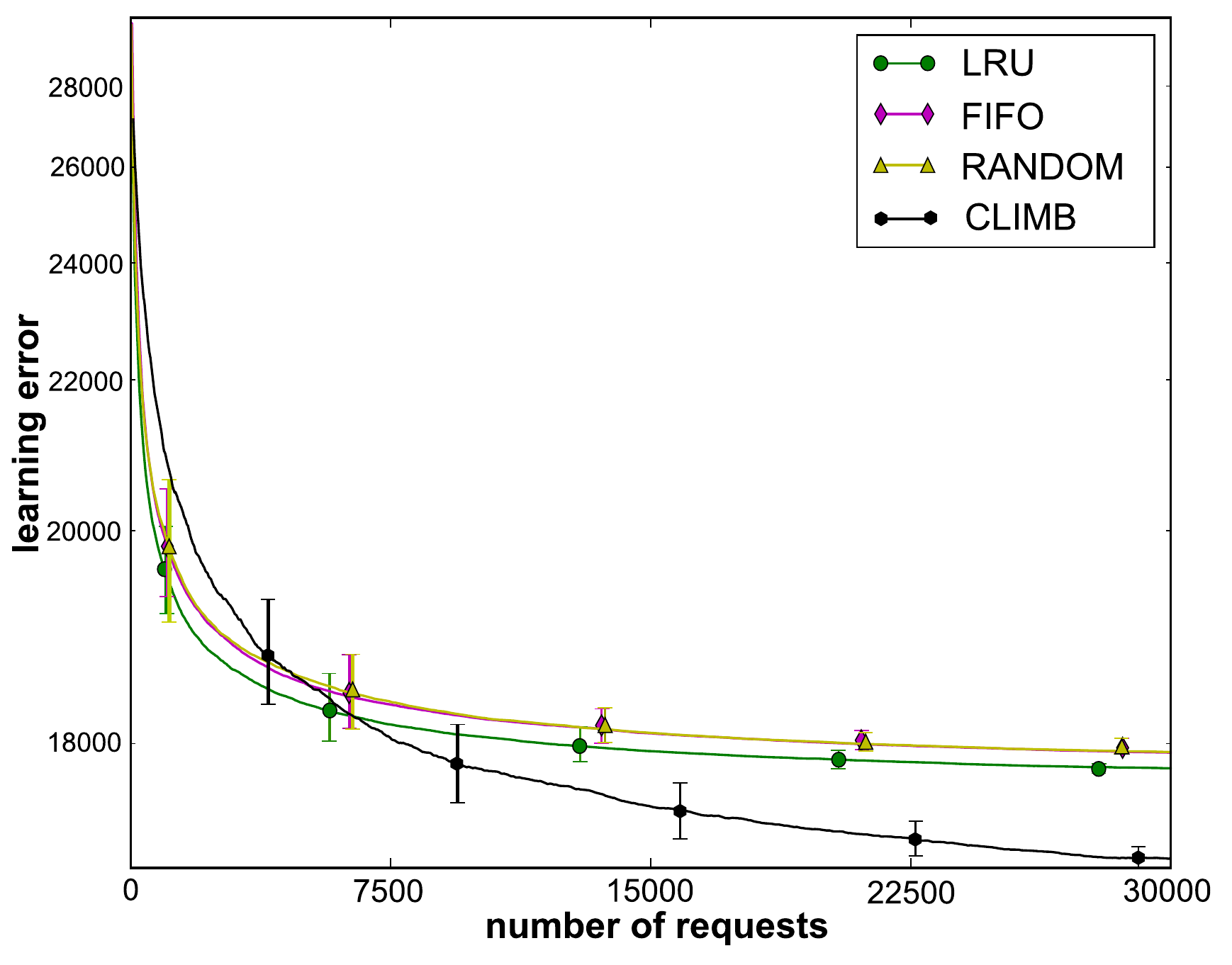}
\caption{{Dynamic learning error of {single-level} caching under IRM arrivals.}}
\label{fig:learning-error-single}
\end{minipage}
\vspace{-0.2in}
\end{figure*}

{Since the $\tau$-distance characterizes  how accurately an
  algorithm learns the popularity distribution, a smaller
  $\tau$-distance should correspond to a larger hit probability.  
  We illustrate how different algorithms perform using a content library size of $n=20,$ and using caches of size $2, 3, 4, 5.$   Figures~\ref{fig:tau-distance-hit-prob-syn-single} and~\ref{fig:tau-distance-hit-prob-syn-multi} compare the $\tau$-distance and hit probabilities of various caching algorithms.  The points on each curve correspond to cache size of $2, 3, 4, 5$ from left to right.  For example, all the points in the dashed pink circle correspond to cache size of $3$.   Since the cache size should be an integer, we partition the cache for LRU($\boldsymbol m$) and A-LRU such that the size of cache $1$ is always $1$, and the remaining cache size is allocated to cache $2$. Note that this is simply for illustration, and A-LRU's highest hit-probability is when it mirrors $2-$LRU.  From Figures~\ref{fig:tau-distance-hit-prob-syn-single} and~\ref{fig:tau-distance-hit-prob-syn-multi}, we can see that the $\tau$-distance and hit probability follow the same rule, i.e., a smaller $\tau$-distance corresponds to a larger hit probability, which is as expected.}
%

\begin{figure*}[ht]
\centering
\begin{minipage}{.32\textwidth}
\centering
\includegraphics[width=1\columnwidth]{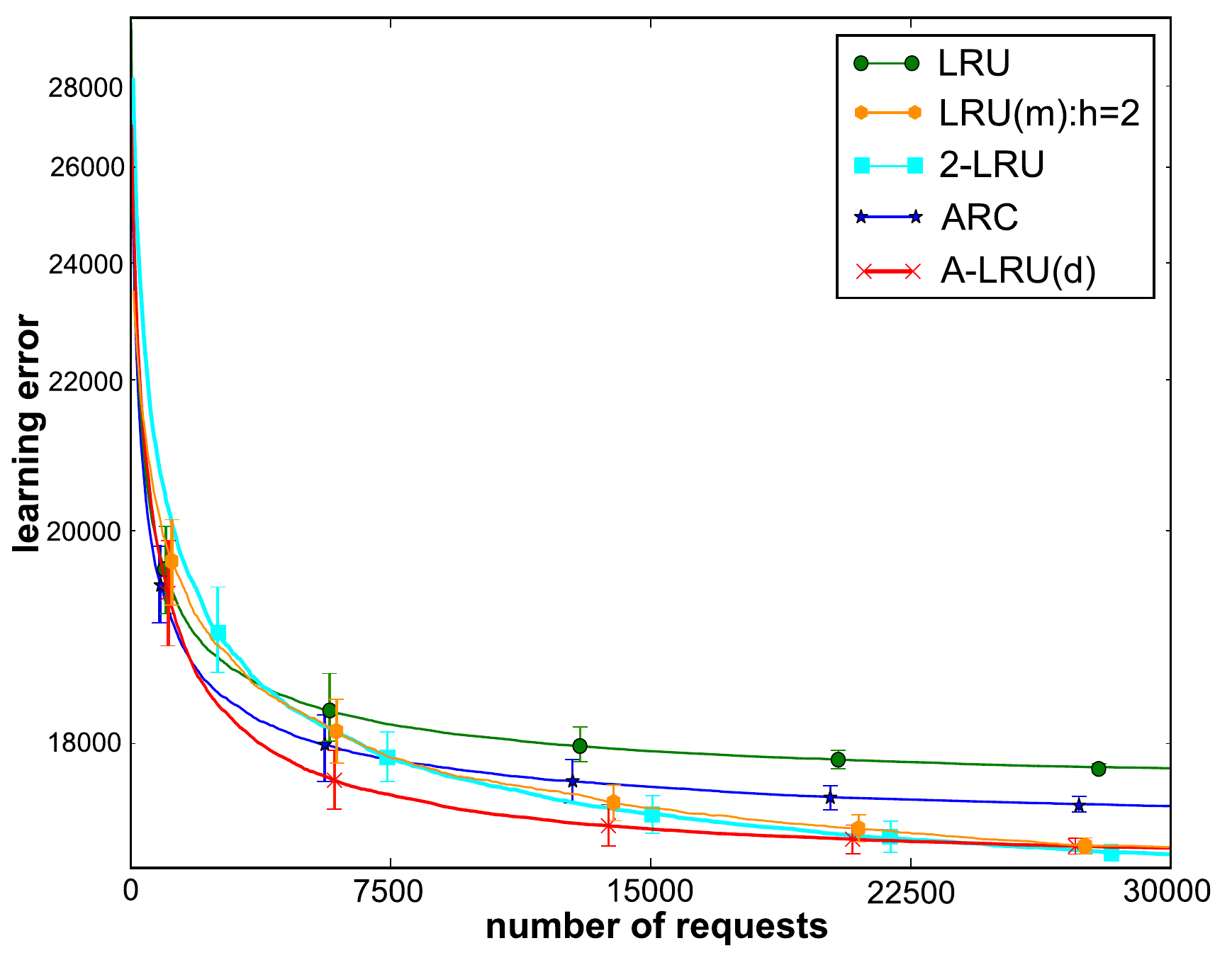}
\caption{{Dynamic learning error of {multi-level} caching under IRM arrivals.}}
\label{fig:learning-error-multi}
\end{minipage}\hfill
\begin{minipage}{.32\textwidth}
\centering
\includegraphics[width=1\columnwidth]{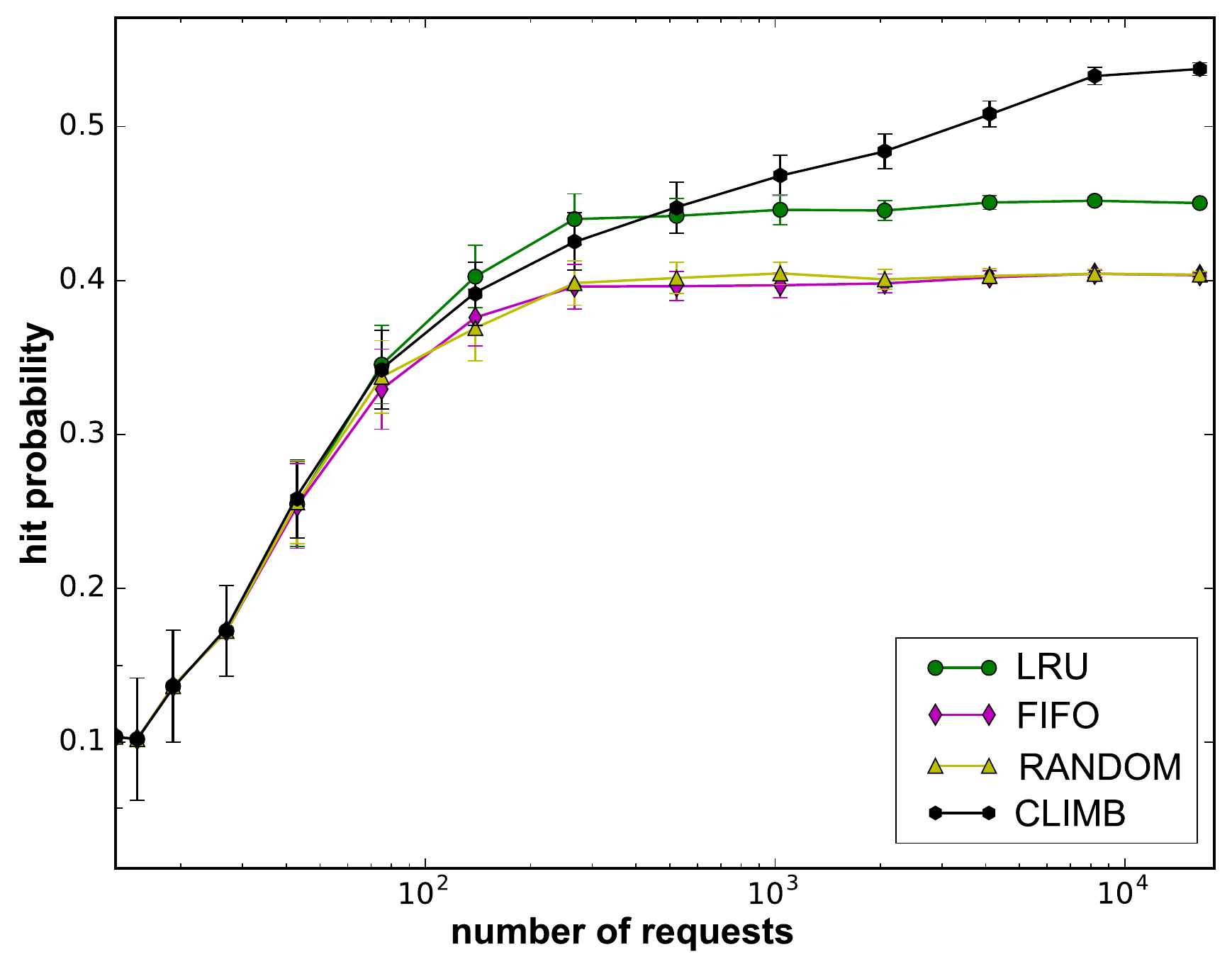}
\caption{{Hit probability of {single-level} caching under IRM arrivals.}}
\label{fig:hit-prob-time-single}
\end{minipage}\hfill
\begin{minipage}{.32\textwidth}
\centering
\includegraphics[width=1\columnwidth]{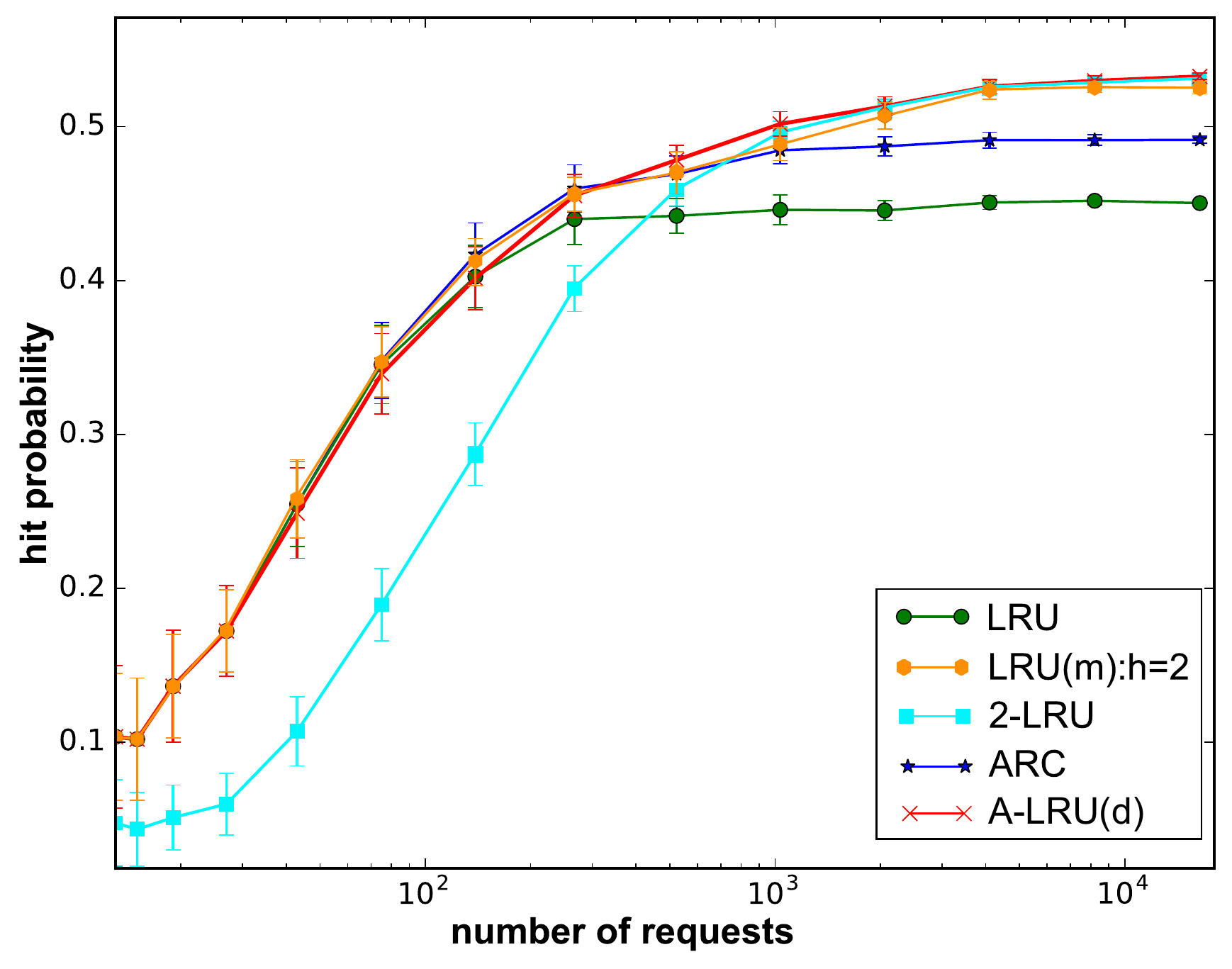}
\caption{{Hit probability of {multi-level} caching under IRM arrivals. }}
\label{fig:hit-prob-time-multi}
\end{minipage}
\vspace{-0.2in}
\end{figure*}

\begin{rem}
The specific parameters in~(\ref{eq:general-kendall}) used to produce Figures~\ref{fig:tau-distance-hit-prob-syn-single} and~\ref{fig:tau-distance-hit-prob-syn-multi}  are as follows.  We consider a Zipf popularity distribution with $\alpha=0.8$; for simplicity, we set the element weights as $w_i=n-i+1$, and the swapping cost $\zeta_i=\log i$ for $i>1,$ and $\zeta_1=0.1$.  Since different choices of weights result in different values of the $\tau$-distance, the y-axis value in Figures~\ref{fig:tau-distance-hit-prob-syn-single} and~\ref{fig:tau-distance-hit-prob-syn-multi}  is only used to show the relative difference between algorithms. 
\end{rem}

\subsection{Learning Error}\label{sec:error-evaluation}
{We consider a cache system with $(n, m)=(20, 4),$ where $n$ is the total number of different items in the library and $m$ is the cache size.  We consider requests following the IRM model, with a Zipf popularity law with parameter $\alpha=0.8$.  We compare the performance of LRU, FIFO, RANDOM, CLIMB, LRU($\boldsymbol m$), $2$-LRU, ARC and A-LRU with respect to the stationary hit probability and learning error.  In particular, we consider a two-level version of A-LRU which is characterized by a parameter $\beta\in[0,1]$ that determines the interpolation rate between LRU and $2$-LRU; the detailed description is in Section~\ref{sec:alru}.  For LRU($\boldsymbol m$), we consider the capacity allocation as $(m_1, m_2)=(1, 3).$ The corresponding stationary hit probability of these algorithms are $0.325$, $0.308$, $0.308$, $0.414$, $0.407$, $0.408$, $0.352$, $0.408,$ respectively.}

{The learning error of these algorithms as a function of the number of requests received is illustrated in Figure~\ref{fig:learning-error-single} for single-level caching algorithms (from Figure~\ref{fig:example} (a)),  and Figure~\ref{fig:learning-error-multi} for multi-level caching algorithms,  (from Figure~\ref{fig:example} (b), (c), (d)),  respectively, where the y-axis is shown in a logarithmic scale.  Note that the results for A-LRU presented in Figure~\ref{fig:learning-error-multi} used $c=500$ and $T=1250.$   We see immediately that FIFO and RANDOM have higher learning error than the other algorithms, regardless of the number of requests, which corresponds to the smallest stationary hit probability.  This shows why their performance is poor,  no matter how long they are trained.  The learning error for LRU decreases fast initially and then levels off, whereas 2-LRU and LRU($\boldsymbol m$) have a slower decay rate, but the eventual error is lower than that of LRU.  This corresponds to faster mixing of LRU but a poorer eventual accuracy ($\tau$-distance) as compared to 2-LRU and LRU($\boldsymbol m$), which are formally characterized in Section~\ref{sec:error}.  ARC has a good performance initially, but it too levels off to an error larger than $2$-LRU.  CLIMB eventually has good performance, but it has a much slower decay rate.  This corresponds to the slow mixing of CLIMB when compared to LRU, FIFO and RANDOM, consistent with our analysis in Theorems~\ref{thm:upper-bound-rand},~\ref{thm:upper-bound-climb} and~\ref{eq:thm-mixing-klru}, Section~\ref{sec:mixing}.   A-LRU learns fast initially, and then smoothly transitions to learning accurately, which captures both the merits of LRU and $2$-LRU, i.e., accurate learning and fast mixing, and is able to attain a low learning error quickly.}

\subsection{{Hit Probabilities}}
{The effects seen in Figures~\ref{fig:learning-error-single} and~\ref{fig:learning-error-multi} are also visible in the evolution of hit probabilities shown in Figures~\ref{fig:hit-prob-time-single} and~\ref{fig:hit-prob-time-multi}, respectively, where the x-axis is on a logarithmic scale.  Here, we choose $(n=150, m=30)$ in order to explore a range of cache partitions for LRU($\boldsymbol m$) and A-LRU from $(0,30)$--$(30,0)$.  We compare the upper envelope of achievable hit probability by LRU($\boldsymbol m$) and A-LRU with various other caching algorithms.  We find that for any given learning time (requests), there is a cache partition such that A-LRU will attain a higher hit probability after learning for that time.  These effects become more pronounced as the partition space (cache size) available for A-LRU increases.  }

{Finally, we characterize the impact of the number of levels on the performance. 
 We compare the hit probability of $k$-LRU for $k=1, 2, 3, 4$ with $(n, m)=(50, 10).$ We observe that as the number of cache levels $k$ increases, the resulting algorithm achieve a higher hit probability (i.e., higher accuracy) at the expense of much larger number of requests (i.e., larger mixing time), which is consistent with the mixing time analysis of $k$-LRU shown in Section~\ref{sec:k-lru-mixing}. Due to space constraints, the plot is shown in Appendix~\ref{app:levels}.}

\section{Trace-based Simulations}\label{sec:trace-sim}

The ideas presented thus far have been based on the hypothesis that the request distribution changes dynamically, and hence an optimal caching algorithm should track the changes at a time scale consistent with the time scale of change.  We now validate this hypothesis and the benefits of using the A-LRU algorithm using two different trace-based simulations.

\subsection{YouTube Trace}
We use a publicly available data trace \cite{zink08} that was extracted from a $2$-week YouTube request traffic dump between June $2007$ and March $2008$. The trace contains a total of $611,968$ requests for $303,331$ different videos.  About $75\%$ of those videos were requested only once during the trace.   There is no information on the video sizes. We therefore assume that the cache size is expressed as the number of videos that can be stored in it.  This should have a low impact if the correlation between video popularity and video size is low.  
We find that $\alpha=0.605$ is the best fit for a Zipf distribution.  However, a detailed inspection shows that this trace exhibits significant non-stationarity, i.e., the popularity distribution is time-varying.  

\begin{figure*}[ht]
\centering
\begin{minipage}{.32\textwidth}
\centering
\includegraphics[width=1\columnwidth]{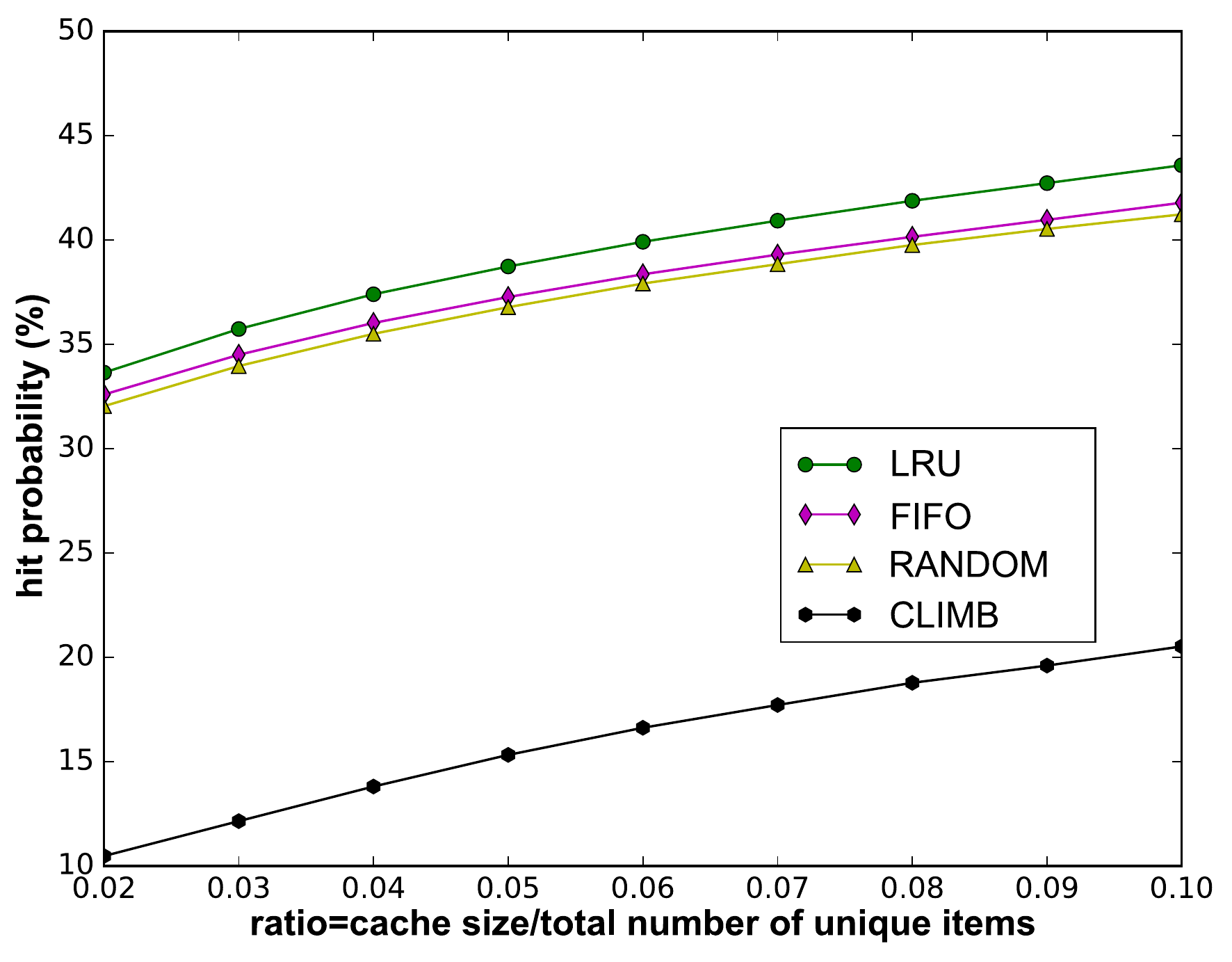}
\caption{Hit probability vs. cache size for various {single-level} caching algorithms with two-week long YouTube trace \cite{zink08}.}
\label{fig:youtube-twoweek-p02-single}
\end{minipage}\hfill
\begin{minipage}{.32\textwidth}
\centering
\includegraphics[width=1\columnwidth]{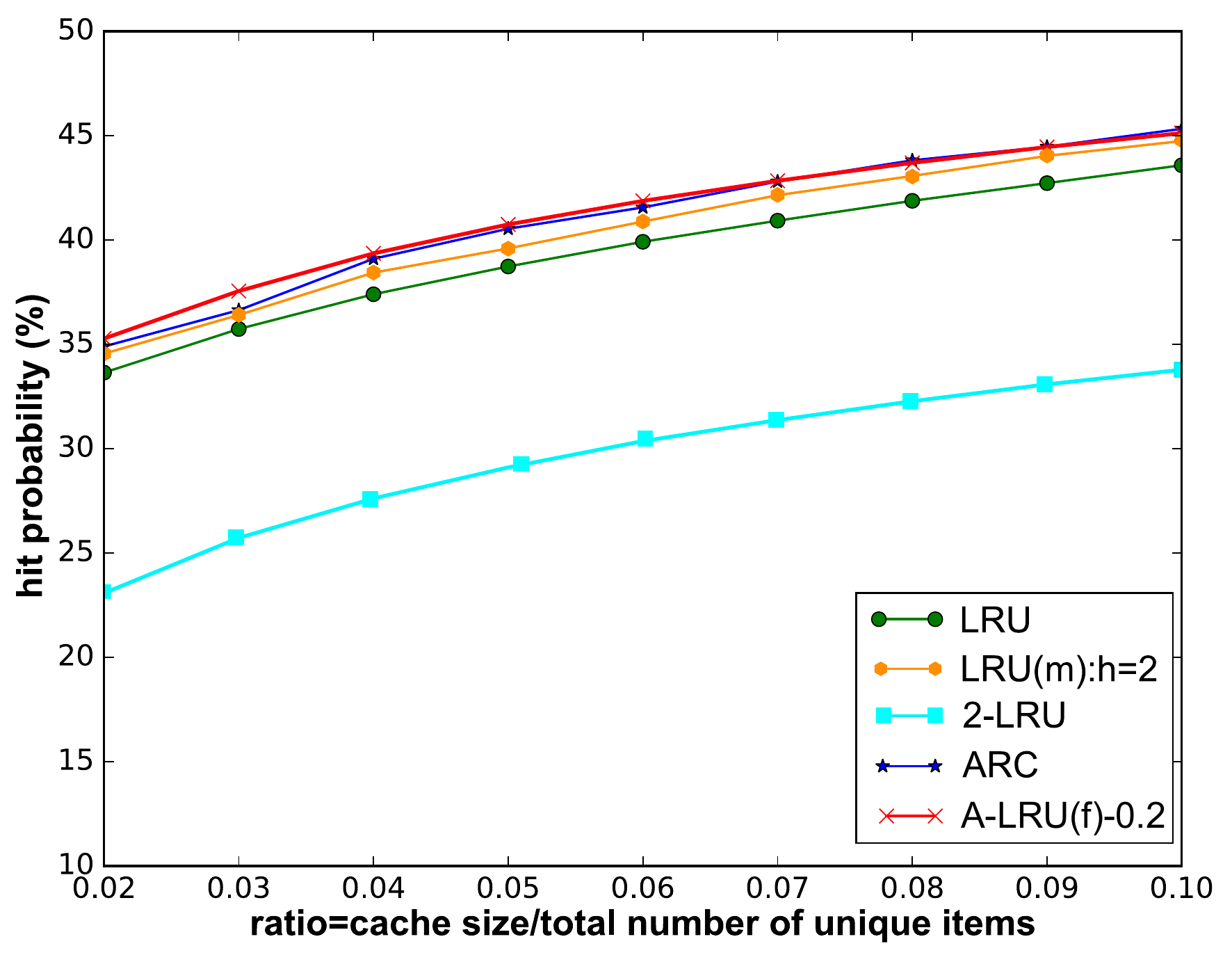}
\caption{Hit probability vs. cache size for various {multi-level} caching algorithms with two-week long YouTube trace \cite{zink08}.}
\label{fig:youtube-twoweek-p02-multi}
\end{minipage}\hfill
\begin{minipage}{.32\textwidth}
\centering
\includegraphics[width=1\columnwidth]{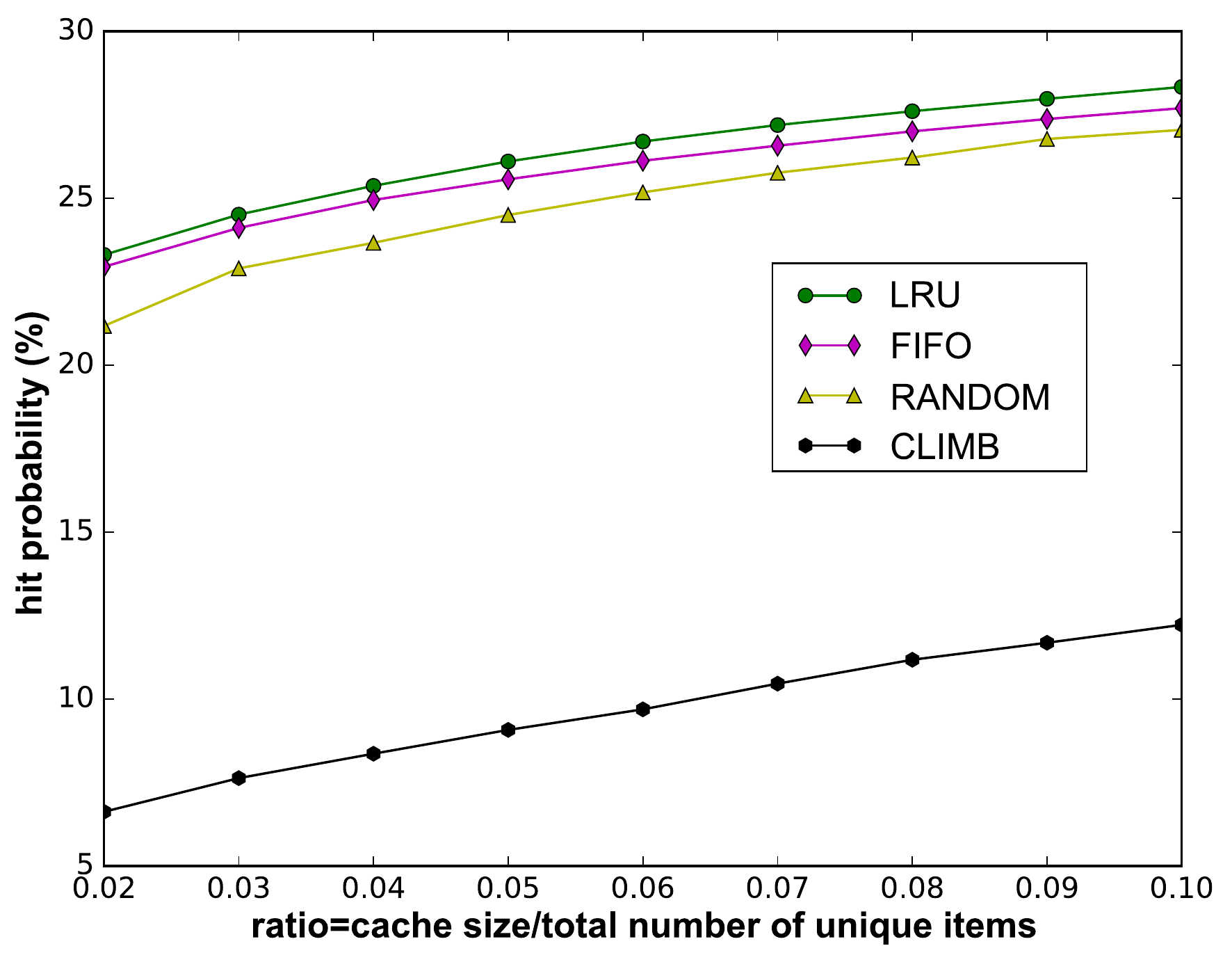}
\caption{Hit probability vs. cache size for various {single-level} caching algorithms with one particular day YouTube trace \cite{zink08}.}
\label{fig:youtube-day9-p05-single}
\end{minipage}
\vspace{-0.2in}
\end{figure*}

We compare the hit performance of different algorithms by varying cache size.  Figures~\ref{fig:youtube-twoweek-p02-single} and~\ref{fig:youtube-twoweek-p02-multi} depicts the hit probability as a function of the cache size when the total number of unique videos is $n=303,331$, and we use the ratios $m/n=0.01, \cdots, 0.10.$  For ease of visualization, we only depict A-LRU with the optimal $\beta,$ which outperforms all the other caching algorithms.  

\begin{figure*}[htbp]
\centering
\begin{minipage}{.32\textwidth}
\centering
\includegraphics[width=1\columnwidth]{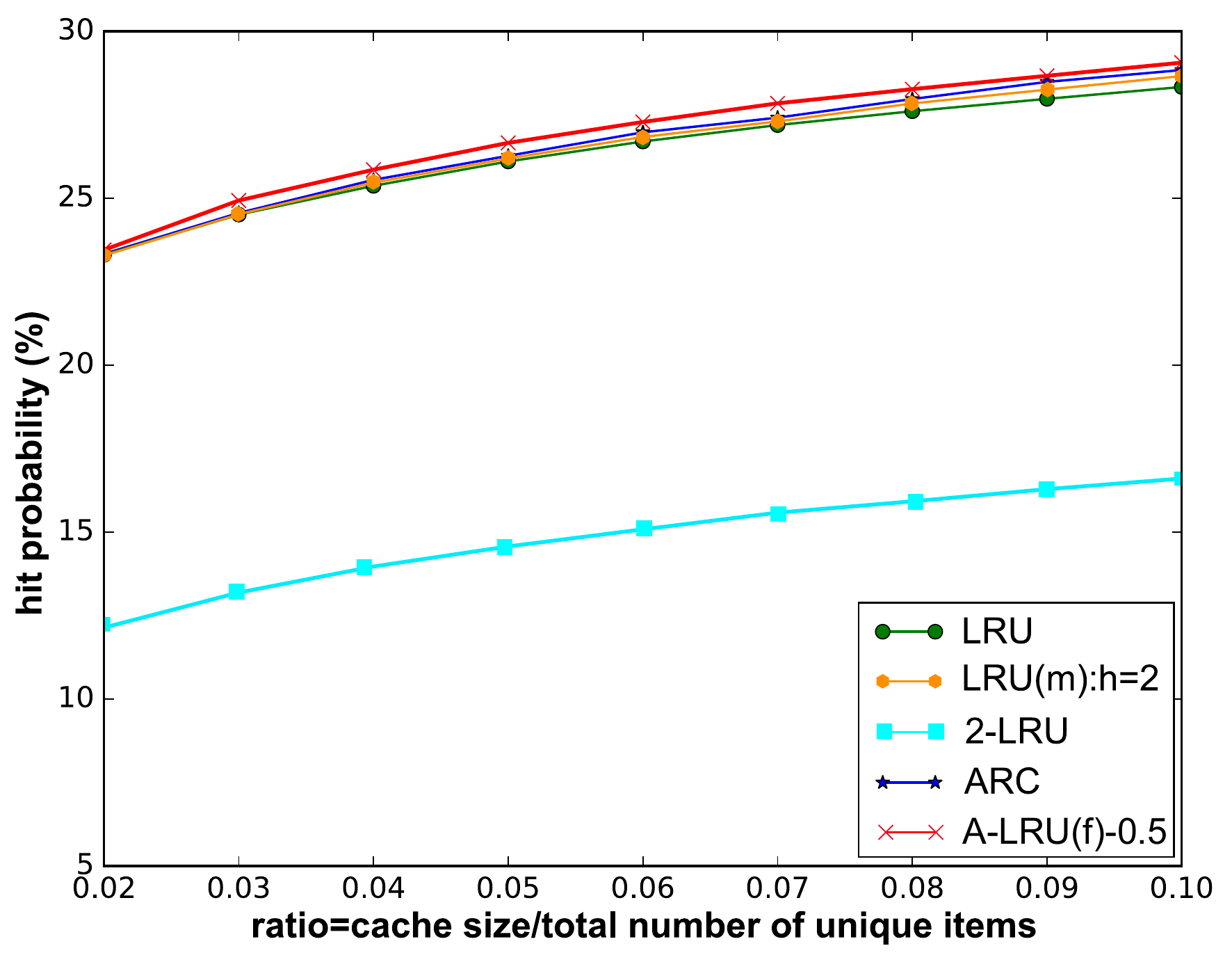}
\caption{Hit probability vs. cache size for various {multi-level} caching algorithms with one particular day YouTube trace \cite{zink08}.}
\label{fig:youtube-day9-p05-multi}
\end{minipage}\hfill
\begin{minipage}{.32\textwidth}
\centering
\includegraphics[width=1\columnwidth]{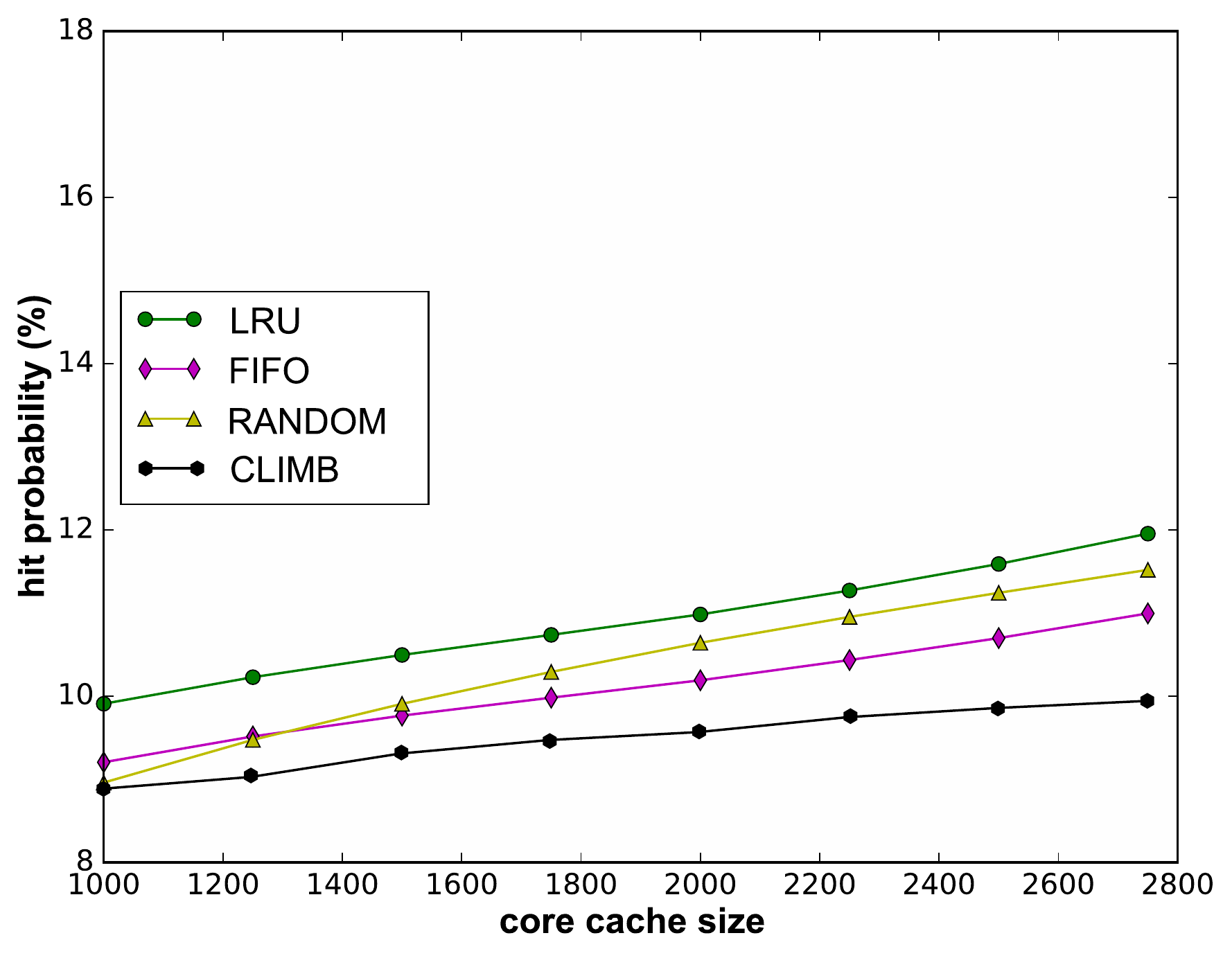}
\caption{Hit probability vs. cache size for various {single-level} caching algorithms with SD Network trace \cite{bianchi13check} for ICN.}
\label{fig:icn-single}
\end{minipage}\hfill
\begin{minipage}{.32\textwidth}
\centering
\includegraphics[width=1\columnwidth]{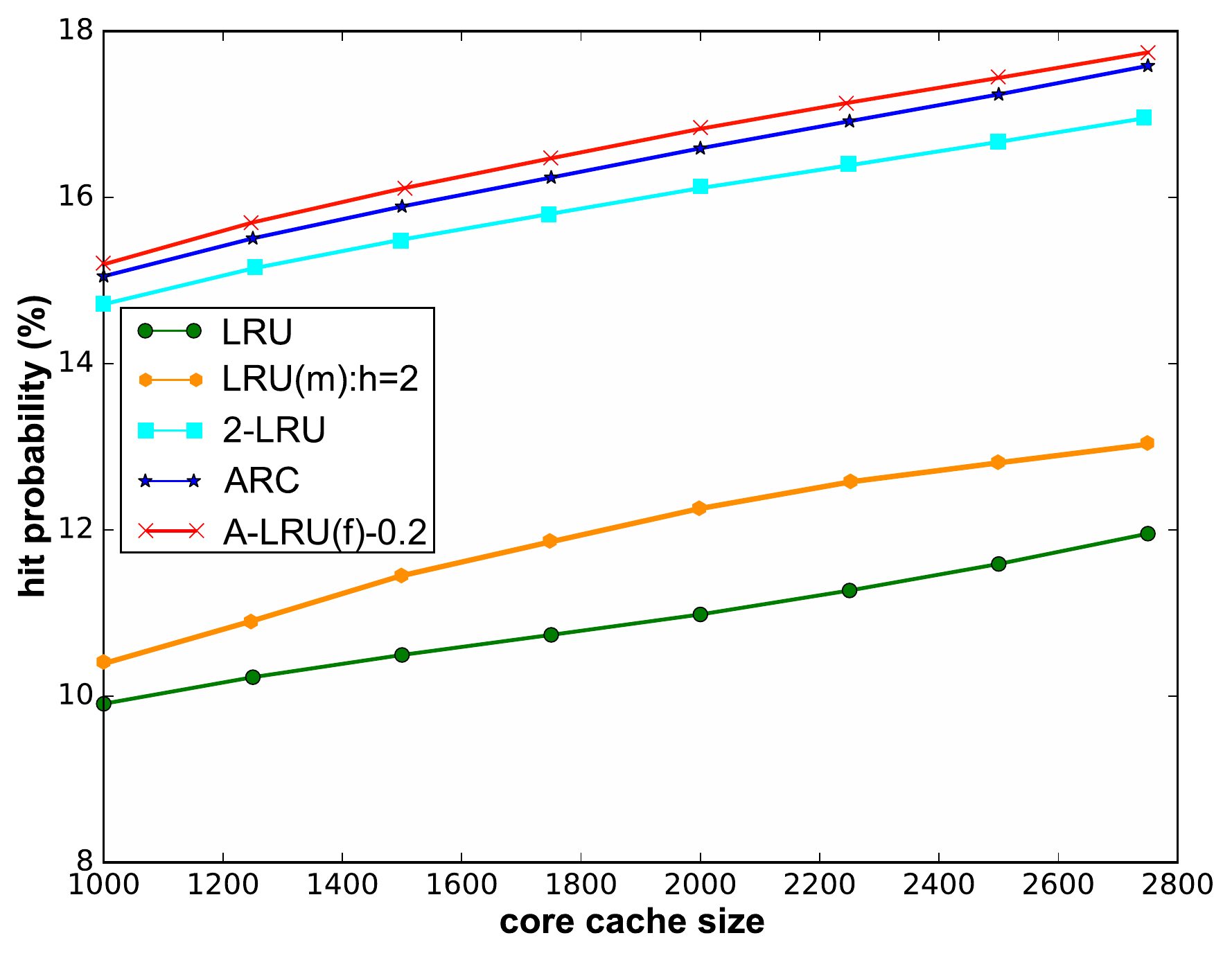}
\caption{Hit probability vs. cache size for various {multi-level} caching algorithms with SD Network trace \cite{bianchi13check} for ICN.}
\label{fig:icn-multi}
\end{minipage}
\vspace{-0.2in}
\end{figure*}

We also conduct experiments on a one-day YouTube trace to illustrate the adaptability of A-LRU. We randomly pick one day from the two-week traces, in which the total number of uniques videos is $3\times 10^5$ and the Zipf distribution parameter $\alpha=0.48$ (but popularity varies with time). Figures~\ref{fig:youtube-day9-p05-single} and~\ref{fig:youtube-day9-p05-multi} depicts the hit probability as a function of the cache ratio.  We see that A-LRU again outperforms all other caching algorithms.

\subsection{ICN Traces}\label{sec:icn-sim}

We run similar experiments using the traces from the IRCache project \cite{bianchi13check}, with attention on data gathered from the SD Network Proxy (the most loaded proxy to which end-users can connect) in Feb. 2013. A detailed study shows that such traces capture regional traffic and exhibit significant non-stationaries due to daily traffic fluctuations. We only considered the traces in the $4$ hour peak traffic period of each day in order to measure the performance expected in the busy hour.   The traces contain $3416817$ to $4121865$ requests for $811827$ to $993711$ different data.  About $67.61\%$ to $69.27\%$ of those data were requested only once during the trace.  We find that $\alpha=0.814$ to $\alpha=0.821$ are the best fits for Zipf distributions.   

Figures~\ref{fig:icn-single} and~\ref{fig:icn-multi} show the overall cache hit probability versus the core cache size. We again observe that there is a significant improvement using A-LRU, especially with $\beta=0.2.$

\section{Conclusion}\label{sec:conclusion}

In this paper, we attempted to characterize the adaptability properties of different caching algorithms when confronted with non-stationary request arrivals.   To begin with, we first considered the stationary distributions of various caching algorithms under a stationary request process, and computed the $\tau$-distance between each one and the optimal content placement in the cache.  We then analyzed the mixing time of each algorithm, to determine how long each one takes to attain stationarity.  By combining both of these metrics, we constructed the \emph{learning error,} which characterizes the tradeoff between speed and accuracy of learning.  The learning error provides insight into the likely performance of each algorithm under non-stationary requests, and using the insights learned we developed a new algorithm, A-LRU, that can adapt to different non-stationary request processes  and consequently has a higher hit probability than any of the standard algorithms that we compared against, under both synthetic and trace-based evaluation.

\appendices
\section{Caching Algorithms}\label{app:algorithms}
\subsection{{Single-Level Caching Algorithms}}
{We consider the conventional algorithms to manage a single cache as shown in Figure~\ref{fig:example} (a), including LRU {\cite{coffman73operating,fill96,king71}}, FIFO {\cite{coffman73operating,king71}}, RANDOM \cite{coffman73operating} and CLIMB \cite{coffman73operating, starobinski01}.  }\\
\noindent{\textbf{LRU:}} {\cite{coffman73operating,fill96,king71}} When there is a request for item $i$, there are two cases: (1) $i$ is not in the cache (cache miss), then $i$ is inserted in the first position in the cache, all other items move back one position, and the item that was in the last position of the cache is evicted; (2) $i$ is in position $j$ of the cache (cache hit), then $i$ moves to the first position of the cache, and all other items that were in positions $1$ to $j-1$ move back one position.\\
\textbf{FIFO:}  {\cite{coffman73operating,king71}} The difference between FIFO and LRU is when a cache hit occurs on an item that was in position $j$. In FIFO, this item does not change its position. \\
\textbf{RANDOM:}  \cite{coffman73operating} The difference between RANDOM and FIFO is when a cache miss occurs, the item is inserted in a random position, and the item that was in this randomly selected position is evicted.\\
\textbf{CLIMB:} \cite{coffman73operating, starobinski01} {The difference between CLIMB and LRU is when a cache hit occurs on an item that was in position $j$. In CLIMB, this item moves up one position to position $j-1$. }

\subsection{{Meta-cache Caching Algorithms}}
${}$\\
\noindent{\textbf{$k$-LRU:}}  $k$-LRU \cite{martina2014,gast16asym} manages a cache of size $m$ by making use of $k-1$ virtual caches, {which only store meta-data to keep track of the recent request history.}  Each cache is ordered such that the item in the $j$-th position of cache $l$ is the $j$-th most-recently-used item among all items in cache $l$. When item $i$ is requested, two events occur: (1) For each cache $l$ in which item $i$ appears, say in position $j$ for cache $l$, then item $i$ moves to the first position of cache $l$ and items in positions $1$ to $j-1$ move back one position; (2) For each cache $l$ in which item $i$ does not appear but appears in cache $l-1$, item $i$ is inserted in the first position of cache $l$, 
all other items of cache $l$ move back one position, and the last item is evicted.

{Note that meta-caches enhance caching performance in terms of hit probability since only popular items that have been requested at least $k$ times can be inserted into real cache. {See Section~\ref{sec:evaluation}.} }

\subsection{{Multi-Level Caching Algorithms}}
${}$\\
We consider a general linear cache network,  which is composed of $h$ caches labeled as $1,\cdots, h$, each with size $m_i\in \mathbb{N}$, $i=1, \cdots, h$.  The total cache size is denoted as $m=\sum_{i=1}^h m_i.$  There are no exogenous requests at caches $2, \cdots, h$.  An item enters the cache network via cache $1$, and will be promoted to a higher index cache whenever there is a cache hit on it.  For this structure, we will consider a class of LRU-based replacement algorithms. We denote it as {LRU($\boldsymbol m$) \cite{gast16asym}}, where $\boldsymbol m=(\boldsymbol m_1, \cdots, \boldsymbol m_h).$
When there is a cache hit on item $k$ that was in position $j$ of cache $i<h$, $k$ moves to the first position of cache $i+1$,  and all other items in cache $i+1$ move back one position, and the last item that was in cache $i+1$ moves to the first position of cache $i$, and all items that were in positions $1$ to $j-1$ of cache $i$ move back one position. When there is a cache hit in position $j$ of cache $h$, item $k$ moves to the first position of cache $h$ and all other items move back one position.  When there is a cache miss on item $k$, it is inserted into the first position of cache $1$ and all other items move back one position, the last item that was in cache $1$ is evicted.

\noindent{\textbf{Adaptive Replacement Cache:}} ARC \cite{megiddo03arc} uses the history of recently evicted items to change its recency or frequency preferences.  Specifically, ARC splits the cache into two parts, $\mathcal{T}_1$ and $\mathcal{T}_2$, which cache items that have been accessed only once, and many times, respectively. Furthermore, ARC maintains two additional lists, $\mathcal{B}_1$ and $\mathcal{B}_2$, to record (LRU-based) eviction history of $\mathcal{T}_1$ and $\mathcal{T}_2$, respectively.  Recency or frequency preferences are adjusted by dynamically changing target sizes of $\mathcal{T}_1$ and $\mathcal{T}_2$ according to eviction histories recorded in $\mathcal{B}_1$ and $\mathcal{B}_2$. In this way, ARC tracks changes in traffic patterns and adjusts the replacement policy to emphasize frequency or recency accordingly.  We refer the reader to \cite{megiddo03arc} for details of dynamic adjustments made in ARC.

\section{Reversibility and Mixing time}\label{app:revers-mixing}
In this section, we provide an argument regarding the relationship between reversibility and mixing time, which was briefly discussed in Section~\ref{sec:reversibility}.

We first define the spectral gap of a Markov chain with state space $\mathcal{S}$ through the Rayleigh quotient and the Dirichlet form.
\begin{definition} \cite{chung05, montenegro06}
For $f, g: \mathcal{S}\rightarrow\mathbb{R},$ let $\mathcal{E}(f, g)=\mathcal{E}_P(f, g)$ denote the Dirichlet form,
\begin{equation}
\begin{aligned}
\mathcal{E}(f, g)&={\langle f, (I-P)g\rangle}_{\pi^*_P}=\sum_{\boldsymbol x,\boldsymbol y}f(\boldsymbol x)(g(\boldsymbol x)-g(\boldsymbol y))P(\boldsymbol x, \boldsymbol y)\pi^*_P(\boldsymbol x).
\end{aligned}
\end{equation}
If $f=g,$ then 
\begin{equation}\label{eq:dirichlet}
\mathcal{E}(f, f)=\frac{1}{2}\sum_{\boldsymbol x,\boldsymbol y}(f(\boldsymbol x)-f(\boldsymbol y))^2P(\boldsymbol x, \boldsymbol y)\pi^*_P(\boldsymbol x).
\end{equation}
\end{definition}
The Rayleigh quotient for any $f:\mathcal{S}\rightarrow\mathbb{R}$, is defined as follows \cite{chung05},
\begin{equation}\label{eq:rayleigh}
R(f)=\frac{\mathcal{E}(f, f)}{\sum_{\boldsymbol x}|f(\boldsymbol x)|^2\pi^*_P(\boldsymbol x)}.
\end{equation}
The spectral gap of the Markov chain with transition matrix $P$ is defined as \cite{chung05}
\begin{equation}\label{eq:spectral-gap}
\gamma^*=\inf_{\substack{f\\ \sum_{\boldsymbol x}f(\boldsymbol x)\pi^*_P(\boldsymbol x)=0}}\frac{R(f)}{2}.
\end{equation}

Next, we discuss the difference between reversible and non-reversible Markov chains, and show how to bound the mixing time of a non-reversible Markov chain by constructing a related reversible Markov chain.  We will use the result later to obtain a bound on the mixing time of the LRU algorithm, which is associated with a non-reversible Markov chain.

Suppose that $P$ is the transition matrix of a non-reversible chain, and $\pi^*_P$ be its corresponding stationary distribution. Consider the time-reversal $P^*$, which is defined by 
\begin{equation}\label{eq:reversal}
\pi^*_P(\boldsymbol x)P^*(\boldsymbol x, \boldsymbol y)=\pi^*_P(\boldsymbol y)P(\boldsymbol y, \boldsymbol x),
\end{equation}
where $\boldsymbol x, \boldsymbol y\in\mathcal{S}.$ 

Then it is easy to check that the additive Markov chain with transition matrix $\frac{P+P^*}{2}$ is reversible \cite{montenegro06}. 
Now,  $\forall\boldsymbol x\in\mathcal{S},$ 
$\pi^*_{P}(\boldsymbol x)=\pi^*_{P^*}(\boldsymbol x),$
where $\pi^*_{P^*}$ is the stationary distribution of Markov chain with transition matrix $P^*$. Therefore, we obtain 
\begin{equation}\label{eq:stationary-relations}
\pi^*_P(\boldsymbol x)=\pi^*_{P^*}(\boldsymbol x)=\pi^*_{\frac{P+P^*}{2}}(\boldsymbol x).
\end{equation}

From Equation~(\ref{eq:dirichlet}), we immediately have \cite{montenegro06} 
\begin{equation}
\mathcal{E}_P(f, f)=\mathcal{E}_{P^*}(f, f)=\mathcal{E}_{\frac{P+P^*}{2}}(f, f),
\end{equation}
Furthermore, by the definition of Rayleigh quotient in Equation~(\ref{eq:rayleigh}) and the spectral gap of a Markov chain in Equation~(\ref{eq:spectral-gap}), we have
\begin{equation}\label{eq:spectral-gap-relations}
\gamma^*_P=\gamma^*_{P^*}=\gamma^*_{\frac{P+P^*}{2}}.
\end{equation}

Therefore, for any non-reversible Markov chain with transition matrix $P$, we can construct a reversible Markov chain with transition matrix $\frac{P+P^*}{2}$ for the spectral gap analysis.  Since these two Markov chains have the same stationary distribution~(\ref{eq:stationary-relations}) and spectral gap~(\ref{eq:spectral-gap-relations}), by~(\ref{eq:eig-mixing}), we can equivalently use the reversible Markov chain  $\frac{P+P^*}{2}$ to bound the mixing time of the non-reversible Markov chain $P$ through applying the existing results on reversible Markov chains.  It is important to note that for this method, one needs the knowledge of the stationary distribution $\pi^*_P(\cdot),$ which is something that we earlier already possess or can accurately bound for the purpose of our analysis of caching algorithms.

\section{Stationary Distribution: LRU}\label{app:station-lru}
The proof of the expression for the steady state probability $\pi_{\text{LRU}}^*(\boldsymbol x)$ is via a probabilistic argument, following \cite{hendricks76}, which essentially uses the Loynes argument 
 to sum the probablities of all possible sample paths to get to a particular state.

{\bf Proof of Theorem~\ref{thm:stationary}:}
\begin{IEEEproof}
We consider the current state $\boldsymbol x,$ and attempt to reconstruct the past history by looking backwards in time.  In order to achieve the current state $\boldsymbol x=(x_1, \cdots, x_m), $ the past history of requests, listed from the most remote to the most recent, must be ordered as follows: \\
($2m$): A last request for item $x_m$ is made;\\
($2m-1$): Requests for item $(x_1, x_2, \cdots, x_{m-1})$ are made;\\
($2m-2$): A last request for item $x_{m-1}$ is made;\\
($2m-3$): Requests for item $(x_1, x_2, \cdots, x_{m-2})$ are made;\\
($2m-4$): A last request for item $x_{m-2}$ is made;\\
($2m-5$): Requests for item $(x_1, x_2, \cdots, x_{m-3})$ are made;\\
$\ldots$\\
($2$): A last request for item $x_2$ is made;\\
($1$): At least one request for item $x_1$ is made. 

The probability that step $2j$ ($2\leq j\leq m)$ occurs with probability $p_{x_j},$ and the probability steps $2j$ and $2j-1$ together is given by 
\begin{equation}
p_{x_j}\sum_{k=0}^{\infty}\left(\sum_{l=1}^{j-1}p_{x_l}\right)^k=\frac{p_{x_j}}{1-\sum_{l=1}^{j-1}p_{x_l}}.
\end{equation}
Note that the last two steps occur with probability $p_{x_2}\sum_{k=1}p_{x_1}^k=p_{x_2}\cdot\frac{p_{x_1}}{1-p_{x_1}}.$  Combing them together, we get the steady state probability of state $\boldsymbol x$ as 
\begin{equation}
\pi_{\text{LRU}}^*(\boldsymbol x)=\frac{\Pi_{i=1}^m p_{x_i}}{(1-p_{x_1})(1-p_{x_1}-p_{x_2})\cdots(1-p_{x_1}-\cdots-p_{x_{m-1}})},
\end{equation}
which gives us the desirable result. 
\end{IEEEproof}

\section{Example for Class and Subclass of Sample Paths}\label{app:example}
Here, we present a more involved example for the definition of \emph{class of sample paths} and \emph{subclass of sample paths} given in Section~\ref{sec:meta-cache}.

\begin{eg}\label{exm1}
When $k=1,$ i.e., the caching algorithm is the base LRU scheme, consider the state $\boldsymbol x=(12)$.  Since there is only one level and two unique items, there are totally $2$ items needed to be fixed to obtain a sample path.  Due to the LRU policy, there is only one valid arrangement $\bm{\hat{x}} = 2\rightarrow 1$ over these two items, so $|\Upsilon(\boldsymbol x)|=1.$  {The unique valid subclass of sample paths $\tilde{\Lambda}( 2\rightarrow 1)$, which is equivalent to} the unique valid class of sample paths $\Lambda( 2\rightarrow 1)$ consists of all sample paths of the form $z\rightarrow 2\rightarrow 1 \cdots \rightarrow 1 \rightarrow 1,$ where $z$ represents any sequence of items drawn from $\mathcal{L}.$ 
\end{eg}


\section{Stationary Distribution: $k$-LRU}\label{app:station-k-lru}
Here, we provide the argument to find the steady state probability of $k$-LRU, given in Equation~(\ref{eq:steady-state-klru}).  Similar to the analysis of LRU, we follow a probabilistic argument.  

{\bf Proof of Theorem~\ref{eq:thm-klru}:}
\begin{IEEEproof}
We consider a particular cache  state $\boldsymbol x=(\boldsymbol x_1,\cdots,\boldsymbol x_k),$ where $\boldsymbol x_j=(x_{(j,1)}, \cdots, x_{(j, m)}),$ and attempt to reconstruct the past history by looking backwards in time.  Note that only the $\boldsymbol x_k$ caches the real data items, while all other levels cache meta-data. 

However, it is not as easy as in the case of LRU to find a particular path to reconstruct the past history.  Essentially, we need to deteremine all the possible paths that result in the state $\boldsymbol x$ under $k$-LRU.  Recall that according to our notation $h(x_{(i,j)})$ is the highest cache level in which the item appears, and that 
\begin{align}
    \mathcal{J}_{x_{(i,j)}}= 
\begin{cases}
    p_{x_{(i,j)}},& \text{if } i= h(x_{(i,j)})\\
    1,              & \text{otherwise.}
\end{cases}
\end{align}
As discussed in Section~\ref{sec:meta-cache},  any sample path $\gamma(\bm{\hat{x}})$  leading to $\boldsymbol x$ must contain an arrangement $\bm{\hat{x}}$ over the set  $\hat{\mathcal{X}}$ in which $h(x_{(i,j)})$ copies of $x_{(i,j)}$ appear (these are the  final requests for that item in that sample path).   {This arrangement is common to all paths in subclass $\tilde{\Lambda}(\bm{\hat{x}})$ and then class $\Lambda(\bm{\hat{x}}).$ }  Also, every class must contain some arrangements of the elements in $\hat{\mathcal{X}}.$  The probability of occurrence of any of these arrangements, which is common to every sample path leading to $\boldsymbol x$ is $\prod_{i=1}^k\left(\prod_{j=1}^{m}\mathcal{J}_{x_{(i,j)}}\right)^i.$ 

Next, consider a particular class of sample paths $\Lambda(\bm{\hat{x}}).$  Denote the sequence of items fixed by the arrangement $\bm{\hat{x}}$ by  $\xi_1, \cdots, \xi_\delta\cdots\xi_{\hat{\mathcal{X}}},$ where $\xi_\delta$ stands for the identity of the $\delta$-th item in this arrangement.  A sample path can be constructed by adding other items from $\mathcal{L}$ between these fixed items in such a way that the end result is $\bm{\hat{x}}$ under $k$-LRU (i.e., it is valid).  Let the probability of requesting these other items for a particular sample path be $\Xi({\gamma{(\boldsymbol x)}}).$  Then the probability of a valid sample path $\gamma(\bm{{x}})$ is simply $\prod_{i=1}^k\left(\prod_{j=1}^{m}\mathcal{J}_{x_{(i,j)}}\right)^i \Xi({\gamma{(\boldsymbol x)}}).$

{Thus, the sum of the probabilities of all the possible sample paths in the subclass $\tilde{\Lambda}(\bm{\hat{x}}),$ is}
\begin{align}
 \sum_{\gamma{(\boldsymbol x)}\in\tilde{\Lambda}(\bm{\hat{x}})} \prod_{i=1}^k\left(\prod_{j=1}^{m}\mathcal{J}_{x_{(i,j)}}\right)^i\Xi({\gamma{(\boldsymbol x)}}).
\end{align}

Following this argument,  {we consider all the possible subclasses of sample path $\tilde{\Lambda}(\bm{\hat{x}})\in\Lambda(\bm{\hat{x}}),$ and all the possible classes of sample path $\Lambda(\bm{\hat{x}})\in\Upsilon(\boldsymbol x),$} and sum all the corresponding probabilities to obtain (\ref{eq:steady-state-klru}).
\end{IEEEproof}

\section{Example for Calculating the Stationary Distribution of $k$-LRU}\label{app:example-k-lru}
Here, we present a more detailed example to illustrate how to calculate the stationary distribution of $k$-LRU given in~(\ref{eq:steady-state-klru}).
\begin{eg}\label{exm3}
Consider $k=1$, i.e., LRU algorithm. There is only one class of sample path, i.e., $|\Upsilon(\boldsymbol x)|=1$ for each state $\boldsymbol x,$ and we pick state $\boldsymbol x=(12)$ as our candidate. The unique class of sample path is $\Lambda( 2\rightarrow 1),$ i.e., the last request for item $2$ should come before the last request for item $1$ in all valid sample paths.  {The unique subclass of sample path is $\tilde{\Lambda}( 2\rightarrow 1),$ i.e., only item $1$ can be requested between item $2$ and item $1$.} Each such sample path $\gamma(\bm{x})$ must satisfy one of the following conditions: 
\begin{itemize}
\item there is no request between $2$ and $1$ in $\gamma(\bm{x}).$  In this case  $\Xi({\gamma{(\boldsymbol x)}})=1;$
\item there is only one request for item $1$ between $2$ and $1$ in $\gamma(\bm{x}).$  In this case $\Xi({\gamma{(\boldsymbol x)}})=p_1;$
\item there are two requests for item $1$ between $2$ and $1$ in $\gamma(\bm{x}).$  In this case $\Xi({\gamma{(\boldsymbol x)}})=p_1^2;$
\item $\cdots$
\item there are $k$ requests of item $1$ between $2$ and $1$ in $\gamma(\bm{x}).$  In this case $\Xi({\gamma{(\boldsymbol x)}})=p_1^k. $
\end{itemize}
Note that only item $1$ can be requested between $2$ and $1$ in all valid sample paths $\gamma(\bm{x})$ using LRU.  Also $k=1,\cdots, \infty.$

Therefore, the steady state probabilities for state $\boldsymbol x=(12)$ under LRU is the sum of probabilities of all sample paths $\gamma(\bm{x}),$ i.e., $\pi_{\text{LRU}}^*(\boldsymbol x=(12))=p_1p_2\sum_{k=0}^{\infty}p_1^k=\frac{p_1p_2}{1-p_1},$ which is consistent with the classical result presented in Theorem~\ref{thm:stationary}.
\end{eg}

\section{Stationary Distribution: LRU($\boldsymbol m$)}\label{app:station-lru-m}
{\bf Proof of Theorem~\ref{thm:thm-lrum}:}
\begin{IEEEproof}
We consider the current state  $\boldsymbol x=(\boldsymbol x_1,\cdots,\boldsymbol x_h),$ where $\boldsymbol x_i=(x_{(i,1)}, \cdots, x_{(i, m_i)}),$ and attempt to reconstruct the past history by looking backwards in time. 

However, it is not as easy as LRU to find a particular path to reconstruct the past history.  Essentially, we need to determine all the possible paths that result in the state $\boldsymbol x$ under LRU($\boldsymbol m$).   As discussed earlier, any sample path $\gamma{(\hat{\boldsymbol x})}$ leading to $\boldsymbol x$ must contain an arrangement $\bm{\hat{x}}$ over the set $\hat{\mathcal{X}}$ in which $l(x_{(i, l)})$ copies of $x_{(i, l)}$ appear (these are the final requests for that item in that sample path). This arrangement is common to all paths in this class $\Lambda(\bm{\hat{x}}).$ Also, every class must contain some arrangement of the elements in $\hat{\mathcal{X}}.$  The probability of occurrence of any of these arrangements, which is common to every sample path leading to $\boldsymbol x$ is $\prod_{i=1}^h\left(\prod_{j=1}^{m_i}p_{x(i,j)}\right)^i.$  

Next, consider a particular class of sample paths $\Lambda(\bm{\hat{x}}).$  Denote the sequence of items fixed by the  $\bm{\hat{x}}$ by  $\xi_1, \cdots, \xi_\delta, \cdots, \xi_{\hat{\mathcal{X}}},$ where $\xi_\delta$ stands for the identity of the $\delta$-th item in this arrangement.  A sample path can be constructed by adding other items from $\mathcal{L}$ between these fixed items in such a way that the end result is $\bm{\hat{x}}$ under LRU($\boldsymbol m$) (i.e., it is valid).  Let the probability of requesting these other items for a particular sample path be $\Xi({\gamma{(\boldsymbol x)}}).$  Then the probability of a valid sample path $\gamma(\bm{{x}})$ is simply $\prod_{i=1}^h\left(\prod_{j=1}^{m_i}p_{x(i,j)}\right)^i \Xi({\gamma{(\boldsymbol x)}}).$ 
 
{Thus, the sum of the probabilities of all the possible sample paths in the subclass $\tilde{\Lambda}(\bm{\hat{x}}),$ is}
\begin{align}
\sum_{\gamma{(\boldsymbol x)}\in\Lambda(\bm{\hat{x}})}\prod_{i=1}^h\left(\prod_{j=1}^{m_i}p_{x(i,j)}\right)^i\Xi({\gamma{(\boldsymbol x)}}).
\end{align} 

Following this argument, {we consider all the possible subclasses of sample path $\tilde{\Lambda}(\bm{\hat{x}})\in\Lambda(\bm{\hat{x}}),$ and all the possible classes of sample path $\Lambda(\bm{\hat{x}})\in\Upsilon(\boldsymbol x),$} and sum all the corresponding probabilities to obtain (\ref{eq:steady-state-lrum}).
\end{IEEEproof}

\section{Hit Probability}\label{sec:hit-prob}
A primary performance measure in caching systems is the hit probability. One can derive the hit probability once we have the stationary distribution.  We illustrate how to compute this standard performance measure in this section, and will present hit probability as a special case of our new more general metrics that we will develop in the next few sections.

Denote the hit probability of algorithm $A$ as $H_m^A$ and let $F_m^A=1-H_m^A$ be the miss probability under IRM.

By the ergodic theorem, the miss probability $F_m^A$ is equal to the stationary probability of a miss. 

\begin{theorem}
Under IRM, we have 
\begin{equation}\label{eq:miss-prob-general}
F_m^A=\sum_{\boldsymbol x\in\mathcal{S}}\left(1-{\sum_{k(A)}\sum_{j=1}^{m_{k(A)}}p_{x_{k(A), j}}}\right)\pi_A^*(\boldsymbol x),
\end{equation}
where $\pi_A^*(\boldsymbol x)$ is given in~(\ref{eq:steadystate}), ~(\ref{eq:steady-state-klru}), 
 and $x_{k(A), j}$ stands for the identity of the item at position $j$ in level $k(A)$ under algorithm $A$. In particular, $k(A)=1$ for all the conventional {single-level} algorithms considered in Section~\ref{sec:sta-distri} with $m_1=m$; $k(A)=k$ for $k$-LRU, as discussed in Section~\ref{sec:meta-cache}, and {$k(A)=1,\cdots, h$ for LRU($\boldsymbol m$), as discussed in Section~\ref{sec:multi-level-cache}.}
\end{theorem}

\section{Mixing time of LRU}\label{app:mixing-lru}
Since the transition matrix $P^{\text{LRU}}$ of LRU is non-reversible, we consider the reversible Markov chain with transition matrix $\frac{P^{\text{LRU}}+P^{\text{LRU},*}}{2},$ where $P^{\text{LRU},*}$ is the time reversal chain of $P^{\text{LRU}},$ and $\pi^*_{P^{\text{LRU}}}=\pi^*_{\frac{P^{\text{LRU}}+P^{\text{LRU},*}}{2}}$. 

{\bf Proof of Theorem~\ref{thm:upper-bound-lru}:}
\begin{IEEEproof}
We first calculate an upper and lower bound on the stationary probability of LRU.   Based on the steady state probability of LRU given in~(\ref{eq:steadystate}), it is obvious that the state that achieves the minimum steady state probability is $\boldsymbol x=(n, n-1, \cdots, n-m+1).$  In other words, the least $m$ popular items are cached in the cache with the least popular item in the first position, and proceeding onwards until the $m$-th cache spot.  Similarly,  the state that achieves the maximum steady state probability is $\boldsymbol x^*=(1, 2, \cdots, m).$ Then we have
\begin{align}
&\pi^{\text{LRU}}(\boldsymbol x)\geq\frac{\prod_{j=n-m+1}^n p_j}{\prod_{j=1}^{m-1}\left(1-\sum_{l=1}^{j}p_{n-l+1}\right)}\triangleq\pi_{\text{min}}^{\text{LRU}},\displaybreak[0]\\
&\pi^{\text{LRU}}(\boldsymbol x)\leq\frac{\prod_{j=1}^m p_j}{\prod_{j=1}^{m-1}\left(1-\sum_{l=1}^{j}p_{l}\right)}\triangleq\pi_{\text{max}}^{\text{LRU}}.
\end{align}

Therefore, the congestion of LRU is upper bounded by
\begin{align}\label{eq:congestion-lru-general}
\rho^{\text{LRU}}&\leq\frac{1}{\pi_{\text{min}}^{\text{LRU}}P_{\text{min}}}\cdot\Gamma^2\cdot(\pi_{\text{max}}^{\text{LRU}})^2=\frac{\Gamma^2\left(\prod_{j=1}^m p_j\right)^2\prod_{j=1}^{m-1}\left(1-\sum_{l=1}^{j}p_{n-l+1}\right)}{p_n\left(\prod_{j=1}^{m-1}\left(1-\sum_{l=1}^{j}p_{l}\right)\right)^2\prod_{j=n-m+1}^n p_j},
\end{align}
where $\Gamma=\Theta(n^m)$ is the number of states.  Then by Lemma~\ref{lem:mixing-methodology}, we have 
\begin{align}\label{eq:mixingtime-cheeger-app-lru}
t_{\text{mix}}^{\text{LRU}}=&O\Bigg(\frac{n^{4m}\left(\prod_{j=1}^m p_j\right)^4\prod_{j=1}^{m-1}\left(1-\sum_{l=1}^{j}p_{n-l+1}\right)^2}{p_n^2\left(\prod_{j=1}^{m-1}\left(1-\sum_{l=1}^{j}p_{l}\right)\right)^4\prod_{j=n-m+1}^n p_j^2}\ln\left(\frac{\prod_{j=1}^{m-1}\left(1-\sum_{l=1}^{j}p_{n-l+1}\right)}{\prod_{j=n-m+1}^n p_j}\right)\Bigg).
\end{align}
\end{IEEEproof}

We consider the specialization of the result in Theorem~\ref{thm:upper-bound-lru} to a Zipf popularity distribution.
 
{\bf Proof of Corollary~\ref{cor:upper-bound-lru}:} 
\begin{IEEEproof}
By the Proof of Theorem~\ref{thm:upper-bound-lru}, we can get $\pi_{\text{min}}^{\text{LRU}}$ and $\pi_{\text{max}}^{\text{LRU}}$ under the Zipf popularity distribution. Then, by~(\ref{eq:congestion-rand-general}), the congestion of LRU is upper bounded as 
\begin{equation}\label{eq:cong-lru}
\begin{aligned}
\rho^{\text{LRU}}=O(n^{(2\alpha+1) m+1}).
\end{aligned}
\end{equation}

From~(\ref{eq:mixingtime-cheeger-app-lru}), we obtain
\begin{equation}\label{eq:upper-bound-lru}
t_{\text{mix}}^{\text{LRU}}=O(n^{(4\alpha+2) m+2}\ln n).
\end{equation}
\end{IEEEproof}

\section{Mixing time of RANDOM and FIFO}\label{app:mixing-rand-fifo}
Since it is easy to verify that these algorithms have reversible Markov chains, we now can characterize the mixing time of RANDOM and FIFO through~(\ref{eq:mixing-reversible}).

{\bf Proof of Theorem~\ref{thm:upper-bound-rand}:}
\begin{IEEEproof}
We first calculate an upper and lower bound on the stationary probability of RANDOM.   Based on the steady state probability of RANDOM given in~(\ref{eq:steadystate}), it is obvious that the state that achieves the minimum steady state probability is $\boldsymbol x=(n, n-1, \cdots, n-m+1).$  In other words, the least $m$ popular items are cached in the cache with the least popular item in the first position, and proceeding onwards until the $m$-th cache spot.  Similarly,  the state that achieves the maximum steady state probability is $\boldsymbol x^*=(1, 2, \cdots, m).$ Then we have
\begin{align}
\pi^{\text{RANDOM}}(\boldsymbol x)\geq\frac{\prod_{i=n-m+1}^n p_i}{\sum_{\boldsymbol x\in\mathcal{S}^\prime} \Pi_{i=1}^m p_{x_i}}\geq \frac{\prod_{i=n-m+1}^n p_i}{\Gamma\prod_{i=1}^m p_i}\triangleq\pi_{\text{min}}^{\text{RANDOM}},
\end{align}
\begin{align}
\pi^{\text{RANDOM}}(\boldsymbol x)\leq\frac{\prod_{i=1}^m p_i}{\sum_{\boldsymbol x\in\mathcal{S}^\prime} \Pi_{i=1}^m p_{x_i}}\leq \frac{\prod_{i=1}^m p_i}{\Gamma\prod_{i=n-m+1}^n p_i}\triangleq\pi_{\text{max}}^{\text{RANDOM}},
\end{align}
where $\Gamma=\Theta(n^m)$ is the number of states. 

Therefore, the congestion of RANDOM is upper bounded by
\begin{align}\label{eq:congestion-rand-general}
\rho^{\text{RANDOM}}&\leq\frac{1}{\pi_{\text{min}}^{\text{RANDOM}}P_{\text{min}}}\cdot\Gamma^2\cdot(\pi_{\text{max}}^{\text{RANDOM}})^2=\frac{\Gamma}{p_n}\left(\frac{\prod_{i=1}^m p_i}{\prod_{i=n-m+1}^n p_i}\right)^3.
\end{align}

Then by Lemma~\ref{lem:mixing-methodology}, we have 
\begin{align}\label{eq:mixingtime-cheeger-app-rand}
t_{\text{mix}}^{\text{RANDOM}}=O\Bigg(\frac{n^{2m}\left(\prod_{i=1}^m p_i\right)^6}{p_n^2\left(\prod_{i=n-m+1}^n p_i\right)^6}\ln\left(\frac{n^m\prod_{i=1}^m p_i}{\prod_{i=n-m+1}^n p_i}\right)\Bigg).
\end{align}
\end{IEEEproof}

We consider the specialization of the result in Theorem~\ref{thm:upper-bound-rand} to a Zipf popularity distribution.
 
{\bf Proof of Corollary~\ref{cor:upper-bound-rand}:} 
\begin{IEEEproof}
By the Proof of Theorem~\ref{thm:upper-bound-rand}, we can get $\pi_{\text{min}}^{\text{RANDOM}}$ and $\pi_{\text{max}}^{\text{RANDOM}}$ under the Zipf popularity distribution. Then, by~(\ref{eq:congestion-rand-general}), the congestion of RANDOM is upper bounded as 
\begin{equation}\label{eq:cong-rand}
\begin{aligned}
\rho^{\text{RANDOM}}=O(n^{(3\alpha+1) m+1}).
\end{aligned}
\end{equation}

From~(\ref{eq:mixingtime-cheeger-app-rand}), we obtain
\begin{equation}\label{eq:upper-bound-rand}
t_{\text{mix}}^{\text{RANDOM}}=O(n^{(6\alpha+2) m+2}\ln n).
\end{equation}
\end{IEEEproof}

\section{Mixing time of CLIMB}\label{app:mixing-climb}
We follow the same arguments as we did in the proof of Theorem~\ref{thm:upper-bound-rand} to characterize the mixing time of CLIMB.

{\bf Proofs of Theorem~\ref{thm:upper-bound-climb}:}
\begin{IEEEproof}
\begin{equation}\label{eq:congestion-climb-general}
\begin{aligned}
&\rho^{\text{CLIMB}}\leq\frac{\Gamma}{p_n}\left(\frac{\prod_{i=1}^m p_i^{m-i+1}}{\prod_{i=n-m+1}^n p_i^{i-n+m}}\right)^3.
\end{aligned}
\end{equation}

Similarly, by Lemma~\ref{lem:mixing-methodology}, we have
\begin{equation}\label{eq:mixingtime-cheeger-app-climb}\small
t_{\text{mix}}^{\text{CLIMB}}=O\Bigg(\frac{n^{2m}\left(\prod_{i=1}^m p_i^{m-i+1}\right)^6}{p_n^2\left(\prod_{i=n-m+1}^n p_i^{i-n+m}\right)^6}\ln\left(\frac{n^m\prod_{i=1}^m p_i^{m-i+1}}{\prod_{i=n-m+1}^n p_i^{i-n+m}}\right)\Bigg).
\end{equation}
\end{IEEEproof}

Again, we consider the Zipf popularity distribution. 

{\bf Proof of Corollary~\ref{cor:upper-bound-climb}:}
\begin{IEEEproof}
Under the Zipf distribution, by~(\ref{eq:congestion-climb-general}), the congestion of CLIMB is upper bounded by 
\begin{equation}\label{eq:cong-climb}
\begin{aligned}
\rho^{\text{CLIMB}}=O(n^{\frac{3\alpha m(m+1)}{2}+m+1}).
\end{aligned}
\end{equation}

From~(\ref{eq:mixingtime-cheeger-app-climb}), we have 
\begin{equation}\label{eq:upper-bound-climb}
\begin{aligned}
&t_{\text{mix}}^{\text{CLIMB}}=O(n^{3\alpha m(m+1)+2m+2}\ln n).
\end{aligned}
\end{equation}
\end{IEEEproof}

\section{Mixing time of $k$-LRU}\label{app:mixing-k-lru}
Since the Markov chain associated with $k$-LRU is non-reversible, we use the form of the stationary distribution of $k$-LRU in Theorem~\ref{eq:thm-klru}, and the mixing time expression of Corollary~\ref{cor:mixing-methodology} to obtain the desired result.

Finally, we consider the Zipf distribution. In order to obtain the result given in Corollary~\ref{cor:upper-bound-klru}, we need first prove the following result. 
\begin{lemma}\label{lem:conv1}
For $0<\alpha<1,$ as $n\rightarrow\infty,$ we have
$\left(\frac{1}{1-\max_{\boldsymbol x\in\mathcal{S}, \delta}\left(\sum_{j\in\chi^{\Lambda^{\text{max}}(\boldsymbol x)}_{(\delta, \delta+1)}} p_j\right)}\right)^{k(k+1)m/2-1}\rightarrow 1.$ 
\end{lemma}
{\bf Proof of Lemma~\ref{lem:conv1}:}
\begin{IEEEproof}
Due to the policy of $k$-LRU, at most $m-1$ unique items can be requested between any two fixed items on the path, i.e., $|\chi^{\Lambda^{\text{max}}(\boldsymbol x)}_{(\delta, \delta+1)}|\leq m-1,$ $\forall\delta.$ Hence $\max_{\boldsymbol x\in\mathcal{S}, \delta}\left(\sum_{j\in\chi^{\Lambda^{\text{max}}(\boldsymbol x)}_{(\delta, \delta+1)}} p_j\right)=\sum_{i=1}^{m-1}p_i,$ then, we have
\begin{align}
&\left(\frac{1}{1-\max_{\boldsymbol x\in\mathcal{S}, \delta}\left(\sum_{j\in\chi^{\Lambda^{\text{max}}(\boldsymbol x)}_{(\delta, \delta+1)}} p_j\right)}\right)^{k(k+1)m/2-1}\nonumber\displaybreak[0]\\
&\leq\left(\frac{1}{1-\sum_{i=1}^{m-1}p_i}\right)^{k(k+1)m/2-1}\nonumber\displaybreak[1]\\
&\leq\left(\frac{1}{1-A\int_1^{m-1}\frac{1}{x^{\alpha}}dx}\right)^{k(k+1)m/2-1}\nonumber\displaybreak[2]\\
&\stackrel{(a)}{=}\left(\frac{1}{1-\left(\frac{1-\alpha}{n^{1-\alpha}}+O\left(\frac{1}{n}\right)\right)\frac{(m-1)^{1-\alpha}-1}{{1-\alpha}}}\right)^{k(k+1)m/2-1}\nonumber\displaybreak[3]\\
&\leq\left(\frac{1}{1-\left(\frac{1-\alpha}{n^{1-\alpha}}+O\left(\frac{1}{n}\right)\right)\frac{(m-1)^{1-\alpha}}{{1-\alpha}}}\right)^{k(k+1)m/2-1}\nonumber\\
&=\left(\frac{1}{1-\left(\frac{m-1}{n}\right)^{1-\alpha}-O\left(\frac{1}{n}\right)}\right)^{k(k+1)m/2-1}\nonumber\\
&\leq\left(1+\frac{\left(\frac{m-1}{n}\right)^{1-\alpha}}{1-\left(\frac{m-1}{n}\right)^{1-\alpha}-O\left(\frac{1}{n}\right)}\right)^{k(k+1)m/2-1}\nonumber\\
&=\left(1+\frac{\left(m-1\right)^{1-\alpha}}{n^{1-\alpha}-\left(m-1\right)^{1-\alpha}-O\left(n^{-\alpha}\right)}\right)^{k(k+1)m/2-1}\nonumber\\
&\stackrel{(b)}{=}1+\frac{\left(m-1\right)^{1-\alpha}}{n^{1-\alpha}-\left(m-1\right)^{1-\alpha}-O\left(n^{-\alpha}\right)}\left(k(k+1)m/2-1\right)\nonumber\\
&-O\left(\frac{\left(m-1\right)^{2-2\alpha}}{\left(n^{1-\alpha}-\left(m-1\right)^{1-\alpha}-O\left(n^{-\alpha}\right)\right)^2}(k(k+1)m/2-1)^2\right)\nonumber\\
&\rightarrow 1, \qquad \text{as $n\rightarrow\infty$, for any $0<\alpha<1.$}
 \end{align}
where $(a)$ holds true since $A=\frac{1-\alpha}{n^{1-\alpha}}+O\left(\frac{1}{n}\right)$ as $n\rightarrow\infty;$ and $(b)$ is based on Taylor expansion.  
\end{IEEEproof}

Given the result in Lemma~\ref{lem:conv1}, we are now ready to characterize the mixing time of $k$-LRU under the Zipf popularity distribution. 

{\bf Proof of Corollary~\ref{cor:upper-bound-klru}:} 
\begin{IEEEproof}
Under the Zipf distribution, we have $p_i=A/i^{\alpha}$ and $A=\frac{1-\alpha}{n^{1-\alpha}}+O\left(\frac{1}{n}\right).$ With the result in Lemma~\ref{lem:conv1} and~(\ref{eq:thm-mixing-klru}),  we have we have (note that we consider the order sense, so we ignore constants in the following)

\begin{align}
&t_{\text{mix}}^{\text{$k$-LRU}}=O\Bigg(\frac{n^{4km+{4(m-1)[k(k+1)m/2-1]}}}{(\frac{A^m}{n^{m\alpha}})^{k(k+1)} \frac{A^2}{n^{2\alpha}}}\left(\frac{A^m}{(m!)^{\alpha}}\right)^{4k}\ln\frac{1}{(\frac{A^m}{n^{m\alpha}})^{k(k+1)/2}}\Bigg)\nonumber\displaybreak[1]\\
&=O\Bigg(A^{4km-mk(k+1)-2}n^{mk(k+1)\alpha+2\alpha+4km+4(m-1)[k(k+1)m/2-1]}\ln\left(\frac{n^{m\alpha}}{A^m}\right)^{k(k+1)/2}\Bigg)\nonumber\displaybreak[3]\\
&=O\Bigg(n^{-(1-\alpha)(4km-mk(k+1)-2)+mk(k+1)\alpha+2\alpha}\cdot n^{4km+4(m-1)[k(k+1)m/2-1]}\ln\frac{n^{k(k+1)m\alpha/2}}{A^{k(k+1)m/2}}\Bigg)\nonumber\\
&=O\Bigg(n^{(k+1)k(2m-1)m+4(k\alpha-1)m+6}\ln n\Bigg). 
\end{align}

\end{IEEEproof}

\section{Mixing time of LRU($\boldsymbol m$)}\label{app:mixing-lrum}
{\bf Proof of Theorem~\ref{thm:upper-bound-lrum}:} 
\begin{IEEEproof}
Consider the stationary probability $\pi_{\text{LRU($\boldsymbol m$)}}^*(\boldsymbol x)$ given in Equation~(\ref{eq:steady-state-lrum}), we first obtain its upper bound $\pi^*_{\text{max}}$ and lower bound $\pi^*_{\text{min}}$. We omit the subscript ${}_{.\text{LRU($\boldsymbol m$)}}$ for brevity. 

First, we characterize $\pi^*_{\text{max}}.$   Recall that we denote the sequence of items fixed by the arrangement $\bm{\hat{x}}$ by  $\xi_1, \cdots, \xi_\delta, \cdots, \xi_{|\hat{\mathcal{X}}|},$ where $\xi_\delta$ stands for the identity of the $\delta$-th item in this arrangement.  A sample path contains other items from $\mathcal{L}$ between these fixed items in such a way that the end result is $\bm{\hat{x}}$ under LRU($\boldsymbol m$), with the probability of requesting these other items being $\Xi({\gamma{(\boldsymbol x)}}).$ 

We obtain $\pi^*_{\text{max}}$ by taking the maximum of each part in Equation~(\ref{eq:steady-state-lrum}): (i) We take the maximum of the product form $\prod_{i=1}^h\left(\prod_{j=1}^{m_i}p_{x(i,j)}\right)^i$ over all states $\boldsymbol x\in\mathcal{S};$ (ii) {We consider the maximum number of subclasses of sample paths and classes of sample paths; } 
and (iii) We maximize the sum of  $\Xi({\gamma{(\boldsymbol x)}})$ over all states, sample paths and classes.

Then we have 
\begin{align}
\pi^*(\boldsymbol x)&\leq\max_{\boldsymbol x\in\mathcal{S}}\left(\prod_{i=1}^h\left(\prod_{j=1}^{m_i}p_{x(i,j)}\right)^i\right)\cdot|\tilde{\Upsilon}_{\Lambda(\bm{\hat{x}})}(\boldsymbol x)| \cdot |\Upsilon(\boldsymbol x)| \cdot\max_{\boldsymbol x\in\mathcal{S}}\left(\max_{\Lambda(\bm{\hat{x}})}\left(\sum_{\gamma{(\boldsymbol x)}\in\Lambda(\bm{\hat{x}})} \Xi({\gamma{(\boldsymbol x)}})\right)\right). 
\end{align}
There are three terms in the above expression.  We may upper bound the first term by using our assumption that $p_1\geq\cdots\geq p_n.$  It is obvious that 
\begin{align}
&\max_{\boldsymbol x}\left(\prod_{i=1}^h\left(\prod_{j=1}^{m_i}p_{x(i,j)}\right)^i\right)=\left(\prod_{j=1}^{m_h}p_j\right)^h\left(\prod_{j=m_h+1}^{m_h+m_{h-1}}p_j\right)^{h-1}\cdots\left(\prod_{j=m-m_1+1}^{m}p_j\right)^1=\prod_{i=1}^h\left(\prod_{k=1+\sum_{j=1}^{i-1}m_{h-j+1}}^{\sum_{j=1}^i m_{h-j+1}}p_k\right)^{h-i+1},
\end{align}
with $\sum_{j=1}^{i-1}m_{h-j+1}=0$ for $i=1.$ 

{The second term is upper bounded by $(\sum_{j=1}^h jm_j)!=(m_1+2m_2+\cdots+hm_h)!,$ and the third term is upper bounded by $\left(\binom{n}{m-1}\right)^{m_1+\cdots+hm_h-1}$ since at most $m-1$ unique items can be requested between any two fixed items on the arrangement.}

We then consider the third term in the expression.  Now, for any $\bm{x}$
\begin{align}\label{eq:third-term}
\sum_{\gamma{(\boldsymbol x)}\in\Lambda(\bm{\hat{x}})} \Xi({\gamma{(\boldsymbol x)}})
= \prod_{\delta=1}^{|\hat{\mathcal{X}}|} \left(\sum_{\zeta=0}^{\infty} \left(\sum_{j\in\chi^{\Lambda(\bm{\hat{x}})}_{(\delta, \delta+1)}} p_j\right) ^{\zeta}\right).
\end{align}

It is clear that this term is maximized when the number of product terms is in maximum, since each term in the product is greater than or equal to one.  Since there are exactly $m$ unique items in each state $\bm{\hat{x}}$ under LRU($\boldsymbol m$), and the items on the $j$-th level needed to be requested $j$ times to get fixed, thus, the number of items that need to be fixed in a sample path leading to a state $\bm{\hat{x}}$ is $\sum_{j=1}^h jm_j.$ Denote $\Lambda^{\text{max}}(\boldsymbol x)=\arg\max_{\Lambda(\bm{\hat{x}})}\left(\sum_{\gamma{(\boldsymbol x)}\in\Lambda(\bm{\hat{x}})} \Xi({\gamma{(\boldsymbol x)}})\right)$ as the class of sample path that achieves the maximum in the third term~(\ref{eq:third-term}).   Let  $\chi^{\Lambda^{\text{max}}(\boldsymbol x)}_{(\delta, \delta+1)}$ be the set of items that can be requested between the $\delta$-th and $(\delta+1)$-th fixed items on this class of sample path.

Then we have 
\begin{align}\label{eq:upper-stationary-prob}
\max_{\boldsymbol x\in\mathcal{S}}\left(\max_{\Lambda(\bm{\hat{x}})}\left(\sum_{\gamma{(\boldsymbol x)}\in\Lambda(\bm{\hat{x}})} \Xi({\gamma{(\boldsymbol x)}})\right)\right)
= &\max_{\boldsymbol x\in\mathcal{S}}\left(\sum_{\gamma{(\boldsymbol x)}\in\Lambda^{\text{max}}(\boldsymbol x)}\Xi(\gamma{(\boldsymbol x))}\right)\nonumber\displaybreak[0]\\
\stackrel{(a)}{=}& \max_{\boldsymbol x\in\mathcal{S}}\left( \prod_{\delta=1}^{\sum_{j=1}^h jm_j-1}\left(\sum_{\zeta=0}^{\infty} \left(\sum_{j\in\chi^{\Lambda^{\text{max}}(\boldsymbol x)}_{(\delta, \delta+1)}} p_j\right) ^{\zeta}\right)\right)\nonumber\displaybreak[1]\\
\stackrel{(b)}{\leq}&\prod_{\delta=1}^{\sum_{j=1}^h jm_j-1}\left(\sum_{\zeta=0}^{\infty}\left(\max_{\boldsymbol x\in\mathcal{S}, \delta}\left(\sum_{j\in\chi^{\Lambda^{\text{max}}(\boldsymbol x)}_{(\delta, \delta+1)}} p_j\right)\right)^{\zeta}\right)\nonumber\displaybreak[2]\\
=&\left(\frac{1}{1-\max_{\boldsymbol x\in\mathcal{S}, \delta}\left(\sum_{j\in\chi^{\Lambda^{\text{max}}(\boldsymbol x)}_{(\delta, \delta+1)}} p_j\right)}\right)^{\sum_{j=1}^h jm_j-1}.
\end{align}
Note that in~(\ref{eq:upper-stationary-prob}), (a) follows from the discussion above, and (b) is true since we take the maximum of the probabilities over all the items that can be requested between any two fixed items over all possible states, sample paths and classes. 

Hence, 
\begin{align}
\pi^*(\boldsymbol x)\leq&\prod_{i=1}^h\left(\prod_{k=1+\sum_{j=1}^{i-1}m_{h-j+1}}^{\sum_{j=1}^i m_{h-j+1}}p_k\right)^{h-i+1} (m_1+\cdots+hm_h)!\cdot\left(\binom{n}{m-1}\right)^{m_1+\cdots+hm_h-1}\nonumber\displaybreak[0]\\
&\cdot \left(\frac{1}{1-\max_{\boldsymbol x\in\mathcal{S}, \delta}\left(\sum_{j\in\chi^{\Lambda^{\text{max}}(\boldsymbol x)}_{(\delta, \delta+1)}} p_j\right)}\right)^{\sum_{j=1}^h jm_j-1}\triangleq \pi^*_{\text{max}}.
\end{align}

Next, we characterize $\pi^*_{\text{min}}$. We obtain $\pi^*_{\text{min}}$ by taking the minimum of each part in Equation~(\ref{eq:steady-state-lrum}): (i) We take the minimum over the product form $\prod_{i=1}^h\left(\prod_{j=1}^{m_i}p_{x(i,j)}\right)^i$ over all states $\boldsymbol x\in\mathcal{S};$  (ii) {We consider the minimum number of subclass of sample path and the minimum number of class of sample path, i.e., $|\tilde{\Upsilon}_{\Lambda(\bm{\hat{x}})}(\boldsymbol x)|=1$ and $|\Upsilon(\boldsymbol x)|=1;$}  and (iii) We minimize the sum of  $\Xi({\gamma{(\boldsymbol x)}})$ over all states, sample paths and classes.

Then we have
\begin{align}
\pi^*(\boldsymbol x)&\geq\min_{\boldsymbol x\in\mathcal{S}}\left(\prod_{i=1}^h\left(\prod_{j=1}^{m_i}p_{x(i,j)}\right)^i\right)\cdot 1
 \cdot  \min_{\boldsymbol x\in\mathcal{S}}\left(\min_{\Lambda(\bm{\hat{x}})}\left(\sum_{\gamma{(\boldsymbol x)}\in\Lambda(\bm{\hat{x}})} \Xi({\gamma{(\boldsymbol x)}})\right)\right).
\end{align}
Again, there are three terms in the above expression. We may lower bound the first term by using our assumption that $p_1\geq\cdots\geq p_n.$
It is obvious that 
\begin{align}
&\min_{\boldsymbol x\in\mathcal{S}}\left(\prod_{i=1}^h\left(\prod_{j=1}^{m_i}p_{x(i,j)}\right)^i\right)=\left(\prod_{j=1}^{m_h}p_{n-j+1}\right)^h\cdots\left(\prod_{j=m-m_1+1}^{m}p_{n-j+1}\right)^1=\prod_{i=1}^h\left(\prod_{k=1+\sum_{j=1}^{i-1}m_{j}}^{\sum_{j=1}^i m_{j}}p_{n+k-m}\right)^i,
\end{align}
with $\sum_{j=1}^{i-1}m_j=0$ for $i=1.$
The second term is already lower bounded by $1$. 

We then consider the third term in the expression.  Again, by~(\ref{eq:third-term}), we know that this term is greater than or equal to one (which happens when all terms equal to one). Since we only consider one class of sample path, we fix the items of the current state (which leads to the first term), and then consider all the possible requests between each two fixed items.  Since we want to lower bound the third term, we consider the case where there are no further requests between any two fixed items, i.e., $ \min_{\boldsymbol x\in\mathcal{S}}\left(\min_{\Lambda(\bm{\hat{x}})}\left(\sum_{\gamma{(\boldsymbol x)}\in\Lambda(\bm{\hat{x}})} \Xi({\gamma{(\boldsymbol x)}})\right)\right)=1.$

Hence, we have 
\begin{align}
\pi^*(\boldsymbol x)\geq\min_{\boldsymbol x\in\mathcal{S}}\left(\prod_{i=1}^h\left(\prod_{j=1}^{m_i}p_{x(i,j)}\right)^i\right)]
=\prod_{i=1}^h\left(\prod_{k=1+\sum_{j=1}^{i-1}m_{j}}^{\sum_{j=1}^i m_{j}}p_{n+k-m}\right)^i\triangleq\pi^*_{\text{min}},
\end{align}
with $\sum_{j=1}^{i-1}m_j=0$ for $i=1.$

Therefore, given that $\Gamma=O(n^m),$ we have
\begin{align}
\rho&\leq\frac{1}{\pi^*_{\text{min}}P_{\text{min}}}\cdot\Gamma^2\cdot(\pi^*_{\text{max}})^2\nonumber\displaybreak[0]\\
&=O\Bigg(\frac{n^{2m}\cdot n^{2(m-1)(m_1+\cdots+hm_h-1)}}{\prod_{i=1}^h\left(\prod_{k=1+\sum_{j=1}^{i-1}m_{j}}^{\sum_{j=1}^i m_{j}}p_{n+k-m}\right)^i\cdot p_n}\Bigg(\prod_{i=1}^h\left(\prod_{k=1+\sum_{j=1}^{i-1}m_{h-j+1}}^{\sum_{j=1}^i m_{h-j+1}}p_k\right)^{h-i+1} \nonumber\displaybreak[1]\\
&\qquad\qquad\qquad\cdot \left(\frac{1}{1-\max_{\boldsymbol x\in\mathcal{S}, \delta}\left(\sum_{j\in\chi^{\Lambda^{\text{max}}(\boldsymbol x)}_{(\delta, \delta+1)}} p_j\right)}\right)^{\sum_{j=1}^h jm_j-1} \Bigg)^2\Bigg).
\end{align}

Then follow Corollary, we have
\begin{align}
t_{\text{mix}}^{\text{LRU($\boldsymbol m$)}}&=O\Bigg(\frac{n^{4m+4(m-1)(m_1+\cdots+hm_h-1)}}{\prod_{i=1}^h\left(\prod_{k=1+\sum_{j=1}^{i-1}m_{j}}^{\sum_{j=1}^i m_{j}}p_{n+k-m}\right)^{2i}\cdot p_n^2} 
 \cdot\left(\prod_{i=1}^h\left(\prod_{k=1+\sum_{j=1}^{i-1}m_{h-j+1}}^{\sum_{j=1}^i m_{h-j+1}}p_k\right)^{h-i+1} \right)^4\nonumber\displaybreak[0]\\
&\cdot\left( \left(\frac{1}{1-\max_{\boldsymbol x\in\mathcal{S}, \delta}\left(\sum_{j\in\chi^{\Lambda^{\text{max}}(\boldsymbol x)}_{(\delta, \delta+1)}} p_j\right)}\right)^{\sum_{j=1}^h jm_j-1} \right)^4\ln\frac{1}{\prod_{i=1}^h\left(\prod_{k=1+\sum_{j=1}^{i-1}m_{j}}^{\sum_{j=1}^i m_{j}}p_{n+k-m}\right)^i}\Bigg).
\end{align}
\end{IEEEproof}

Finally, we consider the Zipf distribution. In order to obtain the result given in Corollary~\ref{cor:upper-bound-lrum}, we need first prove the following result. 
\begin{lemma}\label{lem:conv2}
$\left(\frac{1}{1-\max_{\boldsymbol x\in\mathcal{S}, \delta}\left(\sum_{j\in\chi^{\Lambda^{\text{max}}(\boldsymbol x)}_{(\delta, \delta+1)}} p_j\right)}\right)^{\sum_{j=1}^h jm_j-1}\rightarrow 1,$ for $0<\alpha<1,$ as $n\rightarrow\infty.$
\end{lemma}
{\bf Proof of Lemma~\ref{lem:conv2}:} 
\begin{IEEEproof}
Due to the policy of LRU($\boldsymbol m$), at most $\max\{m_1, \cdots, m_h\}-1\leq m-1$ unique items can be requested between any two fixed items on the path, i.e., $|\chi^{\Lambda^{\text{max}}(\boldsymbol x)}_{(\delta, \delta+1)}|\leq m-1,$ $\forall\delta.$  Hence $\max_{\boldsymbol x\in\mathcal{S}, \delta}\left(\sum_{j\in\chi^{\Lambda^{\text{max}}(\boldsymbol x)}_{(\delta, \delta+1)}} p_j\right)=\sum_{i=1}^{m-1}p_i,$ then, we have
\begin{align}
&\left(\frac{1}{1-\max_{\boldsymbol x\in\mathcal{S}, \delta}\left(\sum_{j\in\chi^{\Lambda^{\text{max}}(\boldsymbol x)}_{(\delta, \delta+1)}} p_j\right)}\right)^{\sum_{j=1}^h jm_j-1}\nonumber\\
&\leq\left(\frac{1}{1-\sum_{i=1}^{m-1}p_i}\right)^{\sum_{j=1}^h jm_j-1}\nonumber\\
&\leq\left(\frac{1}{1-A\int_1^{m-1}\frac{1}{x^{\alpha}}dx}\right)^{\sum_{j=1}^h jm_j-1}\nonumber\\
&=\left(\frac{1}{1-\left(\frac{1-\alpha}{n^{1-\alpha}}+O\left(\frac{1}{n}\right)\right)\frac{(m-1)^{1-\alpha}-1}{{1-\alpha}}}\right)^{\sum_{j=1}^h jm_j-1}\nonumber\displaybreak[0]\\
&\leq\left(\frac{1}{1-\left(\frac{1-\alpha}{n^{1-\alpha}}+O\left(\frac{1}{n}\right)\right)\frac{(m-1)^{1-\alpha}}{{1-\alpha}}}\right)^{\sum_{j=1}^h jm_j-1}\nonumber\displaybreak[1]\\
&=\left(\frac{1}{1-\left(\frac{m-1}{n}\right)^{1-\alpha}-O\left(\frac{1}{n}\right)}\right)^{\sum_{j=1}^h jm_j-1}\nonumber\displaybreak[2]\\
&\leq\left(1+\frac{\left(\frac{m-1}{n}\right)^{1-\alpha}}{1-\left(\frac{m-1}{n}\right)^{1-\alpha}-O\left(\frac{1}{n}\right)}\right)^{\sum_{j=1}^h jm_j-1}\nonumber\displaybreak[3]\\
&=\left(1+\frac{\left(m-1\right)^{1-\alpha}}{n^{1-\alpha}-\left(m-1\right)^{1-\alpha}-O\left(n^{-\alpha}\right)}\right)^{\sum_{j=1}^h jm_j-1}\nonumber\\
&=1+\frac{\left(m-1\right)^{1-\alpha}}{n^{1-\alpha}-\left(m-1\right)^{1-\alpha}-O\left(n^{-\alpha}\right)}\left(\sum_{j=1}^h jm_j-1\right)\nonumber\\
&-O\left(\frac{\left(m-1\right)^{2-2\alpha}}{\left(n^{1-\alpha}-\left(m-1\right)^{1-\alpha}-O\left(n^{-\alpha}\right)\right)^2}\left(\sum_{j=1}^h jm_j\right)^2\right)\nonumber\\
&\rightarrow 1, \qquad \text{as $n\rightarrow\infty$, for any $0<\alpha<1.$}
 \end{align}
 \end{IEEEproof}

Given the result in Lemma~\ref{lem:conv2}, we are now ready to characterize the mixing time of LRU($\boldsymbol m$) under the Zipf popularity distribution. 

{\bf Proof of Corollary~\ref{cor:upper-bound-lrum}:} 
\begin{IEEEproof}
Under the Zipf distribution, we have $p_i=A/i^{\alpha}$ and $A=\frac{1-\alpha}{n^{1-\alpha}}+O(\frac{1}{n}).$  With the result in Lemma~\ref{lem:conv2} and~(\ref{eq:upper-bound-lrum}),  we have (Note that we consider the order sense, so for some constants term, we just ignore them in the followings)
\begin{align}
t_{\text{mix}}^{\text{LRU($\boldsymbol m$)}}
&=O\Bigg(\frac{1}{\frac{A^{2(m_1+\cdots+hm_h)}}{n^{2(m_1+\cdots+hm_h)\alpha}}\cdot \frac{A^2}{n^{2\alpha}}}\cdot n^{4m}\cdot A^{4(m_1+\cdots+hm_h)}\cdot n^{4(m-1)(m_1+\cdots+hm_h-1)}\cdot \ln\frac{1}{\frac{A^{(m_1+\cdots+hm_h)}}{n^{(m_1+\cdots+hm_h)\alpha}}}\Bigg)\nonumber\displaybreak[0]\\
&=O\Bigg(A^{2(m_1+\cdots+hm_h)-2}\cdot n^{2(m_1+\cdots+hm_h)\alpha+2\alpha+4m}\cdot n^{4(m-1)(m_1+\cdots+hm_h-1)}\cdot\ln\frac{1}{\frac{A^{(m_1+\cdots+hm_h)}}{n^{(m_1+\cdots+hm_h)\alpha}}}\Bigg)\nonumber\displaybreak[1]\\
&=O\Bigg(n^{-(1-\alpha)(2(m_1+\cdots+hm_h)-2)}\cdot n^{2(m_1+\cdots+hm_h)\alpha+2\alpha+4m}\cdot n^{4(m-1)(m_1+\cdots+hm_h-1)}\cdot\ln\frac{n^{(m_1+\cdots+hm_h)\alpha}}{A^{(m_1+\cdots+hm_h)}}\Bigg)\nonumber\displaybreak[2]\\
&=O\Bigg(n^{(4m+4\alpha-6)(m_1+2m_2+\cdots+hm_h)+6}\ln n\Bigg),
\end{align}
since $m\geq 1$ and $h\leq m$, the power $(4m+4\alpha-6)(m_1+2m_2+\cdots+hm_h)+6>0$ for any $m\geq 1.$
\end{IEEEproof}

 \begin{figure}
\centering
\includegraphics[width=0.5\columnwidth]{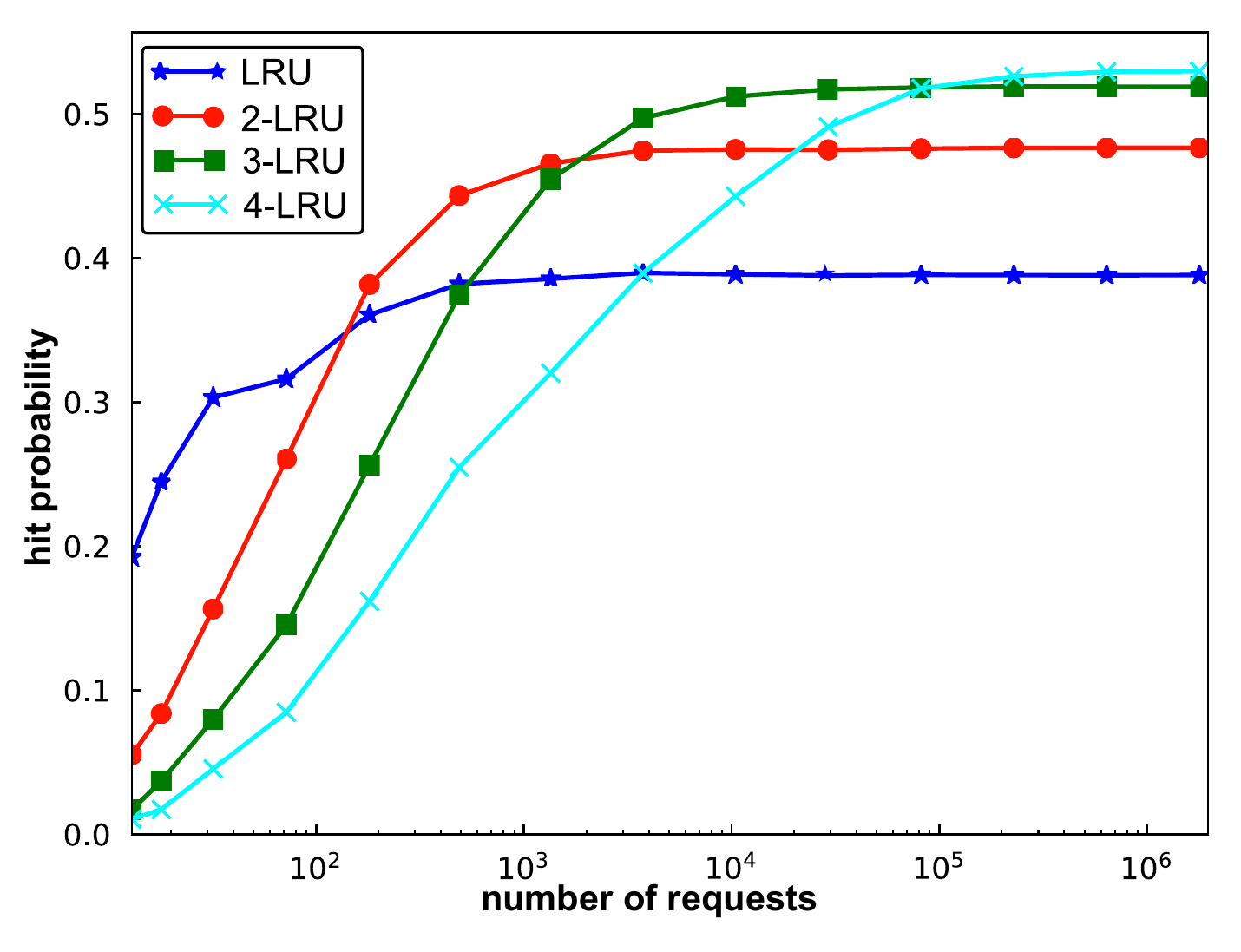}
\caption{Hit probability of meta-caching algorithms under IRM arrivals.}
\label{fig:hit-prob-time-levels}
\vspace{-0.2in}
\end{figure}

\section{Impact of levels on the performance}\label{app:levels}
We characterize the impact of the number of levels on the performance, shown in Figure~\ref{fig:hit-prob-time-levels}. We compare the hit probability of $k$-LRU for $k=1, 2, 3, 4$ with $(n, m)=(50, 10).$ We observe that as the number of cache levels $k$ increases, the resulting algorithm achieve a higher hit probability (i.e., higher accuracy) at the expense of much larger number of requests (i.e., larger mixing time), which is consistent with the mixing time analysis of $k$-LRU shown in Section~\ref{sec:k-lru-mixing}.

%

\bibliographystyle{IEEEtran}
\bibliography{refs}

\end{document}